\documentclass[aps,pre,floatfix]{revtex4}
\usepackage{epsf}
\usepackage{subfigure}
\usepackage{amsmath}
\usepackage{amssymb}
\usepackage{graphicx}
\usepackage{epstopdf}
\def\bPhi{\hbox{\boldmath{$\Phi$}}}
\def\bG{\hbox{\boldmath{$\Gamma$}}}

\begin{document}
%\draft

\newcommand{\be}{\begin{equation}}
\newcommand{\ee}{\end{equation}}
\newcommand{\bea}{\begin{eqnarray}}
\newcommand{\eea}{\end{eqnarray}}

\title{\bf Hamiltonian dynamics, nanosystems, and nonequilibrium statistical mechanics}

\author{Pierre Gaspard}
\affiliation{Center for Nonlinear Phenomena and Complex Systems,\\
Universit\'e Libre de Bruxelles, Code Postal 231, Campus Plaine, 
B-1050 Brussels, Belgium}

\begin{abstract}
An overview is given of recent advances in nonequilibrium
statistical mechanics on the basis of the theory of
Hamiltonian dynamical systems and in the perspective
provided by the nanosciences.
It is shown how the properties of
relaxation toward a state of equilibrium can be derived
from Liouville's equation for Hamiltonian dynamical systems.
The relaxation rates can be conceived in terms of
the so-called Pollicott-Ruelle resonances.
In spatially extended systems, the transport
coefficients can also be obtained from the Pollicott-Ruelle resonances.  
The Liouvillian eigenstates associated with these
resonances are in general singular
and present fractal properties.
The singular character of the nonequilibrium states
is shown to be at the origin of the positive entropy
production of nonequilibrium thermodynamics.
Furthermore, large-deviation dynamical relationships are obtained
which relate the transport properties to the
characteristic quantities of the microscopic dynamics
such as the Lyapunov exponents, the Kolmogorov-Sinai
entropy per unit time, and the fractal dimensions.
We show that these large-deviation
dynamical relationships belong to the same family of
formulas as the fluctuation theorem, as well as a
new formula relating the entropy production
to the difference between an entropy per unit time
of Kolmogorov-Sinai type and a time-reversed
entropy per unit time.  The connections to the
nonequilibrium work theorem
and the transient fluctuation theorem are also discussed.
Applications to nanosystems are described.  
\end{abstract}

\noindent Lecture notes for the International Summer School \\
{\it Fundamental Problems in Statistical Physics XI}\\
(Leuven, Belgium, September 4-17, 2005)

\vskip 0.5 cm

\maketitle

\section{Introduction}
\label{Intro}

Traditionally, statistical mechanics deals with macroscopic systems
containing a number of particles equal to the Avogadro number
$N_{\rm A}=6.0221415 \times 10^{23}$.  However, this number
has an anthropocentric origin since it relates 
the atoms to human artefacts such as the kilogram.
Therefore, the Avogadro number does not represent a fundamental scale
with respect to the atoms (except perhaps in biology).
The question remains of the scale at which a phenomenon
admits a statistical description.
In this regard, statistical mechanics is currently challenged
by the nanosciences which provide many examples of systems
- often out of equilibrium - of size intermediate between
the atoms and the macrosystems and which clearly require
a statistical description.  Examples of such nanosystems are the following:
\begin{itemize}
\item{} Electronic transport in semiconducting quantum dots, 
as well as molecules, conducting polymers, or carbon nanotubes
of a few dozen to hundred of nanometers where the conductance
is given by the Landauer-B\"uttiker formula.
\item{} Quantum corrals of 10-20 nm built with
the scanning tunneling microscope (STM)
by positioning iron atoms
on the surface of copper at 4 K  \cite{Eigler,Heller}.  
\item{} Diffusion of atoms and molecules on surfaces
studied by field ion microscopy (FIM) \cite{Gomer}
and STM \cite{Denmark,Ehrlich,Lauhon-Ho}.
\item{}
Atomic and molecular clusters
with diameters of a few nanometers
such as Na$_n$ where melting
has been studied as a function of the size $n$
 \cite{Haberland}.
\item{} Large molecules such as polymers and RNA molecules
studied thanks to atomic force microscopy (AFM) and optical tweezers \cite{Phys.Today,CRJSTB05}.
\item{} Rotary nanomotors 
of about $10^{-18}$ Watt
have been constructed by isolating
the F$_1$-ATPase protein complex from
mitochondria and gluing an actin filament or a gold bead
to the axis of the motor \cite{Noji,Yasuda}. 
\item{}
Artificial electric motors of the size
of about 300 nm manufactured 
with an axis made of a multiwalled carbon nanotube \cite{Zettl,Bourlon}.
The carbon nanotube plays the role of a bearing
for the motor.  This bearing is characterized by dynamic friction
which dissipates energy.
Energy is also dissipated by dynamic friction in
the telescoping translational motion
of multiwalled carbon nanotube \cite{Cumings-Zettl,Servantie}.
\item{}
The active motion of nanodroplets on surfaces 
covered by tensio-active chemical
reactants \cite{nanodroplet1,nanodroplet2}.  
\item{}
Oscillating reactions at nanometric tips of
field electron (FEM) and field ion (FIM) microscopes \cite{Kruse}.
These chemical clocks are the nanometric analogues
of the macroscopic Belousov-Zhabotinsky oscillating
reaction.  
\end{itemize}

One of the most fascinating recent results from the
nanosciences is that self-organization can already
manifest itself at the scale of a few nanometers
and this under both equilibrium or nonequilibrium conditions.
The Glansdorff-Prigogine dissipative structures
showed that self-organization is possible in macroscopic
systems under nonequilibrium constraints, such as
the Rayleigh-B\'enard convection rolls
and Turing patterns in the Belousov-Zhabotinsky 
chemical reaction \cite{Glansdorff-Prigogine}.
Recent discoveries in nanosciences
show that the structuration and complexification of matter 
already start at the nanoscale.
The formation of micelles of about 10 nm 
in liquid mixtures or nanoclusters of similar sizes at solid-gas
or solid-liquid interfaces are examples of self-organization
at the nanoscale.  The micelles or interfacial nanoclusters already exist
as equilibrium structures of colloidal phases, but their formation
by nucleation from a homogeneous phase
is a typical nonequilibrium process.
It is clear that the biological world has taken advantage
of such structuration mechanisms to build up
self-reproducible compartimentized organisms with an internal metabolism.
The biological metabolism is a nonequilibrium property which
provides the organism with a relative autonomy with respect to its environment.
The wonderful discoveries of biology clearly show that
the complexification starts at the scale just above the size of atoms
by the formation of biomolecules, proteins, RNA, DNA, which themselves
self-assemble into supramolecular structures such as ribosomes, 
nanomotors, membranes, virus, organelles, bacteria, etc.
Similar structurations without metabolism and self-reproduction 
exists at the nanoscale in the inorganic world
with the fullerenes C$_n$, the nanotubes of carbon,
of vanadium oxides, 
or silicate minerals like chrysotile,
and the zeolites as a few examples.

These results provide a perspective very different from the
traditional view which divides the realm of interest
into the microscopic world of atoms, on the one hand, and the macroscopic world
of bulk phases, on the other hand.  The new perspective provided
by the nanosciences urges the development of
statistical physics in order to explain the equilibrium and nonequilibrium properties of the
intermediate structures.  The understanding of the onset
of biological organization is here a future fundamental challenge.
Of immediate interest is the formulation of theoretical
schemes able to deal with equilibrium and nonequilibrium nanosystems.
We are here at the interface between dynamical systems theory and
traditional statistical mechanics.
Indeed, nanosystems may be sufficiently small to be considered
as dynamical systems in which the motion of the particles is described
by Newton's equations and large enough for the onset of statistical behaviors
such as friction, diffusion, viscosity, and sustained nonequilibrium motions
or oscillations.  The link between dynamical systems theory
and statistical mechanics is thus of immediate interest for nanosciences.
Many studies indeed show that statistical behavior starts
to emerge already in relatively small systems containing
dozen, hundred, or thousand particles depending on the
property of concern.  For instance, the transport coefficients can be calculated
in hard-ball systems with a few dozen particles \cite{VG03a,VG03b}
and the Rayleigh-B\'enard convection rolls
can be simulated by molecular dynamics in systems
of several thousand particles \cite{Mareschal}.

Of fundamental interest are the new concepts coming from
dynamical systems theory \cite{ER85} and which have changed our perspective
about irreversibility and the second law of thermodynamics.
Indeed, the study of chaotic systems in the late eighties and early nineties
have introduced new types of relationships within nonequilibrium physics.
These new relationships concern the large deviations or large fluctuations
that the dynamical properties of a system may undergo during
the time evolution.  In the escape-rate formalism, these large-deviation relationships 
have first been discovered between microscopic quantities such as the Lyapunov exponents
and the Kolmogorov-Sinai entropy per unit time and
the transport coefficients, establishing fundamental connections
between the microscopic dynamics and the irreversible properties
of the system \cite{GN90,GB95,DG95,GD95,TVB96,G98,D99}.  
The Lyapunov exponents characterize the sensitivity to initial conditions
of the underlying microscopic dynamics while the Kolmogorov-Sinai
entropy per unit time measures the degree of dynamical randomness
developed by the trajectories of the system during their time evolution.
Thereafter, these large-deviation relationships were extended
to other situations and become known under the name
of the fluctuation theorem
\cite{ECM93,ES94,GC95,K98,C99,LS99,M99,G04a,AG04,AG05,AG06}.  
Historically, this theorem came out of discussions
during the conference organized in Sardinia in July 1991
where the importance and generality of these large-deviation relationships
became evident \cite{Sardinia}.  Today, these relationships have significantly changed
our understanding of the second law of thermodynamics
by showing how to formulate entropy production
in fluctuating systems, in particular, at the nanoscale \cite{Phys.Today}.
In the way, systematic methods have been developed to carry out 
the {\it ab initio} derivation of the transport properties and
the entropy production of nonequilibrium thermodynamics
from the underlying Hamiltonian dynamics.

The purpose of these lecture notes is to describe these new developments.
In Sec. \ref{Dynamics}, the statistical description is formulated
in terms of Liouville's equation for Hamiltonian systems.
We show how to extract the instrinsic relaxation rates from this equation
after suitable assumptions on the dynamics of the system.
Liouville's equation also lead to the master or kinetic equations
which provide intermediate and efficient description of the
 time evolution.  The concepts of Lyapunov exponents and
 Kolmogorov-Sinai entropy per unit time are presented in Sec.
 \ref{Dynamics}. Section \ref{EscRate} is devoted to
 the escape-rate formalism where a first large-deviation
 relationship is derived which relates the transport coefficients
 to the characteristic quantities of the underlying microscopic dynamics.
 In Sec. \ref{Modes}, we carry out the construction of the hydrodynamic modes
 in the case of diffusion and we show how the entropy
 production of nonequilibrium thermodynamics can be derived
  from the underlying Hamiltonian dynamics.
  The properties of the diffusive modes such as their fractal
  character are here given by further large-deviation relationships
  very similar to the one of the escape-rate formalism.
   In Sec. \ref{FT}, we proceed with 
  systems maintained out of equilibrium by fluxes
  of matter and energy and described by master equations.
  The fluctuation theorem and some of its consequences are derived in
  this framework.  The connections to the escape-rate formula \cite{GN90},
the nonequilibrium work relations \cite{C99,Jarzynski}, and
the transient fluctuation theorem \cite{ES94,ES02} are discussed.
  In Sec. \ref{Info}, a new large-deviation relationship is presented
  which relates entropy production to the difference
  between a newly introduced time-reversed entropy per unit time \cite{G04b}
  and the usual entropy per unit time as previously introduced by 
  Shannon, Kolmogorov, and Sinai.  This new relationship provides
  us with a new interpretation of the second law.
 Applications to nanosystems are given in Sec. \ref{Nano}
 where we deal with friction in carbon nanotubes, biological nanomotors, and oscillating
 chemical reactions at the nanoscale.
 Conclusions are drawn in Sec. \ref{Concl}.
 
 \section{Dynamical systems theory}
 \label{Dynamics}
 
 \subsection{Hamiltonian dynamics and the statistical description}
 
 At the microscopic level of description, the natural systems are composed
 of particles with a position, an momentum, and possibly a spin.
 
 Let us consider an {\it isolated system}, which is a system isolated from the external
 world since its initial preparation. Examples are provided
 by atomic and molecular clusters \cite{Haberland}.
 In classical mechanics, the positions and momenta evolve in time $t$
 according to Hamilton's equation
 \be
 \frac{d{\bf r}_a}{dt} = + \frac{\partial H}{\partial {\bf p}_a} 
 \qquad \qquad \frac{d{\bf p}_a}{dt} = - \frac{\partial H}{\partial {\bf r}_a} \label{Hamilton}
  \ee
  The space to which the positions and momenta belong is called
 the phase space: $\bG=({\bf r}_1,{\bf p}_1,{\bf r}_2,{\bf p}_2,...{\bf r}_N,{\bf p}_N)\in{\cal M}$.
 The number of degrees of freedom is the dimension of the position space $f=Nd$.
 The dimension of the phase space is equal to twice this number: 
 $M\equiv {\rm dim}\, {\cal M}=2f=2Nd$.
 For a non-relativistic system without magnetic field, the Hamiltonian function is given by
  \be
  H=\sum_{a=1}^N \frac{{\bf p}_a^2}{2m_a} + U({\bf r}_1,{\bf r}_2,...,{\bf r}_N)
  \label{Hamiltonian.0}
  \ee
  which leads to Newton's equations in terms of the forces ${\bf F}_a = - \partial_{ {\bf r}_a} U$.
  The Hamiltonian time evolution has the property to preserve the phase-space volumes,
 $d\bG = d{\bf r}_1 d{\bf p}_1 d{\bf r}_2  d{\bf p}_2 ... d{\bf r}_N d{\bf p}_N$, 
 which is known as Liouville's theorem. This property originates
 from the unitarity of the underlying quantum mechanics.
Systems invariant under continuous symmetries such as time or spatial translations,
or rotations have constants of motion such as the total energy $E=H$.

The time evolution is symmetric under time reversal
\be
\Theta({\bf r}_1,{\bf p}_1,{\bf r}_2,{\bf p}_2,...{\bf r}_N,{\bf p}_N)=({\bf r}_1,-{\bf p}_1,{\bf r}_2,-{\bf p}_2,...{\bf r}_N,-{\bf p}_N)
\ee
if the Hamiltonian function is an even function of the momenta.  This is the case
for the electromagnetic force ruling the interactions between the atoms and molecules.
 
Hamilton's equations determine the trajectories of the system
 from the initial conditions on the positions and momenta.
 The uniqueness of the trajectories is guaranteed by Cauchy's theorem
 if the Hamiltonian function is sufficiently smooth, which is the
 basis of determinism.  Cauchy's theorem defines a flow in the phase space, $\bG = \bPhi^t(\bG_0)$,
which is the one-parameter Abelian group of time evolution.
 
 However, the Newtonian scheme leaves open
 the determination of the initial conditions which can take any value
 depending on our experimental abilities to manipulate the system.
 The preparation of initial conditions is never the feature of the system left alone
 without external intervention.  Typically, this preparation involves
 a larger system including the system itself with the preparing device,
which is not described by Hamilton's equations (\ref{Hamilton}).  
Of course, there is no doubt that this larger system also admits 
a Newtonian description but again
the preparation of its initial conditions would require a still larger system.
The natural systems for which nearly perfect initial conditions can be
determined are particles such as photons, electrons, atoms, or even molecules
prepared by the formation of beams such as laser beams or molecular beams
or by the confinement and relaxation of particles in traps.
Such devices provide a nearly perfect reproducibility of the initial state.
However, the reproducibility is never perfect and the
initial conditions always differ from one experiment to the next even if
the experimental conditions are under the best possible control
of reproducibility.  Accordingly, a statistical analysis of the initial conditions
is required for instance by looking for the frequency at which
such and such positions and momenta occur during successive
otherwise identical experiments.  The distribution of initial conditions
is thus described by the probability density 
$p_0(\bG) = p_0({\bf r}_1,{\bf p}_1,{\bf r}_2,{\bf p}_2,...{\bf r}_N,{\bf p}_N)$,
which describes the statistical ensemble of initial conditions.

The same problem of initial conditions occurs in the {\it closed
systems} which are systems with a fixed number of particles
but in contact with one or several heat reservoirs, as well as the {\it open systems}
which are systems in contact with one or several particles reservoirs.
Most of the examples listed in the introduction are open systems.
Particles may enter and exit at the boundaries of an open system. The income of particles
depends on the environment and is typically undetermined from
the inside.  The position and momentum of a particle
is thus fixed at the moment of its entrance and its trajectory takes
place inside the system till the exit of the particle.  The exit is determined
by the conditions at the entrance but these later typically differ
from particle to particle so that only a statistical distribution is required
for the description of the influx of particles.  A similar reasoning holds for systems
in contact with heat reservoirs.  These systems should be described by
random (also called stochastic) boundary conditions.  
A deterministic description can be achieved possibly by including
the reservoirs in a larger total system which is isolated.
However, the reservoirs are typically characterized by
a few parameters such as their temperature and chemical potentials,
which suppose statistical conditions at the boundaries.
Therefore, we should expect that the need of a statistical description
is even more important for closed and open systems.

 \subsection{Dynamical instability}

Nevertheless, the success of celestial mechanics and of the space mission
shows that statistical uncertainties on the initial conditions
do not hamper a description in terms of Hamiltonian trajectories.
This question is related to the problem of our ability to predict
the future state of the system from the knowledge of its initial
conditions, which depends on
the stability or instability of the trajectories.
In integrable systems, the constants of motion
limit the phase-space regions where the trajectories evolve
so that nearby trajectories stay close to each other with possible
dephasing in the direction of the flow.  In contrast, chaotic systems
present a sensitivity to initial conditions of exponential type.
Nearby trajectories separate at exponential rates.
The rates of separation are the so-called Lyapunov exponents \cite{ER85}.
There is one positive Lyapunov exponent for each unstable directions in phase space
\be
\lambda_i = \lim_{t\to\infty} \frac{1}{t} \ln \frac{\Vert \delta \bG_i(t) \Vert}{\Vert \delta \bG_i(0) \Vert}
\ee
where $\delta\bG_i(t)$ denotes the infinitesimal separation between the reference
trajectory $\bPhi^t[\bG(0)]$ and the perturbed trajectory $\bPhi^t[\bG(0)+\delta\bG_i(0)]$.
The Lyapunov exponents form a spectrum of exponents ordered from a
maximum to a minimum value.
There are as many Lyapunov exponents as phase-space dimensions $M$.
The symplectic character of Hamilton's equal implies that
the Lyapunov exponents obey the pairing rule that each
negative Lyapunov exponent forms a pair with a corresponding positive
Lyapunov exponent: $\{ \lambda_i, -\lambda_i \}_{i=1}^f$ \cite{G98}.
Liouville's theorem has for consequence that the sum of all
the Lyapunov exponents vanishes: $\sum_{i=1}^{2f} \lambda_i =0$.
Indeed, this sum is the rate of expansion of the phase-space volumes.
Moreover, a pair of Lyapunov exponents vanishes for each constant of motion
because there is no exponential separation of trajectories
in the phase-space direction of the constant of motion and another
exponent vanishes by the pairing rule.  

In a chaotic system, the phase-space distance between a reference trajectory and
a perturbed trajectory increases exponentially at a rate given by the maximum
Lyapunov exponent $\lambda_{\rm max}$.
If the initial conditions are known with a precision $\varepsilon_{\rm initial}$
and if a prediction with a precision $\varepsilon_{\rm final}$ is required,
the time interval of the prediction should not exceed a value called
the Lyapunov time
\be 
t < t_{\rm Lyap} \simeq \frac{1}{\lambda_{\rm max}} \ln \frac{\varepsilon_{\rm final}}
{\varepsilon_{\rm initial}}
\ee
The horizon of prediction is thus of the order of magnitude of the inverse of the
maximum Lyapunov exponent \cite{Nic95}.  Beating the Lyapunov time requires
an increase in the number of decimals or bits of the initial
conditions which goes linearly with the time interval requested.
In systems with many degrees of freedom this problem is enhanced
by the dimensionality of the phase space.  Indeed,
we have an amplification of the initial error
in each unstable phase-space direction at the rate $\lambda_i$
given by the corresponding Lyapunov exponent.
Therefore, the number of decimals known on the set
of initial conditions should grow at a rate proportional 
to the sum of positive Lyapunov exponents.

This growth is slower in systems without exponential sensitivity to initial conditions
such as polygonal billiards.  Elastic reflections on flat walls
lead to a separation of trajectories which is a power of
time.  In a system with $f$ degrees of freedom, 
the number of decimals on the initial conditions
for a prediction after a time interval $t$ should thus grow as
$f \log_{10} t$.  

\begin{figure}[ht]
\centerline{\includegraphics[width=7cm]{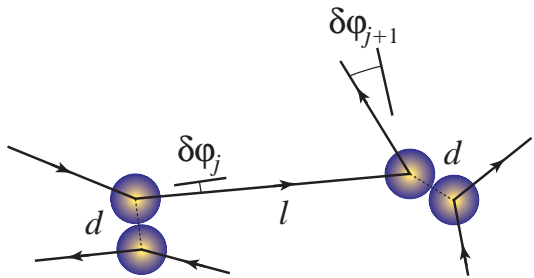}}
\caption{Amplification of some perturbation on the velocity angle $\delta\varphi_j$
for a particle undergoing two successive collisions separated on average with
a distance of the order of the mean free path $l$.
The particles have a diameter $d$. 
\label{fig1}}
\end{figure}

The maximum Lyapunov exponent can be estimated
in systems of interacting particles by
following a reasoning first performed by Krylov \cite{K44}.
We consider air at temperature $T$ and pressure $P$ as a gas of particles
of diameter $d$, the successive collisions
of a particle are separated by a distance of the
order of the mean free path $l \sim 1/(n d^2)$
where $n\simeq P/(k_{\rm B} T)$ is the density of particles.
The intercollisional time $\tau$ is given by the mean free path
divided by the mean velocity $v\sim \sqrt{k_{\rm B}T/m}$: $\tau\sim l/v$.
As shown in Fig. \ref{fig1}, some perturbation on
the velocity angle $\delta\varphi_j$ at the $j^{\rm th}$ collision
is amplified by a factor $l/d$ at the next collision by the defocusing character
of the elastic collisions between spherical particles.
Therefore, the perturbation after $j$ collisions is given by
\be
\delta\varphi_j \sim \delta\varphi_0 \left(\frac{l}{d}\right)^j = \delta\varphi_0 \, {\rm e}^{\lambda_{\rm max} t}
\ee
Since the time for $j$ successive collisions is given by $t\simeq j\tau$,
the maximum Lyapunov exponent is estimated to take
the very high value
\be
\lambda \sim \frac{1}{\tau}\ln\frac{l}{d} \sim 10^{10} \, {\rm digits} \; {\rm sec}^{-1}
\label{Krylov}
\ee
for air at room temperature and pressure.

The systematic numerical calculation of the spectrum of
Lyapunov exponents was undertaken in the eighties and, 
today, it is a routine calculation to obtain the spectrum of Lyapunov exponents
in many-particle systems \cite{PH88,vBD95,DPH96,vBDPD97,vZvBD98,GvB02}.
The chaotic property is thus widespread
among the systems of statistical mechanics.
According to Krylov's formula (\ref{Krylov}), the Lyapunov time of prediction
in such a system is over the order of the intercollisional time, 
i.e., $10^{-10}$ sec/digit in a gas at room temperature and pressure.
This very short time scale is in contrast with the very long Lyapunov times
for chaotic motion in the Solar system.  The shortest known is the
Lyapunov time of $8.5 \; 10^6$ sec/digit for the rotation of Hyperion \cite{G98}.
Therefore, we can understand that the statistical description appears naturally
for the systems of statistical mechanics.

 \subsection{Liouville's equation}
 
 Beyond the time scale of possible trajectory prediction, the statistical ensemble of initial
 conditions tends to spread across large phase-space regions
 and the physical observables no longer take values which can be approximated by
 individual phase-space points, but values obtained by statistical averages
 over the statistical ensemble of points.  The statistical average of a physical observable
 $A(\bG)$ at current time $t$ is given in terms of the ensemble of initial conditions $\bG_0$ by
 \be
 \langle A\rangle_t = \int A(\bPhi^t\bG_0) \; p_0(\bG_0) \; d\bG_0
 = \int A(\bG) \; p_0(\bPhi^{-t}\bG) \; d\bG \equiv \int A(\bG) \; p_t(\bG) \; d\bG
 \label{stat.aver}
 \ee
 where we used Liouville's theorem $d\bG=d\bG_0$ between the initial conditions
 $\bG_0$ and the current phase $\bG=\Phi^t\bG_0$. 
Hence, the probability density evolves in time by the application of
a linear operator called the {\it Frobenius-Perron operator}:
\be
p_t(\bG) = \hat P^t p_0(\bG) \equiv p_0(\bPhi^{-t}\bG)
\ee
which acts by a simple substitution of the current phase-space position
by its preimage under the flow, its preimage giving the initial condition of
the trajectory.  

The flow locally preserves the probability in phase space so that
the probability density obeys the continuity equation
called {\it Liouville's equation} \cite{Nic95}:
\be
\partial_t \, p = - {\rm div}(\dot{\bG} p) = \{ H,p\}  \equiv \hat L p
\label{Liouville.eq}
\ee
where $\{\cdot,\cdot\}$ denotes the Poisson bracket.
Liouville's equation is linear and defined in terms of
the Liouvillan operator
\be
\hat L = \{H, \; \} = \sum_{a=1}^N\left( \frac{\partial H}{\partial {\bf r}_a}\cdot
\frac{\partial }{\partial {\bf p}_a}- \frac{\partial H}{\partial {\bf p}_a}\cdot
\frac{\partial }{\partial {\bf r}_a}\right) 
\ee
If the system is autonomous, the solution of Liouville's equation takes the form
$p_t = {\rm e}^{\hat L t} p_0 = \hat P^t p_0$, 
which shows that the Liouvillian operator is the generator of the 
Frobenius-Perron operator.

Liouville's equation is a partial differential equation and as such requires
boundary conditions to be solved.  Typically, we consider
$N$ particles in a container with elastic reflections on the walls
or $N$ particles on a torus as simulated by molecular dynamics on computers.
However, nothing prevents us to consider other boundary conditions which could
maintain the system in a nonequilibrium steady state \cite{G98}.
This can be achieved by considering appropriate incoming fluxes at the boundaries
of the phase space.  The outgoing fluxes would be fixed by causality.
The total probability within the boundaries $\partial{\cal M}$ of the phase space $\cal M$ obeys the balance equation 
\be
\frac{d}{dt} \int_{\cal M} d\bG \; p = -\int_{\partial{\cal M}} d{\bf A} \cdot \dot{\bG} \; p 
= \underbrace{-\int_{\partial{\cal M}_{\rm in}} d{\bf A} \cdot \dot{\bG} \; p}_{>0} \underbrace{-\int_{\partial{\cal M}_{\rm out}} d{\bf A} \cdot \dot{\bG} \; p}_{<0}
\label{balance.proba}
\ee
where $d{\bf A}$ denotes an infinitesimal element of the area of the boundary.
The vector $d{\bf A}$ points toward the exterior of $\cal M$ and is normal
to its boundary.
The boundary is naturally divided into a part where the vector field $\dot{\bG}$
points toward the interior of $\cal M$ and another to the exterior 
$\partial{\cal M} = \partial{\cal M}_{\rm in} \cup \partial{\cal M}_{\rm out}$.
A nonequilibrium steady state is obtained for {\it flux boundary conditions} with $p\vert_{\partial{\cal M}_{\rm in}}\neq 0$. Such boundary conditions are commonly used in molecular-dynamics simulations
\cite{MKBCN87}.  Instead, we may assume {\it absorbing boundary conditions}
with $p\vert_{\partial{\cal M}_{\rm in}}= 0$.  In this case, the measure 
of the phase-space region $\cal M$ monotonically
decreases, which corresponds to the
escape of trajectories out of this region.  Contrary to a first impression,
such absorbing boundary conditions lead to nontrivial results of great interest for
nonequilibrium considerations as shown in the escape-rate formalism.

 \subsection{The ergodic properties}

If the evolution operator of an autonomous system
is defined on a Hilbert space of phase-space functions,
we can define a unitary group of time evolution
\be
p_t = \hat U^t p_0  = \exp(-i\hat G t ) \; p_0 
\ee
with the Hermitian generator given in terms of the Liouvillian operator by $G = i \hat L$ \cite{CFS82}.
The resolvent of the Hermitian operator is defined by $\hat R(z) \equiv (z-\hat G)^{-1}$
where the variable $z$ can be interpreted as a frequency.
The unitary operator can be recovered from the resolvent
by integration in the plane of the complex variable $z$.
In the following, we introduce the complex variable $s=-iz$,
which has the interpretation of minus a relaxation or decay rate since $\exp(-izt)=\exp(st)$.
We have that ${\rm Im}\, s = -{\rm Re}\, z$ so that a real frequency 
is an imaginary decay rate and ${\rm Re}\, s = {\rm Im}\, z$
so that an imaginary frequency is a real decay rate.

The spectrum of the unitary evolution operator $\hat U^t$
and of its generator $\hat G$ belongs to the axis ${\rm Im}\, z=0$ or ${\rm Re}\, s=0$
of the real frequencies.
This spectrum may include discrete eigenvalues as well as a continuous spectrum.
The ergodic properties determine some of the features of the spectrum.

Around 1887, Boltzmann introduced the property of {\it ergodicity}
which is today defined by requiring that the time average of an observable $A(\bG)$
is equal to its statistical average with respect to some unique invariant probability measure $\mu$
of density $\Psi_0$:
\be
\lim_{T\to\infty} \frac{1}{T} \int_0^T A(\bPhi^t\bG_0) dt = \int A(\bG) \Psi_0(\bG) d\bG = \langle A \rangle
\label{ergodicity.dfn}
\ee
for $\mu$-almost all initial conditions $\bG_0$. We notice that the probability density $\Psi_0$ is a stationary solution of Liouville's equation (\ref{Liouville.eq}) so that it is an eigenstate of the generator, $\hat G\Psi_0=z_0\Psi_0$ or $\hat L\Psi_0=s_0\Psi_0$, corresponding to the vanishing eigenvalue, $z_0=0$ or $s_0=0$.  Since this invariant density is unique, the multiplicity of 
this vanishing eigenvalue is equal to one in the spectrum of an ergodic system.
If $A=I_{\cal A}$ is the indicator function of a phase-space domain ${\cal A}\subset{\cal M}$,
its statistical average defines the probability measure of this domain: 
$\langle A\rangle =\langle I_{\cal A}\rangle = \mu({\cal A})$.
The invariant measure $\mu$ corresponds to a statistical ensemble of initial
conditions which are distributed in the phase space according to the probability density
$\Psi_0$.  Therefore, the introduction of an invariant probability measure
is equivalent to the introduction of a statistical ensemble of initial conditions.
The property of ergodicity guarantees the existence and uniqueness of this invariant measure.
If the boundary conditions do not impose nonequilibrium constraints
on the system, the invariant probability measure defines the state
of thermodynamic equilibrium.

In 1902, Gibbs introduced the property of {\it mixing} which supposes
that each time correlation function between two observables $A$ and $B$
decay at long times as \cite{Gibbs}
\be
\langle A(t) B(0) \rangle \equiv  \int A(\bPhi^t\bG) \, B(\bG) \, \Psi_0(\bG) \, d\bG
\to_{t\to\infty} \langle A\rangle \, \langle B \rangle
\label{mixing.obs}
\ee
The average $\langle\cdot\rangle$ is defined with respect to the invariant measure $\mu$
of density $\Psi_0$ as in Eq. (\ref{ergodicity.dfn}).
If $A$ and $B$ are the indicator functions for the corresponding sets $\cal A$ and $\cal B$,
the condition (\ref{mixing.obs}) can be rewritten as
\be
\lim_{t\to\infty}\mu(\bPhi^{-t}{\cal A}\cap{\cal B})= \mu({\cal A}) \, \mu({\cal B})
\label{mixing}
\ee
which expresses the statistical independency between the occurrence
of the event $\cal A$ at time $t$ and the event $\cal B$ at the initial time $t=0$.
For this to be the case, the spectrum may not include other discrete frequencies
than $z_0=0$, otherwise the time evolution would present undamped oscillations.  
Therefore the mixing property is equivalent
to supposing that the eigenvalue $z_0=0$ or $s_0=0$ is the only one
of the discrete spectrum appearing on the real frequency axis.
The rest of the real frequency axis contains a continuous spectrum.
The mixing property is fundamental for nonequilibrium statistical mechanics
since it guarantees the decay of the time correlation functions.
The mixing property implies the ergodicity (see Refs.  \cite{D99,CFS82}).

 \subsection{The Pollicott-Ruelle resonances}

Beside these properties, the unitary group and its spectrum contains little
information on the characteristic times of the decay or relaxation of the system.
In order to obtain the relaxation rates, we have to perform an analytic continuation
of the resolvent toward the lower half-plane 
of the complex variable $z$ and can pick up the contributions from several
complex singularities which can be poles, branch cuts, or else.
The poles are called the Pollicott-Ruelle resonances \cite{P85,P86,R86a,R86b}.
The sum of the contributions from the resonances, the branch cuts, etc...
gives us an expansion
which is valid for positive times $t>0$ and which defines
the forward semigroup:
\be
\langle A \rangle_t = \int A(\bG) \; \exp(\hat L t) \; p_0(\bG) \; d\bG
\simeq \sum_{\alpha} \langle A \vert \Psi_{\alpha} \rangle \; \exp(s_{\alpha}t) \;
\langle \tilde\Psi_{\alpha}\vert p_0 \rangle + \cdots
\label{forward.expansion}
\ee
The coefficients of this expansion are given by
\bea
\langle A \vert \Psi_{\alpha} \rangle &=& \int A(\bG)^* \, \Psi_{\alpha}(\bG)\, d\bG \label{Psi_a}\\
\langle \tilde\Psi_{\alpha}\vert p_0 \rangle &=& \int \tilde\Psi_{\alpha}(\bG)^* \, p_0(\bG) \, d\bG
\label{tilde_Psi_a}
\eea
in terms of the right- and left-eigenvectors of the Liouvillian operator:
\be
\hat L  \Psi_{\alpha} = s_{\alpha}  \Psi_{\alpha}
\qquad \qquad
\hat L^{\dagger} \tilde\Psi_{\alpha}= s_{\alpha}^*  \tilde\Psi_{\alpha}
\label{eigen}
\ee
The densities $\Psi_{\alpha}(\bG)$ and $\tilde\Psi_{\alpha}(\bG)$
are in general mathematical distributions of Gelfand-Schwartz types \cite{G98}.
Accordingly, the observable $A(\bG)$ and the initial probability density $p_0(\bG)$
must be sufficiently smooth functions for the integrals (\ref{Psi_a}) and (\ref{tilde_Psi_a}) to exist.
In Eq. (\ref{forward.expansion}), the dots denote the contributions beside the simple
exponentials due to the resonances.
These contributions may include Jordan-block structures
if a resonance has a multiplicity $m_{\alpha}$ higher than unity.
In this case, the exponential decay is modified by a power-law
dependence on time as $t^{m_{\alpha}-1} {\rm e}^{s_{\alpha} t}$ \cite{G98}.

On the other hand, the analytic continuation to the upper half-plane
gives a similar expansion
valid for negative times $t<0$, which defines the backward semigroup.
The time-reversal symmetry implies that a singularity located in the upper half-plane at 
$-z_{\alpha}=-is_{\alpha}$ corresponds to each complex singularity in the lower half-plane
of the variable $z_{\alpha}=is_{\alpha}$ and the corresponding eigenstates
are related by the time-reversal operator $\Theta$.

The analytic continuation has the effect of breaking the time-reversal symmetry
and shows that the semigroups are necessarily restricted to one of the two semi-axes of time.

Several important mathematical questions should be answered in order to obtain such expansions:
\begin{itemize}
\item{} What is the spectrum of complex singularities of the resolvent?
\item{} What is the nature of the right- and left-eigenvectors, $\Psi_{\alpha}$ and $\tilde\Psi_{\alpha}$, in each term of the expansion?
For which class of observables $A$ is the right-eigenvector $\Psi_{\alpha}$ defined?
For which class of initial probability density $p_0$ is the left-eigenvector $\tilde\Psi_{\alpha}$ defined?
\item{} Once each term is well defined, does the whole series converge for some classes of observables $A$ and initial probability densities $p_0$?
\end{itemize}

The spectrum of resonances necessarily includes the discrete spectrum of eigenvalues
on the axis of real frequencies and, in particular, the eigenvalue $s_0=0$ if it exists.
Accordingly, the property of {\it ergodicity} guarantees the existence of a unique invariant state 
$\Psi_0$ associated with an eigenvalue $s_0=0$ of multiplicity $m_0=1$.
The property of {\it mixing} implies that $s_0=0$ is the only eigenvalue
on the axis ${\rm Re} s =0$. The measure in Eq. (\ref{mixing}) can be expanded
in terms of the Pollicott-Ruelle resonances, 
which shows that the mixing property holds if there is no eigenvalue
with ${\rm Re}\, s_{\alpha}=0$ other than $s_0=0$.
The non-trivial Pollicott-Ruelle resonances of the forward semigroup
are then located away from the real-frequency axis with ${\rm Re} \; s_{\alpha} < 0$.

There exist most interesting situations in which the leading Pollicott-Ruelle
resonance has a non-vanishing real part ${\rm Re} \; s_0 < 0$.  This happens
for open systems with escape as presented in Sec. \ref{EscRate}, 
as well as for spatially periodic systems as shown in Sec. \ref{Modes}.
In open systems with escape, the leading Pollicott-Ruelle resonance
defines the escape rate $\gamma \equiv -s_0$ corresponding to 
the associated eigenvector
\be
\hat P^t \Psi_0 =  {\rm e}^{s_0 t} \Psi_0 =  {\rm e}^{-\gamma t} \Psi_0
\label{Psi_0}
\ee
which defines a conditionally invariant measure \cite{G98}.

Important results have been obtained for axiom-A systems which is defined as the class of
dynamical systems having the properties that:
\begin{itemize}
\item{} Their non-wandering set $\Omega$ is hyperbolic;
\item{} Their periodic orbits are dense in $\Omega$.
\end{itemize}
The nonwandering set contains all the points for which any neighborhood $\cal U$
of these points has recurrent nonempty intersections ${\cal U}\cap\bPhi^t{\cal U}$ 
at arbitrarily long times $t$.
For these systems, Pollicott and Ruelle have proved 
that the spectrum close to the real axis contains resonances (poles)
which are independent of the observable and initial probability
density within whole classes of smooth enough functions \cite{P85,P86,R86a,R86b}.
Furthermore, it is possible to
show that the Pollicott-Ruelle resonances of an axiom-A system
are given in terms of its unstable periodic orbits.
The periodic-orbit theory is based on a trace formula 
which expresses the trace of the Frobenius-Perron operator
as a so-called Zeta function which is a product
over all the unstable periodic orbits of factors
involving only the period and the instability eigenvalues
of each periodic orbit \cite{CvEc91,G98}.
The Pollicott-Ruelle resonances are thus intrinsic to the system itself if the dynamics
is hyperbolic as it is the case for axiom-A systems.  Therefore, this is a strong
theoretical argument showing that irreversible properties such as the relaxation or decay rates 
do not result from approximations and can be obtained as the generalized eigenvalues
of the Liouvillian dynamics.

A simple example is provided by the Hamiltonian flow in an inverted harmonic potential.
The Hamiltonian of this system can be written in the form \cite{G98}:
\be
H  =  \lambda \; x \; y 
\label{Hxy}
\ee
where $y$ is the conjugate momentum to $x$ and $\lambda$ is the Lyapunov exponent of the unstable equilibrium point at $x=y=0$.
The $x$-coordinate corresponds to the unstable manifold and the $y$-coordinate to the stable one.
The solution of Hamilton's equations are 
$\bPhi^t(x,y)  =  \left({\rm e}^{+\lambda t} x, {\rm e}^{-\lambda t} y\right)$.
In the long-time limit $t\to +\infty$, $\exp(-\lambda t)$ is a small parameter 
in terms of which we can carried out a Taylor expansion 
of the statistical averages (\ref{stat.aver}) according to
\be
\langle A\rangle_t  =  {\rm e}^{-\lambda t} \; \int \; dx' \; dy \; 
p_0\left({\rm e}^{-\lambda t}x',y\right) \ A\left(x', {\rm e}^{-\lambda t}y\right) 
= \sum_{l,m=0}^{\infty}  \; \langle A\vert\Psi_{lm}\rangle \; {\rm e}^{-\lambda (l+m+1)t}
\; \langle \tilde\Psi_{lm}\vert p_0\rangle
\ee
to obtain the decomposition (\ref{forward.expansion}) of the forward semigroup \cite{G98}.
Therefore, the eigenvalues or Pollicott-Ruelle resonances are
given by the integer multiples of the Lyapunov exponent, 
$s_{lm} = - \lambda \; (l+m+1)$ with $l,m=0,1,2,3,...$ 
and the eigenstates can be identified as 
\be
\Psi_{lm}(x,y) = \frac{1}{m!} \; x^l \; \left(-\partial_y\right)^m  \delta(y) \qquad \qquad
\tilde\Psi_{lm}(x,y) = \frac{1}{l!} \; y^m \; \left(-\partial_x\right)^l  \delta(x) \label{RL.eigen}
\ee
The eigenstates are given by the derivatives of the Dirac distribution.  
The support of the right-eigenstates
$\Psi_{lm}$ is the unstable manifold $y=0$, while the left-eigenstates
$\tilde\Psi_{lm}$ have the stable manifold $x=0$ for support.

For non-hyperbolic systems such as intermittent maps or bifurcating equilibrium points,
branch cuts appear which are associated with power-law decays \cite{GNPT95,GT01,TG02}.
The concept of Liouvillian resonance extends to quantum systems \cite{JP97,JP02,P04}.

 \subsection{Time-reversal symmetry and its breaking}
 
 The above construction based on the analytic continuation toward complex
 Liouvillian frequencies breaks the time-reversal symmetry.
 The phenomenon of spontaneous symmetry breaking is well known.
Typically, the solutions of an equation have a lower symmetry than the equation itself.
This is in particular the case for the time-reversal symmetry of Newton's equations
\cite{G05}.

At the level of the flow, the time-reversal symmetry of Newton's equations is expressed by
\be
{\mathbf\Theta}\circ {\mathbf\Phi}^t \circ {\mathbf\Theta} = {\mathbf\Phi}^{-t}
\ee
which means that if the phase-space curve 
\be
{\mathcal C}=\{ {\mathbf\Gamma}_t={\mathbf\Phi}^t({\mathbf \Gamma}_0): t\in{\mathbb R} \}
\ee
is a solution of Newton's equations, then the time-reversed curve
\be
{\mathbf\Theta}({\mathcal C})=\{ \tilde{\mathbf\Gamma}_{t'}={\mathbf\Phi}^{t'}\circ{\mathbf\Theta}({\mathbf\Gamma}_0): {t'}\in{\mathbb R} \}
\ee
starting from the time-reversed initial conditions ${\mathbf\Theta}({\mathbf\Gamma}_0)$ is also a solution of Newton's equations.  The further question is to know if the solution ${\mathcal C}$ of Newton's equations is or is not identical to its time-reversal image 
${\mathbf\Theta}({\mathcal C})$.  If we have a {\it time-reversal symmetric solution:}
\be
{\mathcal C}={\mathbf\Theta}({\mathcal C})
\ee
the solution has inherited of the time-reversal symmetry of the equation.  However,
the solution does not have the time-reversal symmetry of the equation if we have a
{\it time-reversal non-symmetric solution:} 
\be
{\mathcal C}\neq {\mathbf\Theta}({\mathcal C})
\ee
in which case we can speak of the breaking of the time-reversal symmetry by the solution $\mathcal C$.

This reasoning extends to the solutions of Liouville's equation such as the eigenstates $\Psi_{\alpha}$
associated with the resonances.  In general, the equilibrium invariant state $\Psi_0$ of a time-reversal symmetric Hamiltonian corresponding to the eigenvalue $s_0=0$ has the time-reversal symmetry:
\be
\mu_{\rm eq}({\cal A}) = \mu_{\rm eq}(\Theta{\cal A}) 
\ee
However, time reversal maps the genuine resonances onto the anti-resonances
so that the corresponding eigenstates are expected to break the symmetry.
Indeed, the eigenstates $\Psi_{\alpha}$
are smooth in the unstable phase-space directions but singular in the stable directions
as the simple example (\ref{Hxy}) shows with Eq. (\ref{RL.eigen}).
Time reversal maps the stable directions onto the unstable ones so that
the eigenstates of the resonances with ${\rm Re}\, s_{\alpha}\neq 0$ 
typically break the time-reversal symmetry.

In nonequilibrium steady states, it is very important to make the observation that
the invariant probability measure itself breaks in general the time-reversal symmetry \cite{G98}:
\be
\mu_{\rm neq}({\cal A}) \neq \mu_{\rm neq}(\Theta{\cal A}) 
\label{mu.noneq}
\ee
This is due to the fact that the particles are ingoing the system with
a probability distribution which is statistically uncorrelated.
However, the outgoing probability is highly correlated down to very fine
phase-space scales.  The state obtained by time reversal would have
a highly correlated incoming distribution which is uncorrelated by the internal dynamics.
This is never the case because the incoming probability distribution is prepared
by the environment which in general ignore the internal dynamics.
Accordingly, the time-reversal symmetry is explicitly broken 
by nonequilibrium boundary conditions.

In conclusion, the solutions of Newton's or Liouville's equations typically break the time-reversal
symmetry notwithstanding the fact that the equation itself possesses the symmetry.
This phenomenon is known as spontaneous symmetry breaking and it here provides
an elegant explanation for irreversibility which is therefore not incompatible with the symmetry
of the basic equation.
 
 \subsection{From Liouville's equation to master equations}
 \label{hierarchy}
 
 In this section, we show how Liouville's equation can be applied to closed
 and open systems in contact with heat or particle reservoirs.
 
 We first consider a closed subsystem in contact with one or several heat baths of Hamiltonian
 \be
 H=H_{\rm s}(\bG_{\rm s})+ H_{\rm b}(\bG_{\rm b}) + U(\bG_{\rm s},\bG_{\rm b})
 \ee
 where $\bG_{\rm s}$ denote the positions and momenta of the system
 and $\bG_{\rm b}$ those of the baths.  Moreover, we assume that the total Hamiltonian
 takes the form (\ref{Hamiltonian.0}) with interaction potentials
 depending only on positions.
 The heat baths play the role of thermostats for the subsystem.
 Liouville's equation (\ref{Liouville.eq}) holds for the probability density
$p_t(\bG_{\rm s},\bG_{\rm b})$ of the total system.
A reduced description is obtained for the system in terms of the
 reduced probability density
 \be
 f(\bG_{\rm s}) \equiv \int d\bG_{\rm b} \; p(\bG_{\rm s},\bG_{\rm b})
 \ee
 which obeys the equation
 \be
 \partial_t \, f = \{ H_{\rm s}, f \} - \sum_{a=1}^{N_{\rm s}} \frac{\partial}{\partial{\bf p}_{a{\rm s}}}\cdot \int d\bG_{\rm b} \; {\bf F}_{a{\rm s}}(\bG_{\rm s},\bG_{\rm b}) \, p(\bG_{\rm s},\bG_{\rm b})
 \label{Liouville.reduced}
 \ee
 The last terms represent the average of the forces acting on the particles of the system
 due to its interaction with the baths. An example of such a system is Brownian motion in which case
 the system reduces to a single heavy particle and the bath is composed of
 the $N$ particles of the surrounding fluid.  
In the limit of an arbitrarily heavy Brownian particle, the mean force on the Brownian particle can be
approximated by a force due to the friction of the Brownian particle in the fluid and
by a diffusive term originating from the fluctuating Langevin force:
\be
\int dx_1 dx_2 \cdots dx_N \sum_{a=1}^{N}  \; {\bf F}_{a} \, p \simeq -\frac{\zeta}{m}{\bf p} f -\zeta k_{\rm B} 
T \frac{\partial f}{\partial{\bf p}}
\ee
with the notation $x_a = ({\bf r}_a,{\bf p}_a)$.
The friction coefficient is given by Kirkwood formula
\be
\zeta = \frac{1}{3k_{\rm B} T} \int_0^{\infty} \langle {\bf f}(0)\cdot{\bf f}(t)\rangle \; dt
\ee
where ${\bf f}=\sum_{a=1}^N {\bf F}_a$ is the fluctuating force on the Brownian particle \cite{Kirkwood}.
In this example, Eq. (\ref{Liouville.reduced}) becomes the Fokker-Planck equation for Brownian motion,
which has many applications, in particular, in the nanosciences:
 \be
\partial_t \, f = -\frac{{\bf p}}{m}\cdot\frac{\partial f}{\partial{\bf r}} 
-{\bf F}^{\rm (ext)}\cdot\frac{\partial f}{\partial{\bf p}}
+ \frac{\partial}{\partial{\bf p}}\cdot\left( \frac{\zeta}{m}{\bf p} f\right) +\zeta k_{\rm B} 
T \frac{\partial^2 f}{\partial{\bf p}^2}
\label{Fokker-Planck.eq}
\ee

The previous reasoning can be generalized to the case of a system such as a chain of particles coupled 
to two heat baths at two different temperatures to get
 \be
 \partial_t \, f = \{ H_{\rm s}, f \} 
 + \frac{\partial}{\partial{\bf p}_{\rm R}}\cdot\left( \frac{\zeta_{\rm R}}{m}{\bf p}_{\rm R} \, f\right) +\zeta_{\rm R} k_{\rm B} 
T_{\rm R} \frac{\partial^2 f}{\partial{\bf p}_{\rm R}^2}
+ \frac{\partial}{\partial{\bf p}_{\rm L}}\cdot\left( \frac{\zeta_{\rm L}}{m}{\bf p}_{\rm L} \, f\right) 
+\zeta_{\rm L} k_{\rm B} 
T_{\rm L} \frac{\partial^2 f}{\partial{\bf p}_{\rm L}^2}
\label{2baths.eq}
\ee
where the index R (resp. L) refers to the contact with the right-hand (resp. left-hand) heat bath.
The canonical equilibrium state is solution if both temperatures are equal $T_{\rm R}=T_{\rm L}$.

Albeit similar, the situation is more complicated for an open subsystem in contact with one or several 
particle reservoirs, which here play the role of chemiostats as well as of thermostats.
(We can possibly add extra heat baths or thermostats which exchange heat but no particles
with the subsystem.)  We suppose that the subsystem and the reservoirs compose
an isolated total system of volume $\cal V$ containing $\cal N$ identical particles of mass $m$.
The subsystem itself is one part of the total system of volume $V$.  This volume
may be delimited by an external potential which separates the particles between
the subsystem and the reservoirs.  We consider particles with binary interaction for simplicity.
For the total system, Liouville's equation becomes
\be
\partial_t \, p = \sum_{a=1}^{\cal N} \left( -\frac{{\bf p}_a}{m}\cdot\frac{\partial p}{\partial{\bf r}_a} 
-{\bf F}_a^{\rm (ext)}\cdot\frac{\partial p}{\partial{\bf p}_a} \right) -\sum_{1\leq a < b \leq {\cal N}} 
{\bf F}_{ab}\cdot\left( \frac{\partial p}{\partial{\bf p}_a}-\frac{\partial p}{\partial{\bf p}_b}\right)
\ee
where ${\bf F}_a^{\rm (ext)}$ is the external force on the particle No. $a$ and
${\bf F}_{ab}$ is the binary-interaction force on the particle No. $a$ caused by the particle No. $b$.
The particles are identical so that the probability density $p(x_1,...,x_{\cal N})$
should be totally symmetry under the ${\cal N}!$ permutations of the indices.
We introduce the distribution functions
\be
f_N(x_1,x_2,...,x_N) \equiv \frac{{\cal N}!}{({\cal N}-N)!} \int_{{\cal V}-V} dx_{N+1} ...\int_{{\cal V}-V} dx_{\cal N} \; p(x_1,x_2,...,x_{\cal N})
\label{f_N}
\ee
The probabilities $P(N)$ that the system in the volume $V$ contains $N$ particles 
are thus given by the integrals of the functions (\ref{f_N}) over the volume $V^N$
divided by $N!$. They are normalized according to $\sum_{N=0}^{\cal N} P(N) = 1$
as a consequence of the symmetry of the probability density and 
$I_V(x)+I_{{\cal V}-V}(x)=1$.

A hierarchy of equations can be derived for the distribution functions (\ref{f_N})
which are similar but different from the BBGKY equations \cite{Ba75}:
\bea
\partial_t \; f_N = \sum_{a=1}^N\left(-\frac{{\bf p}_a}{m}\cdot\frac{\partial f_N}{\partial{\bf r}_a} 
-{\bf F}_a^{\rm (ext)}\cdot\frac{\partial f_N}{\partial{\bf p}_a}\right)
-\sum_{1\leq a<b\leq N}{\bf F}_{ab}\cdot\left(\frac{\partial f_N}{\partial{\bf p}_a}
-\frac{\partial f_N}{\partial{\bf p}_b}\right) \nonumber \\
- \sum_{a=1}^N\frac{\partial}{\partial{\bf p}_a}\cdot \int_{{\cal V}-V} dx_{N+1}  \; {\bf F}_{a,N+1} \, f_{N+1}
- \Phi_N
\label{hierarchy.eqs}
\eea
where the fluxes of particles
\be
\Phi_N(x_1,...,x_N)\equiv \frac{{\cal N}!}{({\cal N}-N)!} \int_{{\cal V}-V} dx_{N+1} ...\int_{{\cal V}-V} dx_{\cal N} \; \sum_{a=N+1}^{\cal N} \frac{{\bf p}_a}{m}\cdot\frac{\partial p}{\partial{\bf r}_a} 
\label{Phi_N}
\ee
take into account the transitions between configurations with different numbers $N$ of
particles in the volume $V$.  The terms with the mean forces are similar to the terms
encountered for Brownian motion and are due to the Langevin fluctuating force
of the environment on the particles in $V$.  The flux term is specific to open systems
and should describe the transitions $N\rightleftharpoons N+1$
in the number of particles in the volume $V$.  
A hierarchy of differential equations for the probabilities $P(N)$
is naturally associated with Eqs. (\ref{hierarchy.eqs})-(\ref{Phi_N}).
In Eqs. (\ref{hierarchy.eqs})-(\ref{Phi_N}), the flux term includes the fluxes
between the system and the different reservoirs.
If the reservoirs are very large, their discharge is very slow and
a quasi-stationary nonequilibrium state will establish itself,
which is specified by the temperatures 
and the chemical potentials of the reservoirs.
Contrary to the case of the BBGKY equations \cite{Ba75}, we should here not truncate
the hierarchy because the number $N$ of particles in the volume $V$
is a fluctuating variable that is ruled by the complete set of equations.
In order to close the set of equations, we should instead 
express the fluxes as well as the external forces
in terms of the functions $f_N$ themselves by using kinetic
assumptions about the energy and particle exchanges
with the environment.  For the external forces, the treatment is similar
as for Brownian motion and introduces of the friction 
and Langevin forces (related by the fluctuation-dissipation theorem).
For the fluxes, we are led to kinetic assumptions as done for 
the Poisson process (see below).

The present framework can be extended to chemical reactions
in systems with several species of particles.
The system of volume $V$ can be conceived as a nanoreactor
or as the catalytic site of a protein complex.
In particular, equations similar to Eqs. (\ref{hierarchy.eqs})-(\ref{Phi_N}) can be used
to model a nanomotor fuelled by ATP.  In this case, the volume $V$ represents
the spatial domain occupied by the protein including the binding site of ATP which can contain
for instance zero or one fuel molecule.

By using a Markovian approximation,
we are led to master equations such as the Fokker-Planck equation,
the Pauli master equation \cite{Pauli}, or the chemical master equation by Nicolis and coworkers
\cite{NP71,N72,NP77} (see Sec. \ref{FT}).
A master equation is a closed rate equation for the probabilities
to find the system in some coarse-grained states $\omega$
which are defined by integrating the probability density $p_t(\bG)$ over the
corresponding phase-space cells ${\cal C}_{\omega}$.
The symbol $\omega$ denotes for instance a state with given numbers
of molecules for the $c$ different species present in the volume $V$:
$\omega=(N_1,N_2,...,N_c)$.  A finer description would be obtained
by specifying also the phases of these particles in the system $V$.
The Fokker-Planck equation can be obtained in the limit of continuous state variables.
If the master equation provides a faithful representation of the time evolution,
the lowest eigenvalues of the master equation should correspond
to the Pollicott-Ruelle resonances of the underlying deterministic dynamics.

In conclusion, equations (\ref{hierarchy.eqs})-(\ref{Phi_N}) can be used
to model many nonequilibrium systems in nanosciences,
especially, in nanohydrodynamics, chemical and biochemical reactions
in nanosystems such as biological motors.
In this way, master equations as Eqs. (\ref{hierarchy.eqs})-(\ref{Phi_N}) can naturally be derived
from the underlying Hamiltonian microscopic dynamics.
 
 \subsection{The concept of entropy}

The concept of entropy is clearly extra mechanical.
It has been introduced at the beginning of the
XIXth century by the pioneering work of Carnot and Clausius.
In thermodynamics, entropy
is used to express the second law of thermodynamics:
The changes of the entropy
in a system are due to the exchanges $d_{\rm e} S$ with its environment
and to a possible internal production $d_{\rm i} S$ 
which vanishes at equilibrium and is positive out of equilibrium
 \be
 dS = d_{\rm e} S + d_{\rm i} S \qquad\mbox{with}\qquad d_{\rm i}S \geq 0
 \label{entropy.thermo}
 \ee
The second law allows us to distinguish between
 equilibrium and nonequilibrium systems.
 The pressure, the temperature, and the chemical potentials
are uniform at equilibrium.
Each bath or reservoir in contact with a nanosystem
is in general much larger than the system itself and may be assumed
in equilibrium and thus characterized by a temperature and
as many chemical potentials as there are species of particles 
(including the electrons in electronic transport phenomena).
The nanosystem between the reservoirs is in a nonequilibrium state
if the reservoirs have different temperatures or chemical potentials.
For instance, the temperatures of the baths and reservoirs are fixed in 
Eq. (\ref{2baths.eq}) by the terms of coupling to the heat baths.
Similarly, the fluxes (\ref{Phi_N}) depend on the chemical potentials
of the particle reservoirs and fix them in Eqs. (\ref{hierarchy.eqs}).

 Since Carnot and Clausius, an idea
 of partition is associated with the concept of entropy.
 Indeed, the original aim of thermodynamics is to answer
 the question of extracting work by manipulating a gas
 in a steam engine.  A steam engine is composed
 not only of the gas but also of a piston which is
a larger degree of freedom than those of the gas particles
and by which the external manipulation is performed.
Accordingly, entropy concerns the properties
of system when it is manipulated by larger
and external degrees of freedom.
This reasoning suggests that coarse graining is
naturally associated with the concept of entropy
which is the property of a system {\it with respect
to certain types of external manipulations}.
The manipulations we are referring to include
the active preparation of initial conditions, but exclude
mere observations that do not interact with the particles.
From this viewpoint, the partitioning or
coarse graining we refer to is not subjective.
For two centuries of study of entropy and thermodynamics,
we have learned that systems present general
properties which are essential to know for their
external manipulation and which go beyond
the simple motion of the particles from given initial
conditions.  The progress of this last decade
has emphasized this trend.

The status of entropy with respect to the underlying 
Newtonian mechanics was clarified at the end
of the XIXth century and the beginning
of the XXth century.
If we associate an entropy to a microstate $\bG$ we should not
be surprised that the fluctuations spoil a hypothetical monotonic increase
of the entropy and that apparent violations of the second law
would manifest themselves.  This has already been pointed out
by Maxwell when he invented his famous demon.
Boltzmann made a specific proposal for a microscopic
definition of entropy in terms of his famous $H$-function
in dilute gases.  Finally, Gibbs introduced
his coarse-grained entropy and he showed that we should
expect an increase of this quantity from its initial value in
the case of mixing systems.  Both Boltzmann and Gibbs
definitions of entropy refers to a statistical description
and lead to the interpretation of entropy as
a measure of the disorder of the statistical state of the system.

Gibbs' formulation explicitly refers to the probability
distribution $p(\bG)$ at the current time $t$ \cite{Gibbs}.
This probability distribution corresponds
to an equivalent statistical ensemble of phase-space points.
The entropy characterizes the disorder in 
this statistical ensemble.
Since  the different possible microstates $\bG$ 
form a continuum,
a partition of the phase space is required
in order to proceed with the definition of 
discrete coarse-grained states and the counting of their complexions.
The phase space of the system is 
partitioned into disjoint cells ${\cal C}_{\omega}$ forming the partition 
${\cal P}=\{ {\cal C}_1,{\cal C}_2,...,{\cal C}_M \}$.  
These cells can be considered as the sets of microstates 
of the total system corresponding to the states 
$\omega\in\{ 1,2,...,N\}$ of some measuring device 
observing and possibly manipulating the system
for the preparation of some initial condition.  
The coarse-grained states corresponding 
to the events $\bG\in {\cal C}_{\omega}$
are now discrete and can be counted.
We can see in the statistical ensemble how many
copies of the system belong to a given cell ${\cal C}_{\omega}$.
For this purpose, let us observe the $\cal N$ first copies
and count the numbers ${\cal N}_{\omega}$
of copies in the state $\omega$.
The observation of the $\cal N$ first copies
generates a sequence of coarse-grained states
$\omega_1\omega_2... \omega_{\cal N}$
and ${\cal N}=\sum_{\omega} {\cal N}_{\omega}$.
The frequency of occurrence of the state $\omega$ gives
the corresponding probability
\be
\lim_{{\cal N}\to\infty} \frac{{\cal N}_{\omega}}{\cal N} = P(\omega) = \int_{{\cal C}_{\omega}} d\bG \, p(\bG)
\ee
The copies being statistically independent,
the number of possible sequences of successive states is given by
\be
\frac{{\cal N}!}{\prod_{\omega}{\cal N}_{\omega}!} \sim \exp\left[-{\cal N}\sum_{\omega}
P(\omega)\ln P(\omega)\right]
\ee
where Stirling's formula ${\cal N}!\simeq ({\cal N}/{\rm e})^{\cal N}$ was used.
Typically, this number grows exponentially at a rate
given in terms of the entropy
\be
S \equiv - k_{\rm B} \sum_{\omega} P(\omega)\ln P(\omega)
\label{entropy.def}
\ee
where Boltzmann's constant gives the proper
physical units to the entropy.
This entropy provides a measure of the disorder
in the statistical ensemble described by the probability
distribution $\{ P(\omega)\}$.  Indeed, the entropy vanishes
if a single state $\omega$ is present in the ensemble and the
entropy reaches its maximum possible value
if the probabilities $P(\omega)=1/\Omega$ are uniformly distributed
over all the states $\omega$ as 
for a microcanonical equilibrium ensemble on an energy shell.
In this latter case, we recover Boltzmann's formula $S = k_{\rm B} \ln \Omega$
from Eq. (\ref{entropy.def}).

The definition (\ref{entropy.def}) is consistent with the thermodynamic value
of the entropy at equilibrium
if the phase-space cells are identified as the quantum states.
In the classical-quantum correspondence,
the quantum states correspond to phase-space cells
of volume $\Delta\bG=(2\pi\hbar)^{3N}$.
The probabilities are thus given by $P(\omega) \simeq p(\bG) \; \Delta \bG$,
so that the entropy becomes
\be
S \simeq k_{\rm B} \ln \frac{1}{\Delta\bG} - k_{\rm B} \int_{\cal M} d\bG \; p(\bG) \ln p(\bG) + O(\Delta\bG)
\label{entropy}
\ee
For a dilute gas of identical particles with a Maxwell distribution of velocities,
this entropy gives the Sackur-Tetrode formula since $\cal M$
denotes the phase space of non-identical phases which 
is a fraction $1/N!$ of ${\mathbb R}^{6N}$.
Without quantum mechanics which fixes the smallest possible
phase space volume $\Delta\bG$, the entropy could become
arbitrarily large in the limit $\Delta\bG \to 0$.
This is not surprising because the probability density $p(\bG)$
means that the random variables $\bG$ are continuous
and are distributed over a continuum of possible values,
so that this distribution has randomness on arbitrarily
small scales contrary to the case of a discrete random variable.
The second term in Eq. (\ref{entropy}) is often called
the differential entropy in the mathematical literature
or Gibbs' fine-grained entropy.  However, it does not
provide a complete measure of disorder without
the first term.  

Different kinds of coarse graining can be envisaged.
If we describe the system in terms of macrovariables
such as the energy, the states $\omega$
correspond to phase-space cells such that
$E< H(\bG)<E+\Delta E$
where $H$ is the Hamiltonian function.
In a dilute gas we can consider the numbers
of molecules in cells of the position
and momentum space (the so-called $\mu$-space
instead of the $\Gamma$-space).
Fixing these numbers of molecules also amounts
to define domains ${\cal C}_{\omega}$ in the phase space.
We can also expand the entropy (\ref{entropy})
in terms of the distribution functions of one, two, three,... particles
\cite{Nettleton} and truncate this so-called Nettleton-Green expansion 
to get a coarse-grained entropy. These different definitions 
can be summarized with Eq. (\ref{entropy.def}).

The expression (\ref{entropy})
is not useful for nonequilibrium
systems because it remains constant in time
if the terms of order $\Delta\bG$ beyond
the two explicitly given are neglected.
In contrast, the entropy (\ref{entropy.def})
evolves in time in a nontrivial way.
Indeed, Gibbs' mixing property (\ref{mixing}) implies that the probabilities
$P_t(\omega)$ converge to their equilibrium values $P_{\rm eq}(\omega)$
as $t\to\infty$, which allows us to understand 
why the coarse-grained entropy (\ref{entropy.def})
can increase from its initial value up to its asymptotic equilibrium value. 
Since the equilibrium distribution is broader than the initial one,
we should expect that $S_{\rm eq} > S_0$.

Recent work has shown that, indeed,
the entropy increases.  A monotonic increase
is even possible to obtain if the partition of the phase space
is chosen with respect to the dynamics \cite{NN88}.
The way the entropy increases can be calculated
thanks to the expansion (\ref{forward.expansion}),
which shows that the
rate of increase of the entropy is directly determined
by the Pollicott-Ruelle resonances which are intrinsic to the system.
We should thus expect that the entropy production rate calculated
in this way is a property of the system which is
independent to a large extend of the way
the partition is chosen.  This expectation is confirmed
by the detailed calculation of the entropy
production in specific systems (see below).

 \subsection{Dynamical randomness and entropies per unit time}
 \label{dyn.entropies}
 
 If the standard thermodynamic entropy is a measure of the disorder
 in the probability distribution describing the system at the current time,
 a further concept has been introduced in order to characterize the temporal
 disorder of a process evolving in time.
The concept of entropy per unit time has been introduced in the context of random processes by 
Shannon \cite{S48} and, later, by Kolmogorov \cite{K59} and Sinai \cite{S59} in dynamical systems theory.  
The phase space of the system is here again 
partitioned into disjoint cells ${\cal C}_{\omega}$ forming the partition 
${\cal P}=\{ {\cal C}_1,{\cal C}_2,...,{\cal C}_N \}$.  
The symbolic sequence $\pmb{\omega}=\omega_0\omega_1...\omega_{n-1}$
defines a {\it path} or {\it history} which is a set of trajectories visiting the cells
$ {\cal C}_{\omega_0}{\cal C}_{\omega_1} ...{\cal C}_{\omega_{n-1}}$ at the successive times $t_k=k\tau$ ($k=0,1,...,n-1$): $\bPhi^{t_k}\bG \in {\cal C}_{\omega_k}$.
The system is observed in a stationary state described by the invariant probability measure $\mu$.

The multiple-time probability to observe the system in the successive coarse-grained states $\omega_0\omega_1...\omega_{n-1}$ at regular time intervals $\tau$ is given by
\begin{equation}
\mu(\pmb{\omega})=\mu(\omega_0\omega_1...\omega_{n-1}) = \mu\left( {\cal C}_{\omega_0}\cap \bPhi^{-\tau}{\cal C}_{\omega_1}\cap ...\cap\bPhi^{-(n-1)\tau}{\cal C}_{\omega_{n-1}}\right)
\label{proba}
\end{equation}
where $\mu$ is an invariant measure of the time evolution 
$\bPhi^t$ and is assumed to be the stationary probability of the nonequilibrium steady state.
The standard entropy per unit time of the partition $\cal P$ 
is defined as the mean decay rate of the multiple-time probability 
(\ref{proba}) as \cite{ER85,CFS82}
\begin{equation}
h({\cal P}) \equiv \lim_{n\to\infty} - \frac{1}{n\tau} \sum_{\pmb{\omega}} \mu(\pmb{\omega})
\ln \mu(\pmb{\omega}) = \lim_{n\to\infty} - \frac{1}{n\tau} \sum_{\omega_0\omega_1...\omega_{n-1}} \mu(\omega_0\omega_1...\omega_{n-1})  \ln \mu(\omega_0\omega_1...\omega_{n-1}) 
\label{dyn.entr}
\end{equation}

The entropy per unit time characterizes the dynamical randomness 
of the time evolution observed with the measuring device. 
The entropy per unit time can be defined for stochastic processes as
well as for deterministic dynamical systems.  
For stochastic processes such as the Ornstein-Uhlenbeck process,
the entropy per unit time depends in general 
on the size $\varepsilon$ of the cells used 
to make the partition as well as on the
sampling time $\tau$.  Accordingly, a concept of 
$(\varepsilon,\tau)$-entropy per unit time has been introduced
which characterizes the randomness 
as a function of the parameters $\varepsilon$ and $\tau$ \cite{G98,GW93}.

For stochastic processes with discrete state variables 
as well as for deterministic dynamical systems, the entropy
per unit time presents a supremum.
The supremum of the dynamical entropy (\ref{dyn.entr}) 
over all the possible partitions $\cal P$ defines 
the Kolmogorov-Sinai (KS) entropy per unit time 
\be
h_{\rm KS} = {\rm Sup}_{\cal P} h({\cal P})
\ee
which is thus a quantity intrinsic to the system \cite{ER85,CFS82,K59,S59}.

In isolated dynamical systems, the KS entropy is given by the sum of positive Lyapunov exponents
according to Pesin's theorem \cite{P77,ER85}.  We shall present a proof of this result in the next section.
This fundamental result shows that dynamical instability implies dynamical randomness.
The concept of entropy per unit time appears to be fundamental because it is the one which
characterizes randomness.  The name ``chaos" explicitly refers to a property of randomness so that
it is natural to define the chaotic systems 
as deterministic systems with a positive entropy per unit time \cite{G98}.  
We notice that dynamical instability is not enough 
for dynamical randomness as the example (\ref{Hxy}) shows.
Moreover, the entropy per unit time can be defined 
for systems without Lyapunov exponent,
allowing to extend the concept of dynamical chaos.  

The dynamical entropy does not vanish at equilibrium in chaotic systems contrary to the entropy production and no proportionality exists between these quantities.  The entropy per unit time is nevertheless the temporal analogue of the standard thermodynamic entropy.
Typical bulk equilibrium phases are spatially extended 
and present a positive entropy per unit volume $s=S/V$.
This quantity can be defined by Eq. (\ref{dyn.entr}) if $\omega_n$ 
denotes the local state of the particles in cells separated by the 
distance $\tau$.  For instance, in a spin chain, $\omega_n$ 
may represent the state of a spin.  The entropy per unit volume is known to
characterize the spatial disorder of the equilibrium phase: 
the successive spins point in random orientations along the chain.
The entropy per unit time is the analogue with space 
replaced by time. A process of time evolution is recorded and displayed
along the time axis as the successive pictures of movies.  
At successive times, the system is in different states 
$\omega_n$ which form a random sequence of events.
The entropy per unit time characterizes the temporal disorder of the movies.
For a coin-tossing process, head and tail both have probabilities equal to one half.
In this case, the multiple-time probability (\ref{proba}) decays as $1/2^n$
so that the entropy per unit time is equal to $h=\ln 2$.  For a game of dices, it is equal to $h=\ln 6$.

In nonequilibrium processes, we have already pointed out that the invariant measure
may break the time-reversal symmetry.  It is therefore interesting to introduce
a concept of time-reversal entropy per unit time in order to characterize
the dynamical randomness of the time-reversed trajectories of the process \cite{G04b}.
We consider the {\it time-reversed path} (or {\it time-reversed history)}:  
$\pmb{\omega}^{\rm R}=\omega_{n-1}...\omega_1\omega_0$.  We are interested by the probability 
$\mu(\pmb{\omega}^{\rm R})\equiv \mu(\omega_{n-1}...\omega_1\omega_0)$ 
of occurrence of the time-reversed path in the process taking place 
in the nonequilibrium stationary process.  
 We define the {\it time-reversed entropy per unit time} as \cite{G04b}
\begin{equation}
h^{\rm R}({\cal P}) \equiv \lim_{n\to\infty} - \frac{1}{n\tau} \sum_{\pmb{\omega}} \mu(\pmb{\omega})
\ln \mu(\pmb{\omega}^{\rm R}) =\lim_{n\to\infty} - \frac{1}{n\tau} \sum_{\omega_0\omega_1...\omega_{n-1}} \mu(\omega_0\omega_1...\omega_{n-1})  \ln \mu(\omega_{n-1}...\omega_1\omega_0) 
\label{TR.dyn.entr}
\end{equation}
We emphasize that the average is taken with respect to the probability of the forward path.  We can say that the time-reversed entropy per unit time characterizes the dynamical randomness of the time-reversed paths in the forward process of the nonequilibrium steady state.
We shall see in Sec. \ref{Info} that both dynamical entropies can be combined to form the entropy production of nonequilibrium processes.

 \section{The escape-rate formalism}
 \label{EscRate}
 
 The purpose of the escape-rate theory of transport is to relate the transport coefficients 
 to the characteristic quantities of motion which are
the Lyapunov exponents, the Kolmogorov-Sinai entropy, and the fractal dimensions
\cite{GN90,GB95,DG95,GD95,TVB96,G98}.
 Thanks to the escape-rate formalism, relationships can be established
in full respect for the Hamiltonian character of the microscopic dynamics and Liouville's theorem
in contrast with other approaches.  The escape-rate formalism uses the
result that each transport coefficient can be obtained from the random walk of
its associated Helfand moment which is the centroid of the conserved quantity corresponding
to the transport property of interest.  We can set up a first-passage problem
for the diffusive motion of the centroid.  Trajectories of the whole system of particles
are supposed to escape from a certain domain of phase space.  This domain
is defined by imposing threshold value for the centroid or Helfand moment.
The number of trajectories which are still inside this so-defined 
phase-space domain decreases with time when a trajectory reaches
the boundary of this domain.  An equivalent way of presenting
the problem is to consider absorbing boundary conditions as discussed
here above after Eq. (\ref{balance.proba}).  The decay is typically exponential,
which defines an escape rate $\gamma$.  The escape rate is proportional
to the transport coefficient.  On the other hand, the escape rate is given in terms
of the Lyapunov exponents, the Kolmogorov-Sinai entropy, and the fractal dimensions.  
This results into chaos-transport relationships as explained in this section.

\subsection{The transport coefficients and their Helfand moment}
 
It is known since the sixties that
the coefficients of transport properties such as diffusion, viscosity, or heat conductivity
 can be obtained from the fluctuation properties of the Helfand moments $G^{(\alpha)}$ or 
the corresponding microscopic currents $J^{(\alpha)}=dG^{(\alpha)}/dt$ \cite{Gr52,Ku57,He60}.  
A transport coefficient is given as the integral of the time autocorrelation function of 
the associated microscopic current by the Green-Kubo formula \cite{Gr52,Ku57} 
or, equivalently, as the rate of linear growth of the variance of the Helfand
moment \cite{He60}
\be
\alpha=\int_0^{\infty} \langle J_0^{(\alpha)} J_t^{(\alpha)}\rangle \; dt = \lim_{t\to\infty}
\frac{1}{2t} \ \langle(G_t^{(\alpha)}-G_0^{(\alpha)})^2\rangle  \label{coeff}
\ee 
The Helfand moments associated with the different transport properties are given in Table
I.  Each one of them represents the centroid of the conserved quantity which is locally transported.
For the diffusion of a Brownian particle, the Helfand moment is simply
the position of the particle.  For viscosity, it is the center of momenta of
all the particles in the fluid.  And so forth.

\vskip 0.5 cm
\hrule \vskip1pt \hrule height1pt
\vskip 0.2 cm

\centerline{TABLE I. Helfand's moments of different transport properties.}

\vskip 0.2 cm
\hrule 
\vskip 0.1 cm

$$
\vcenter{\openup1\jot 
\halign{#\hfil&\qquad#\hfil\cr
{\it transport property}  &  {\it moment}  \cr & \cr
self-diffusion  & $G^{\rm ({\cal D})} \ = \ x_i$ \cr
shear viscosity & $G^{(\eta)} \ = \ \frac{1}{\sqrt{Vk_{\rm B}T}}
\ \sum_{i=1}^{N} \ x_i \ p_{iy} $ \cr
bulk viscosity ($\psi = \zeta + \frac{4}{3} \eta$) & $G^{(\psi)} \ = \ 
\frac{1}{\sqrt{Vk_{\rm B}T}}\ \sum_{i=1}^{N} \ x_i \ p_{ix} $ \cr
heat conductivity & $G^{(\kappa)} \ = \ \frac{1}{\sqrt{Vk_{\rm B}T^2}}
\ \sum_{i=1}^{N} \ x_i \ (E_i \ - \ \langle E_i \rangle)$ \cr
electric conductivity & $G^{\rm (e)} \ = \ \frac{1}{\sqrt{Vk_{\rm B}T}}
\ \sum_{i=1}^{N} \ eZ_i \ x_i$ \cr }}
$$

\vskip 0.1 cm
\hrule height1pt \vskip1pt \hrule 
\vskip 0.5 cm

The Einstein-like formula in Eq. (\ref{coeff}) shows that the Helfand moments have a diffusive
motion.  A first passage problem can be set up by imposing absorbing boundary conditions in
the space of variation of the Helfand moment \cite{DG95}.  
These absorbing boundary conditions select all
the trajectories of the whole system for which the Helfand moment never reaches the
absorbing boundaries:
\be
-\frac{\chi}{2} \leq G_t^{(\alpha)} \leq + \frac{\chi}{2} 
\label{eq:abc}
\ee
However, most of the trajectories reach the absorbing boundaries and
escape out of the domain delimited by these boundaries.  Accordingly, the trajectories
forever trapped inside the absorbing boundaries form a set of zero probability 
and composed of unstable trajectories.  This set is called the repeller and 
is characterized by an escape rate which can be
evaluated by using a diffusive equation for the motion 
of the Helfand moment $g = G_t^{(\alpha)}$
\be
\frac{\partial p}{\partial t} = \alpha \; \frac{\partial^2 p}{\partial g^2} 
\label{eq:diffHelfand}
\ee
with the corresponding absorbing boundary conditions $p(g=\pm\chi/2,t)=0$.
The solution of this equation is given by
\be
p(g,t) = \sum_{j=1}^{\infty} \; a_j \ \exp(-\gamma_jt)
\ \sin\left(\frac{j\pi g}{\chi}+\frac{j\pi}{2}\right) \qquad
\mbox{with} \qquad 
\gamma_j = \alpha \left(\frac{j\pi}{\chi}\right)^2
\label{eq:soldiff}
\ee
The escape rate is the slowest decay rate which is thus related to the
transport coefficient $\alpha$ according to
\be 
\gamma \simeq \gamma_1 = \alpha \left(\frac{\pi}{\chi}\right)^2  \qquad \hbox{for} \quad
\chi\to\infty 
\label{eq:esc}
\ee

\begin{figure}[ht]
\centering
\includegraphics[width=9cm]{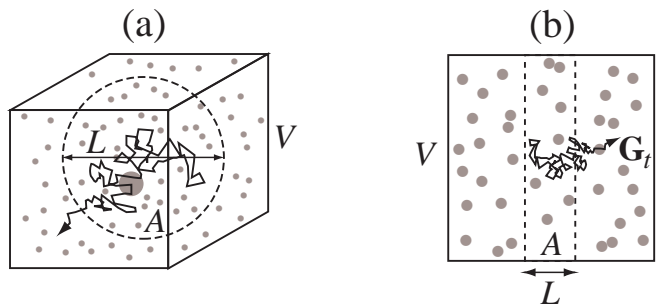}

\caption{Schematic illustration of the application of the escape-rate theory to 
(a) the diffusion of a Brownian
particle escaping out of a sphere $A$ of diameter $L$ in a fluid of volume $V$ 
and (b) the viscosity in a fluid
of volume $V$ in which the Helfand moment ${\bf G}_t$ 
escapes out of a domain $A$ of width $L$.}
\label{fig2}
\end{figure}

In the case of the Brownian motion in a fluid inside a container, the Helfand moment
is the position of the Brownian particle.  The particle is supposed to escape
if its position reaches for instance a sphere $A$ of diameter $L$ in the physical space
as illustrated in Fig. \ref{fig2}a.  This condition corresponds to a certain domain
in the phase space of the whole system composed of the Brownian particle
and the particles of the surrounding fluid.

In the case of viscosity, the Helfand moment is the center of momenta of all
the particles of the fluid.  This Helfand moment or centroid performs
a random walk as illustrated in Fig. \ref{fig2}b.  Here again, 
a phase-space domain corresponds to the condition that this centroid is found
in the domain $A$ of width $L$ in the physical space.

\subsection{Escape and dynamical randomness}

\begin{figure}[ht]
\centering
\includegraphics[width=12cm]{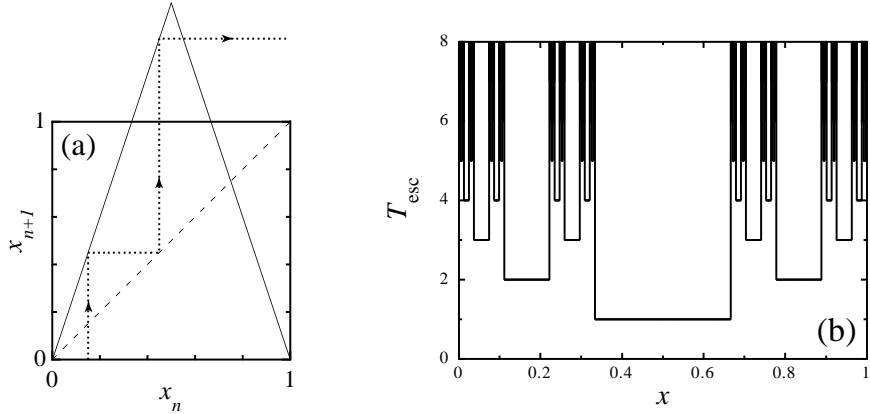}
\caption{(a) First return plot of the one-dimensional map
$x_{n+1} = s \left(\frac{1}{2}- \Big\vert x_n - \frac{1}{2}\Big\vert\right)$
with a slope $s$ larger than two.  (b) Time
taken by a trajectory to escape out of the unit interval under the dynamics of the 
map (a) versus the initial condition $x$ of the trajectory.  We observe that this escape
time varies in an extremely complex way and that it becomes infinite for each initial
condition taken on the Cantor set of trapped trajectories. For this map,
the escape rate (\ref{esc.dfn}) is given by $\gamma=\ln s -\ln 2$,
the Lyapunov exponent (\ref{lyap.dfn}) by $\lambda(\beta)=\ln s$,
the Ruelle topological pressure (\ref{topo.pressure}) by 
$P(\beta)=\ln 2 - \beta \ln s$, the entropy per unit time (\ref{h(beta)})
by $h(\beta)=\ln 2$, and the generalized fractal dimension (\ref{gen.fractal.dim.dfn}) by
$d_q=(\ln 2)/(\ln s)$, as can be checked using the corresponding definitions presented in the text.}
\label{fig3}
\end{figure}

The escape rate turns out to be related to the characteristic quantities
of the underlying dynamics such as the Lyapunov exponents,
the Kolmogorov-Sinai entropy per unit time,
and the fractal dimensions.  These relationships can be obtained
even for one-dimensional maps such as the one depicted in Fig. \ref{fig3}a.
The time for the escape of trajectories depends on their initial condition as shown
in Fig. \ref{fig3}b.  This is a highly complicated function with vertical asymptotes
on a Cantor set.  For these initial conditions, the escape time is infinite
so that the trajectory remains trapped in the unit interval.
The Cantor set is of zero Lebesgue measure but contains
uncountably many trajectories.  
In this simple system, the partition into the two intervals $[0,0.5[$
and $[0.5,1]$ to which the symbols $0$ and $1$ are assigned
allows us to associate a {\it symbolic sequence} 
$\omega_0\omega_1\omega_2...\omega_n...$
(also called a {\it path} or a {\it history})
with each trajectory.  Furthermore, this correspondence is one-to-one for
the trapped trajectories of the Cantor set, in which case we speak about
a generating partition.  Along a given trajectory,
an infinitesimal interval will be expanded by the so-called stretching factor
$\Lambda({\pmb{\omega}}) = \Lambda(\omega_0\omega_1\omega_2...\omega_{n-1})$,
which is given by the product of the slopes 
$\Lambda_0=+s$ and $\Lambda_1=-s$ for the map of Fig. \ref{fig3}a.
A family of invariant probability measures can be defined as
\be
\mu_{\beta}(\pmb{\omega}) \equiv 
\frac{\vert \Lambda(\pmb{\omega})\vert^{-\beta}}{\sum_{\pmb{\omega}} \vert \Lambda(\pmb{\omega})\vert^{-\beta}}
\label{mu.beta}
\ee
in the limit $n\to\infty$.  These probability measures are normalized to unity.
The different values of the parameter $\beta$ are not of equal interest for
the Liouvillian dynamics.  Indeed, an ensemble of initial conditions
uniformly distributed in a small interval $[x,x+\Delta x]$
with a probability density $1/\Delta x$ will broaden under the effect
of the dynamics.  After $n$ iterates, the probability density would drop
to $(\vert\Lambda(\pmb{\omega})\vert\Delta x)^{-1}$, which shows
that $\beta=1$ is the value used by the Liouvillian dynamics.
The escape rate can be obtained by noting that the length
of the intervals which have not escaped till the $n^{\rm th}$ iterate
is given by $\sum_{\pmb{\omega}} \vert \Lambda(\pmb{\omega})\vert^{-1}$
so that the escape rate is given by
\be
\gamma = \lim_{n\tau\to\infty} \; -\frac{1}{n\tau} 
\ln \sum_{\pmb{\omega}} \frac{1}{\vert \Lambda(\pmb{\omega})\vert}
\label{esc.dfn}
\ee
Here again, the value $\beta=1$ is selected in agreement with
the fact that the escape rate is the leading Liouvillian eigenvalue.

Nevertheless, the other values may also be of interest.
In particular, we notice that the value $\beta=0$ leads to a measure
which gives the same probability weight on each trajectory independently
of its instability.  Therefore, the value $\beta=0$ is interesting for counting
the trajectories and characterizes the dynamics from a topological point of view.

The average Lyapunov exponent with respect to the invariant measure (\ref{mu.beta}) is defined by
\be
\lambda(\beta) = \lim_{n\to\infty} \frac{1}{n\tau} \sum_{\pmb{\omega}} \mu_{\beta}(\pmb{\omega})
\ln \vert \Lambda(\pmb{\omega})\vert
\label{lyap.dfn}
\ee
where $\tau=1$ in the special case of a map.
The mean positive Lyapunov exponent of the classical dynamics is given by the value
at $\beta=1$:  $\lambda\equiv \lambda(1)$.
We can here introduce the so-called {\it Ruelle topological pressure} \cite{R78}
\be
P(\beta) \equiv \lim_{n\tau\to\infty} \; \frac{1}{n\tau} 
\ln \sum_{\pmb{\omega}} \frac{1}{\vert \Lambda(\pmb{\omega})\vert^{\beta}}
\label{topo.pressure}
\ee
which contains both the escape rate and the mean Lyapunov exponent as
\be
\gamma = - P(1) \qquad\qquad
\lambda = - P'(1)
\ee
Now, if we calculate the entropy per unit time (\ref{dyn.entr}) of the invariant measure (\ref{mu.beta}),
we obtain from the preceding definitions that \cite{R78}
\be
h(\beta) = \beta \; \lambda(\beta) + P(\beta)
\label{h(beta)}
\ee
When restricting this relation to the value $\beta=1$ and supposing that
the partition is generating the entropy per unit time becomes the
Kolmogorov-Sinai entropy, $h(1)=h_{\rm KS}$, and Eq. (\ref{h(beta)}) becomes
the fundamental escape-rate formula \cite{ER85,KG85}:
\be
\gamma = \lambda - h_{\rm KS}
\ee

Properties such as the topological pressure, the KS entropy, and the Lyapunov exponents
are known as large-deviation dynamical properties because they concern
probabilities or phase-space volumes with exponential behavior in time
and which are thus rapidly suppressed in time.  These large-deviation properties
are at the basis of the thermodynamic formalism \cite{R78}.

Generalized fractal dimensions $d_q$ of parameter $q$
can also be defined for the Cantor set with the natural
probability measure $\mu=\mu_1$ by the condition
\be
\sum_{\pmb{\omega}} \frac{\mu(\pmb{\omega})^q}{\ell(\pmb{\omega})^{(q-1)d_q}} \sim 1 \qquad
\mbox{for}\quad n\to\infty
\label{gen.fractal.dim.dfn}
\ee
where $\ell(\pmb{\omega})\sim \vert\Lambda(\pmb{\omega})\vert^{-1}$ denotes
the size of the interval associated with the trajectories of symbolic sequence $\pmb{\omega}$.  
With the aforementioned definitions, the generalized fractal dimension $d_q$
is given as the root of the following equation \cite{G98}:
\be
P\left[ q + (1-q) d_q \right] = - q \; \gamma
\label{gen.fractal.dim.press}
\ee
The Hausdorff dimension $d_0=d_{\rm H}$ is simply the root of the pressure function, $P(d_{\rm H})=0$,
while the information dimension is given by $d_1=d_{\rm I}=h_{\rm KS}/\lambda$ \cite{Y82}.
The escape-rate formula can be rewritten in terms of the information dimension as 
$\gamma = \lambda (1-d_{\rm I})$. Further material can be found in Refs. \cite{G98,D99}.

The preceding results extend to Hamiltonian systems with two degrees of freedom
and a single positive Lyapunov exponent.  For these systems, the previous
reasoning holds if we consider the $x$-axis of the one-dimensional map of Fig. \ref{fig3}
as the stable or unstable direction.  For the forward time evolution,
the conditionally invariant probability measure (\ref{Psi_0})
is smooth in the unstable direction but fractal in the stable direction.
The aforementioned fractal dimensions thus characterize its structure in the stable direction.
By time-reversal symmetry, a similar structure appears in the unstable direction for
the backward time evolution and an invariant measure can even be defined
for the set of forever trapped trajectories which is a fractal of dimension $D_q=2d_q+1$
in the three-dimensional energy shell \cite{GR89}.  In this regard, $d_q$ is a partial dimension
while $D_q$ is the total one.

In systems with a higher-dimensional phase space and several positive
Lyapunov exponents, $\Lambda(\pmb{\omega})$
represents the expansion in all the unstable directions so that $\lambda$
should be replaced by the sum of positive Lyapunov exponents:
\be
\gamma = \sum_{\lambda_i>0} \lambda_i - h_{\rm KS} = \sum_{\lambda_i>0} \lambda_i (1-d_i) 
\label{ER.formula}
\ee
where $d_i$ denotes a partial information dimension associated
with each unstable direction \cite{LY85}.

The invariant measure or the conditionally invariant measure are here defined
on a fractal set of unstable trajectories (or their stable manifolds) called
the {\it fractal repeller}.  The escape rate is the leading Pollicott-Ruelle
resonance $s_0=-\gamma$ for the dynamics on this fractal repeller.

If the system is progressively isolated by switching off the escape of
trajectories, the escape rate vanishes, the partial dimensions take
the unit value, and we obtain
the famous {\it Pesin identity} \cite{P77}
\be
h_{\rm KS} = \sum_{\lambda_i>0} \lambda_i 
\ee
Consequently, we can conclude that the dynamical instability characterized by
the Lyapunov exponents is directly responsible for the
dynamical randomness of the time evolution and the positive
KS entropy.  This result shows that the KS entropy is a concept
which is here equivalent to the property characterized by
the Lyapunov exponent with the advantage that it can be
extended to systems in which the Lyapunov exponent
are not defined or irrelevant to the fundamental property 
of randomness.  The escape-rate formula (\ref{ER.formula})
generalizes Pesin's identity to open systems with escape.

\subsection{The chaos-transport formula}

If we combine the escape-rate formula (\ref{ER.formula}) with
the proportionality (\ref{eq:esc}) of the escape rate to the
transport coefficient, we obtain the following
large-deviation relationships between the transport coefficients 
and the characteristic quantities of chaos \cite{DG95}
\be
\boxed{\alpha \ = \ \lim_{{\chi,V} \to \infty} \ \left( \frac{\chi}{\pi}
\right)^2 \ \biggl(\sum_{\lambda_i>0} \
\lambda_i  -  h_{\rm KS}\biggr)_{\chi} \ =
\ \lim_{\chi,V\to\infty} \ \left( \frac{\chi}{\pi} \right)^2  \
\sum_{\lambda_i>0} \ \lambda_i (1-d_i)\Big\vert_{\chi}}
\label{eq:escformula}
\ee
In the limit $\chi\to\infty$, the partial information codimensions can be replaced by the
Hausdorff codimensions in hyperbolic systems.  
This formula has already been applied to diffusion in periodic and random
Lorentz gases \cite{GB95,vBLD00},
reaction-diffusion \cite{CG01}, as well as viscosity \cite{VG03a,VG03b}.

\begin{figure}[ht]
\centering
\includegraphics[width=14cm]{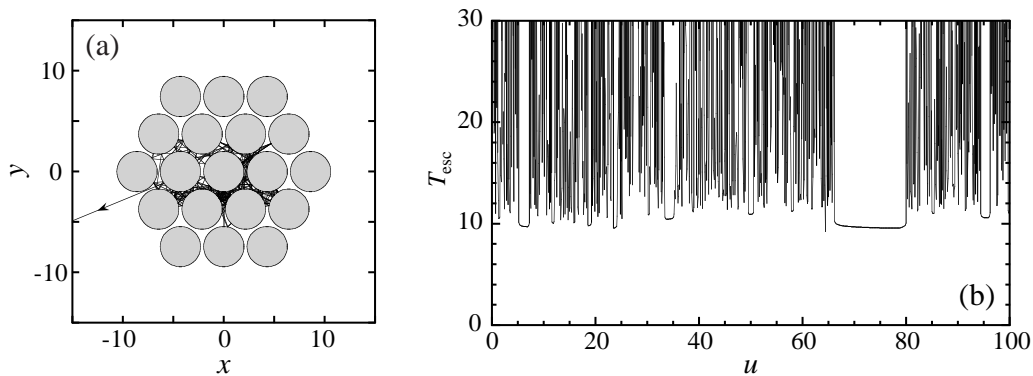}
\caption{Open Lorentz gas: (a) configuration of the system with a typical trajectory; (b) time taken by the
particle to escape versus the initial angle $\theta=0.75\pi-u\pi 10^{-10}$ of velocity of the particle starting
from the central disk.  The escape time becomes infinite on each stable manifold of a trapped trajectory.}
\label{fig4}
\end{figure}

Detailed results have been obtained for diffusion in the open
two-dimensional periodic Lorentz gas (see Fig. \ref{fig4}a) \cite{GB95}.  
In this system, the particles have a
deterministic diffusive motion induced by the elastic collisions 
on immobile scatterers until the particles
escape upon reaching absorbing boundaries in the form of 
a hexagon, a circle, or two parallel lines separated by
a large distance $L$.  The absorbing boundary conditions 
select a fractal repeller of trapped trajectories which
can be numerically evidenced by plotting the time to escape 
as a function of initial conditions (see Fig.
\ref{fig4}b).  In this two-degree-of-freedom system, 
there is only one unstable direction and, thus, one
positive Lyapunov exponent and associated codimension.  
Accordingly, Eq. (\ref{eq:escformula}) shows that the
diffusion coefficient is given in terms of
the positive Lyapunov exponent and the fractal
dimension of the underlying set of trapped
trajectories \cite{GN90,GB95}:
\be
{\cal D} \ = \ \lim_{L \to \infty} \ \left( \frac{L}{\pi}
\right)^2 \ \left(
\lambda  -  h_{\rm KS}\right)_{L} \ =
\ \lim_{L\to\infty} \ \left( \frac{L}{\pi} \right)^2  \
\lambda (1-d)\Big\vert_L
\label{esc.diff}
\ee
where $d$ is either the information or the Hausdorff partial dimension 
associated with the unstable direction.  

 \section{Construction of the hydrodynamic modes}
 \label{Modes}
 
 \subsection{Diffusion in spatially periodic systems}
 
We now consider dynamical systems which are spatially extended
and which can sustain a transport process of diffusion.
We moreover assume that the system is invariant under
a discrete Abelian subgroup of spatial translations $\{{\bf a}\}$.
The invariance of the Liouvillian time evolution
under this subgroup means that the Perron-Frobenius operator  $\hat P^t = \exp(\hat L t)$
commutes with the translation operators $T^{\bf a}$,
whereupon these operators admit common eigenstates:
\be
\left\{
\begin{array}{l}
\hat P^t\Psi_{\bf k} = \exp(s_{\bf k}t)\Psi_{\bf k}\\
\hat T^{\bf a}\Psi_{\bf k} = \exp(i{\bf k}\cdot{\bf a})\Psi_{\bf k}
\end{array}
\right.
\ee
where $\bf k$ is called the wavenumber or Bloch parameter \cite{G98}.
The Pollicott-Ruelle resonances $s_{\bf k}$ are now functions
of the wavenumber.  The leading Pollicott-Ruelle resonance which vanishes 
with the wavenumber defines the dispersion relation of diffusion:
\be
s_{\bf k} =-{\cal D}{\bf k}^2 + O({\bf k}^4)
\label{disp.rel.diff}
\ee
The diffusion coefficient is given by the Green-Kubo formula:
\be
{\cal D} = \int_0^{\infty} \langle v_x(0)v_x(t)\rangle \, dt
\ee
where $v_x$ is the velocity of the diffusing particle.
The associated eigenstate is the hydrodynamic mode of diffusion $\Psi_{\bf k}$.

\subsection{Fractality of the hydrodynamic modes of diffusion}

In strongly chaotic systems, the diffusive eigenstate
$\Psi_{\bf k}$ is a Gelfand-Schwartz distribution which 
cannot be depicted except by its cumulative function
\be
F_{\bf k}(\theta) = \int_0^{\theta}  \, \Psi_{\bf k}(\bG_{\theta'}) \, d\theta' 
\label{cumul0}
\ee
defined by integrating over a curve $x_{\theta}$ in the phase space (with $0\leq \theta < 2\pi$).

\begin{figure}[ht]
\centerline{\includegraphics[width=11cm]{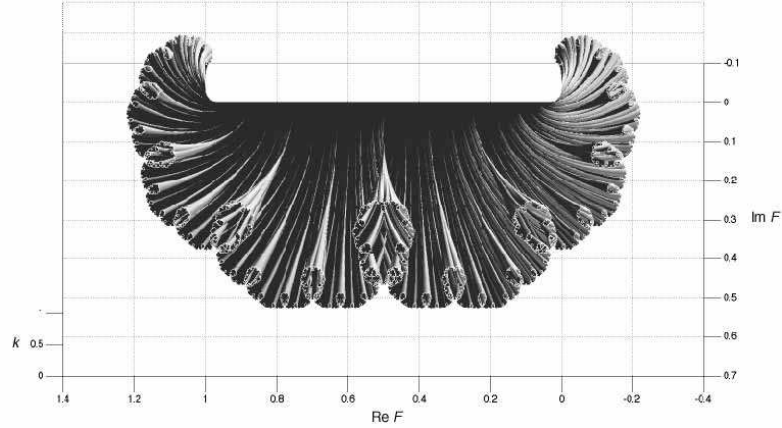}}
\caption{The diffusive modes of the multibaker map represented
by their cumulative function depicted in the complex plane
$({\rm Re}\, F_k, {\rm Im}\, F_k)$ versus the wavenumber $k$.
The multibaker map is defined by
$\phi(l,x,y)=\left( l-1, 2x,\frac{y}{2}\right)$
for $0 \leq x \leq 1/2$ and 
$\phi(l,x,y)=\left( l+1, 2x-1,\frac{y+1}{2}\right)$
for $1 \geq x > 1/2$ with $l\in \mathbb{Z}$ \cite{G98,TG95}.
The leading Pollicott-Ruelle resonance is $s_k = \ln \cos k$.
The topological pressure is here given by
$P(\beta) = (1-\beta)\ln 2$ because the map
is uniformly expanding by a factor $2$. 
According to Eq. (\ref{formula}),
the Hausdorff dimension is $D_{\rm H} = (\ln 2)/[\ln(2 \cos k)]$
\cite{GDG01}.
\label{fig5}}
\end{figure}

\begin{figure}[ht]
\centerline{\includegraphics[width=15cm]{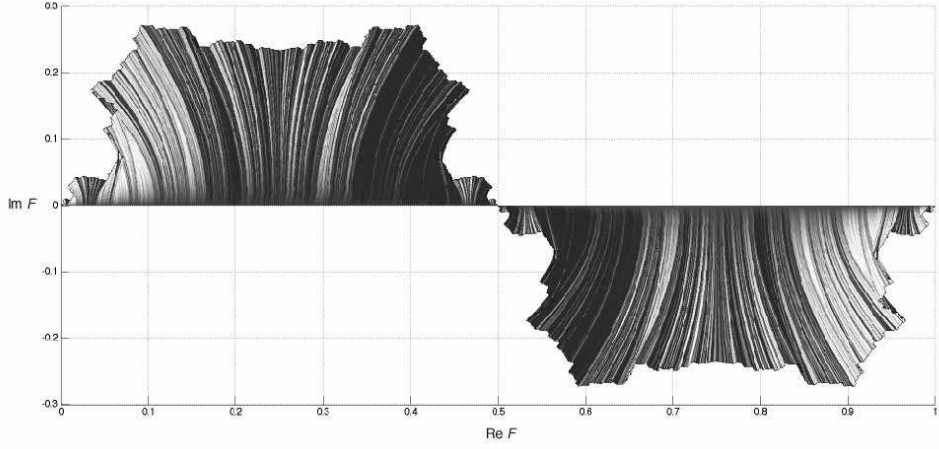}}
\caption{The diffusive modes of the periodic hard-disk Lorentz gas represented
by their cumulative function depicted in the complex plane
$({\rm Re}\, F_{\bf k}, {\rm Im}\, F_{\bf k})$ versus the wavenumber ${\bf k}$ 
which points out of the page
\cite{GCGD01}.
\label{fig6}}
\end{figure}

We notice that the eigenstate $\Psi_{\bf k}$ of the forward semigroup 
can be obtained by applying the time evolution operator over an arbitrary long time
from an initial function which is spatially periodic of wavenumber $\bf k$
so that the cumulative function (\ref{cumul0}) is equivalently defined by \cite{GCGD01}
\be
F_{\bf k}(\theta) = \lim_{t\to\infty} \frac{\int_0^{\theta} d\theta' \exp \left[ i{\bf k}\cdot({\bf r}_t-{\bf r}_0)_{\theta'}
\right]}{\int_0^{2\pi} d\theta' \exp \left[ i{\bf k}\cdot({\bf r}_t-{\bf r}_0)_{\theta'}
\right]}
\label{cumul}
\ee
This function is normalized to take the unit value at $\theta=2\pi$.
For vanishing wavenumber, the cumulative function is equal to
$F_{\bf k}(\theta)=\theta/(2\pi)$, which is the cumulative function
of the microcanonical uniform distribution in phase space.
For nonvanishing wavenumbers, the cumulative function becomes
complex.  The cumulative functions of the diffusive modes
are depicted in Fig. \ref{fig5} for the multibaker model of diffusion 
and in Fig. \ref{fig6} for the hard-disk Lorentz gas.
These cumulative functions form fractal curves in the complex
plane $({\rm Re}\, F_{\rm k}, {\rm Im}\, F_{\rm k})$.
The Hausdorff dimension $D_{\rm H}$ of these fractal curves
can be calculated as follows.
If we write Eq. (\ref{cumul}) in terms of
the symbolic dynamics and if we notice that
the denominator behaves as $\exp(s_{\bf k} t)$,
we get:
\be
F_{\bf k}(\theta) = \lim_{n\to\infty} 
\; \sum_{\pmb{\omega}} \frac{1}{\vert\Lambda(\pmb{\omega})\vert}
\; {\rm e}^{i{\bf k}\cdot ({\bf r}_{n\tau}-{\bf r}_0)_{\pmb{\omega}}}
\; {\rm e}^{-s_{\bf k} n\tau} = \lim_{n\to\infty} 
\; \sum_{\pmb{\omega}} \Delta F_{\bf k}(\pmb{\omega})
\label{approx.cumul}
\ee
which constitutes an approximation of the cumulative function
as a polygonal curve formed by a sequence of complex vectors.
The Hausdorff dimension of this curve is given by
the condition
\be
\sum_{\pmb{\omega}} \vert \Delta F_{\bf k}(\pmb{\omega})\vert^{D_{\rm H}} \sim 1 
\qquad \mbox{for} \quad n\to\infty
\ee
This condition can be rewritten in terms of the topological pressure
(\ref{topo.pressure}) to obtain the formula:
\be
\boxed{P(D_{\rm H}) = D_{\rm H} \; {\rm Re}\; s_{\bf k}}
\label{formula}
\ee
for the Hausdorff dimension \cite{GCGD01}.  

There is no escape of particles from the system so that its escape rate
vanishes and $\gamma =-P(1)=0$.
The Hausdorff dimension can thus be expanded in powers of the wavenumber as
$D_{\rm H}({\bf k}) = 1 + ({\cal D}/\lambda) {\bf k}^2 + {\cal O}({\bf k}^4)$
so that the diffusion coefficient is obtained from the
Hausdorff dimension and the Lyapunov exponent by the formula

\be
{\cal D} = \lambda \; \lim_{{\bf k}\to 0} \frac{D_{\rm H}({\bf k})-1}{{\bf k}^2}
\label{diffdim}
\ee
This formula has been verified for different dynamical systems sustaining
deterministic diffusion \cite{GCGD01}.

We notice that the large-deviation formula (\ref{diffdim})
has a form very similar to the chaos-transport relationship (\ref{esc.diff}) of
the escape-rate formalism.  Indeed, if we consider that the two successive
ascending and descending nodes of the mode of
wavenumber $\bf k$ are located at the absorbing boundary separated
by a distance $L$, the wavenumber is equal to $k=\pi/L$
and the dispersion relation of diffusion gives the escape rate
${\rm Re}\; s_{\bf k} \simeq - {\cal D}(\pi/L)^2 \simeq -\gamma$.
On the other hand, we find
the Lyapunov exponent as well as a fractal dimension
in the right-hand side of both Eqs. (\ref{esc.diff}) and (\ref{diffdim}).
This shows the close connection with the escape-rate formalism.

\subsection{Nonequilibrium steady state}
\label{NESS}

\begin{figure}[ht]
\centerline{\includegraphics[width=8cm]{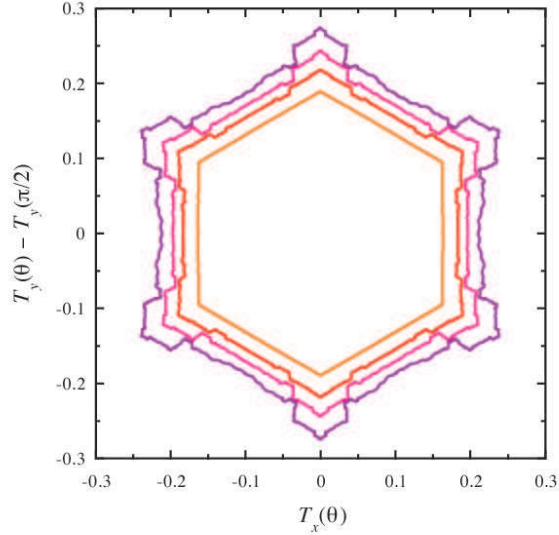}}
\caption{Periodic hard-disk Lorentz gas: Curves of the cumulative functions 
$T_{\bf g}(\theta) \equiv \int_0^{\theta} \Psi_{\bf g}({\mathbf\Gamma}_{\theta'}) \; d\theta'$
of the nonequilibrium steady state
for triangular lattices of hard disks of unit radius with
an intercenter distance of $d=2.001,2.1,2.2,2.3$.
The plot shows the generalized Takagi function 
for a gradient in the $y$-direction versus the one for
a gradient in the $x$-direction \cite{G00}.}
\label{fig7}
\end{figure}

The nonequilibrium steady state corresponding to a gradient of concentration
maintained between two reservoirs of particles at different
chemical potentials can be obtained from the diffusive modes
at small wavenumbers according to
\be
\Psi_{\bf g} = 
-i \; {\bf g}\cdot\frac{\partial\Psi_{\bf k}}{\partial{\bf k}}\Big\vert_{{\bf k}=0} 
\label{Psi_g-Psi_k}
\ee
where $\bf g$ is the gradient of concentration \cite{TG95}.
This leads to the nonequilibrium steady state described by the density:
\be
\Psi_{\bf g}({\mathbf\Gamma}) = {\bf g}\cdot \left[ {\bf r}({\mathbf\Gamma}) 
+ \int_0^{-\infty} {\bf v}({\mathbf\Phi}^t{\mathbf\Gamma}) dt \right]
\label{eq:NESS}
\ee
which is singular.  
These densities admit cumulative functions 
also known as generalized Takagi functions \cite{G98,D99,TG95}. 
These functions are typically self-similar.
An example is depicted in Fig. \ref{fig7} for the hard-disk periodic Lorentz gas \cite{G00}.
These cumulative functions are smooth in the unstable phase-space directions
but singular in the stable directions, which comes from the breaking
of the time-reversal symmetry in the forward semigroup.
This breaking of time-reversal symmetry is here
transferred from the diffusive mode to the nonequilibrium steady state.

The nonequilibrium steady state (\ref{eq:NESS}) implies Fick's law.
Indeed, for a steady state (\ref{eq:NESS}) with a gradient $g$ in the $x$-direction,
the average velocity $v_x$ or current is given by
\be
\langle v_x \rangle_{\rm neq} = \langle v_x \Psi_{{\bf e}_x} \rangle_{\rm eq}
= - g \int_0^{\infty} \langle v_x(0) v_x(t) \rangle_{\rm eq} \; dt = -{\cal D} \; g
\label{GK}
\ee
where we have used the result that $\langle v_x x\rangle_{\rm eq}=0$.
Accordingly, we recover the Green-Kubo formula for the diffusion
coefficient $\cal D$ and Fick's law \cite{G98}.

The expression (\ref{eq:NESS}) for the nonequilibrium steady state can be
obtained by a direct reasoning.  Let us consider
 diffusion in an open Lorentz gas between two chemiostats
or particle reservoirs at phase-space densities $p_{\pm}$ separated by a distance $L$.
The phase-space density inside the system can only take either the value $p_-$
corresponding to the reservoir on the left-hand side or $p_+$
from the reservoir on the right-hand side.  In order to determine which value,
we have to integrate the trajectory backward in time until the time of entrance in the system, $T({\mathbf\Gamma})<0$.
The value is $p_-$ (resp. $p_+$) if the particle enters from the left-hand (resp. right-hand) side.
If the gradient of phase-space concentration is denoted by ${\bf g}$, we can write the
invariant density of the nonequilibrium steady state in the form:
\be
p_{\rm neq}({\mathbf\Gamma}) = \frac{p_++p_-}{2} + {\bf g}\cdot \left[ {\bf r}({\mathbf\Gamma}) + \int_0^{T({\mathbf\Gamma})}
{\bf v}\left( {\mathbf\Phi}^t{\mathbf\Gamma}\right) dt \right]
\label{NESS.prob.dens}
\ee
Indeed, the integral of the particle velocity $\bf v$ backward in time until the time of entrance
gives the position of entrance ${\bf r}({\mathbf\Phi}^{T({\mathbf\Gamma})}{\mathbf\Gamma}) = \pm L/2$ 
minus the current position ${\bf r}({\mathbf\Gamma})$ which cancels
the first term in the bracket.  With the gradient ${\bf g} = (p_+-p_-){\bf e}_x/L$, 
we end up with the result that 
$p_{\rm neq}({\mathbf\Gamma}) = p_{\pm}$
whether the trajectory enters from the left- or right-hand side \cite{G98}.

The density of this nonequilibrium steady state is therefore
a piecewise constant function with its discontinuities
located on the unstable manifolds of the fractal repeller of the escape-rate formalism
of Sec. \ref{EscRate}.  Indeed, trajectories on the unstable manifold of the
fractal repeller remain trapped between both reservoirs under the backward time evolution.
The fractal repeller of the escape-rate formalism therefore controls the
structure of the nonequilibrium steady state.  Its invariant density is very different from the
one of the thermodynamic equilibrium but it remains absolutely continuous
with respect to the Lebesgue measure as long as the reservoirs
are separated by a finite distance $L$. Furthermore, we notice that the nonequilibrium steady state
defined by $\mu_{\rm neq}({\cal A})=\int_{\cal A} p_{\rm neq}(\bG)\, d\bG$
with the probability density (\ref{NESS.prob.dens}) breaks the time-reversal symmetry in the sense of Eq. (\ref{mu.noneq}).

In the limit where the reservoirs are separated by an arbitrarily large distance $L$,
the time of entrance goes to infinity: $T({\mathbf\Gamma})\to\infty$.
If we perform this limit while keeping constant the gradient $\bf g$, 
the invariant density (\ref{NESS.prob.dens}) is related to
the density (\ref{eq:NESS}) after substraction of the mean density $(p_++p_-)/2$.
Therefore, we reach the conclusion that the discontinuities on the
unstable manifolds of the fractal repeller of the escape-rate formalism
gives the singular character to the nonequilibrium steady state.
There is thus a deep connection between the fractal repeller of the escape-rate formalism, the
fractal structure of the hydrodynamic modes, and the singular character of the nonequilibrium steady state.

\subsection{Entropy production and irreversibility}
\label{entrop.prod}

The study of the hydrodynamic modes of diffusion as eigenstates of the Liouvillian
operator has shown that these modes are typically given in terms of singular distributions
without density function.  Since the works by Gelfand, Schwartz, and others in the fifties
it is known that such distributions acquire a mathematical meaning if they are evaluated
for some test functions belonging to certain classes of functions.  The test function used
to evaluate a distribution is arbitrary although the distribution is not.
Examples of test functions are the indicator functions of the cells 
of some partition of phase space.  In this regard, the singular character
of the diffusive modes justifies the introduction of a coarse-graining procedure.
This reasoning goes along the need to carry out a coarse graining
in order to understand that entropy increases with time, as understood
by Gibbs \cite{Gibbs} at the beginning of the XXth century.

We consider a spatially periodic dynamical system
as used for the simulation of Brownian motion by molecular dynamics
with periodic boundary conditions.
The particles of the fluid surrounding the Brownian particle
are reinjected at the boundary opposite to the one
of their exit, while the Brownian particle
continues its random walk on the lattice made of the
images upon spatial transitions of the system around the origin.
The phase-space region ${\cal M}_{\bf l}$ corresponding
to the image of the system at the lattice vector $\bf l$ 
is partitioned into cells $\cal A$.
If the underlying dynamics has Gibbs' mixing property, 
the probabilities that
the system is found in the cells $\cal A$ at time $t$ 
converge to their equilibrium value
at long times:
\be
\mu_t({\cal A})=\int_{\cal A} p_t(\bG)\;  d\bG \to_{t\to\infty} \mu_{\rm eq}({\cal A})
\ee
The knowledge of the Pollicott-Ruelle resonances $\pm s_{\alpha}$
of the forward or backward semigroups (\ref{forward.expansion})
allows us to calculate the approach to the equilibrium values.

As a consequence of Gibbs' mixing property,
the coarse-grained entropy also converges toward its equilibrium value $S_{\rm eq}$
\be
S_t({\cal M}_{\bf l} \vert \{{\cal A}\}) = - k_{\rm B} \sum_{\cal A} \mu_t({\cal A}) \ln  \mu_t({\cal A})
\to_{t\to\infty} S_{\rm eq}
\ee
at rates given by the leading Pollicott-Ruelle resonance (\ref{disp.rel.diff}).

The time variation of the coarse-grained entropy over a time interval $\tau$
\be
  \Delta^{\tau}S = S_t({\cal M}_{\bf l}\vert \{{\cal A}\}) - S_{t-\tau}({\cal M}_{\bf l}\vert \{{\cal A}\})
 \label{entr.var}
\ee
can be separated into the {\it entropy flow}
\be
  \Delta_{\rm e}^{\tau}S \equiv
  S_{t-\tau}(\bPhi^{-\tau}{\cal M}_{\bf l}\vert \{{\cal A}\})-S_{t-\tau}({\cal M}_{\bf l}\vert \{{\cal A}\})
\label{entr.flow.def}
\ee
and the {\it entropy production}      		
\be
  \Delta_{\rm i}^{\tau}S \equiv \Delta^{\tau}S - \Delta_{\rm e}^{\tau}S
  = S_t({\cal M}_{\bf l}\vert\{ {\cal A}\})-S_t({\cal M}_{\bf l}\vert\{ \bPhi^{\tau}{\cal A}\})
  \label{entr.prod.def}
\ee
The entropy production can be calculated using the decomposition
of the time evolution in terms of the diffusive modes \cite{G98,GDG00,DGG02}.  
The idea of this {\it ab initio} derivation is the following.
First of all, we define a {\em lattice Fourier transform} as
\begin{equation}
G(\bG,{\bf l}) = \frac{1}{|{\cal B}|}\int_{\cal B}d{\bf k}
\; e^{i{\bf k}\cdot{\bf l}}\; \tilde G(\bG,{\bf k})
\label{1}
\end{equation}
where the wavenumber $\bf k$ belongs the first Brillouin zone $\cal B$ of the reciprocal
lattice and $|{\cal B}|$ denotes its volume.
We suppose that the initial probability distribution is a small perturbation $R_0$
with respect to the equilibrium one:
\begin{equation}
p_0(\bG,{\bf l}) =p_{\rm eq}\; \left[1+R_0(\bG,{\bf l})\right]
\label{2}
\end{equation}
The time evolution of this perturbation can be
written in terms of the lattice vector ${\bf d}(\bG,t)$
of displacement of the Brownian particle on the lattice
over the time interval $t$ as
\be 
R_t(\bG,{\bf l}) = \frac{1}{\vert{\cal B}\vert} \int_{\cal B} d{\bf k} \;
F_{\bf k} \; {\rm e}^{i {\bf k}\cdot \left[ {\bf l}+{\bf d}(\bG,t)\right]}
\ee
The measure of a cell ${\cal A}\subset{\cal M}_{\bf l}$ at time $t$ can be evaluated as
\be
\mu_t({\cal A}) = \int_{\cal A} p_t(\bG,{\bf l})\; d\bG
=\mu_{\rm eq}({\cal A})+\frac{1}{\vert{\cal B}\vert} \int_{\cal B} d{\bf k} \;
F_{\bf k} \; {\rm e}^{i {\bf k}\cdot{\bf l}} \; C({\bf k},t) \; {\rm e}^{s_{\bf k} t} \; \chi_{\bf k}({\cal A},t)
\equiv \mu_{\rm eq}({\cal A})+\delta\mu_t({\cal A})
\ee
where we have introduced the {\it hydrodynamic measure}
\be
\chi_{\bf k}({\cal A},t) = \mu_{\rm eq}({\cal M}) \frac{ \int_{\cal A} d\bG \; p_{\rm eq} \; {\rm e}^{i{\bf k}\cdot{\bf d}(\bG,t)}}{ \int_{\cal M} d\bG \; p_{\rm eq} \; {\rm e}^{i{\bf k}\cdot{\bf d}(\bG,t)}}
\label{hydro.meas}
\ee
as well as
\be
\frac{ \int_{\cal M} d\bG \; p_{\rm eq} \; {\rm e}^{i{\bf k}\cdot{\bf d}(\bG,t)}}{ \int_{\cal M} d\bG \; p_{\rm eq}}
= C({\bf k},t) \; {\rm e}^{s_{\bf k} t}
\ee
where $C({\bf k},t)$ is a subexponential function of time $t$ while the rate of
exponential decay is given by the dispersion relation of diffusion
\be
s_{\rm k} = \lim_{t\to\infty} \frac{1}{t}\ln \frac{ \int_{\cal M} d\bG \; p_{\rm eq} \; {\rm e}^{i{\bf k}\cdot{\bf d}(\bG,t)}}{ \int_{\cal M} d\bG \; p_{\rm eq}}
\ee
In the limit $t\to\infty$, Eq. (\ref{hydro.meas}) for the hydrodynamic measure
is similar to Eqs. (\ref{cumul0})-(\ref{cumul}) for the cumulative function
of the hydrodynamic mode.  The difference is that we here integrate
the singular density $\Psi_{\bf k}$ over a plain phase-space cell $\cal A$
while it was integrated over a phase-space curve in Eqs. (\ref{cumul0})-(\ref{cumul}).
In the long-time limit, the hydrodynamic measure converges to
a value which is invariant under the time evolution.
This condition is expressed by the equation
\be
{\rm e}^{s_{\bf k}\tau} \; \chi_{\bf k}({\cal A}) =
{\rm e}^{i{\bf k}\cdot{\bf d}(\bG_{\cal A},\tau)}\; \chi_{\bf k}(\bPhi^{-\tau}{\cal A})
\label{23b}
\ee
where ${\bf d}(\bG_{\cal A},\tau)$ is the displacement over the time interval $\tau$ of
a point $\bG_{\cal A}$ representative of the cell $\cal A$.

The hydrodynamic measure can now be expanded
as a series of the wavenumber
\be
\chi_{\bf k}({\cal A}) = \mu_{\rm eq}({\cal A}) + i {\bf k}\cdot{\bf T}({\cal A})+O({\bf k}^2)
\label{wvnbr.exp}
\ee
The first-order term corresponds to the generalized Takagi function of
the nonequilibrium steady state but here evaluated over a plain phase-space cell $\cal A$
instead of a curve as in Fig. \ref{fig7}. As a consequence of Eq. (\ref{23b}), this Takagi measure obeys
\be
{\bf T}({\cal A}) = {\bf T}(\bPhi^{-\tau}{\cal A})+{\bf d}(\bG_{\cal A},\tau)
\label{Tkg.bis}
\ee

If the phase-space region ${\cal M}_{\bf l}$ is partitioned into cells $\{{\cal A}_j\}$
which are sufficiently small, the displacements ${\bf d}_j$ over a time interval $\tau$
of the trajectories issued from these
cells obey the first sum rule:
\be
\sum_j {\bf d}_j \mu_{\rm eq}({\cal A}_j) = 0
\label{sum.rule.1}
\ee
while the coefficients ${\bf T}({\cal A}_j)$ satisfy the second sum rule:
\be
\sum_j\left[ {\bf d}_j {\bf T}({\cal A}_j) +  {\bf T}({\cal A}_j) {\bf d}_j  
+ {\bf d}_j {\bf d}_j \mu_{\rm eq}({\cal A}_j)\right] = 2 \; {\cal D} \; \tau \; \mu_{\rm eq}({\cal M})\; {\bf 1}
\label{sum.rule.2}
\ee
where $\bf 1$ denotes the identity matrix of the physical space.
We notice that this sum rule is a form of the Green-Kubo formula.

Now, the entropy production (\ref{entr.prod.def}) becomes
\be
\Delta_{\rm i}^{\tau}S=\frac{1}{2}
\sum_{\bPhi^{\tau}{\cal A}_j\subset{\cal M}_{\bf l}} 
\frac{[\delta\mu_t(\bPhi^{\tau}{\cal A}_j)]^2}{\mu_{\rm eq}({\cal A}_j)} 
- \frac{1}{2}
\sum_{{\cal A}_j\subset{\cal M}_{\bf l}} \frac{[\delta\mu_t({\cal A}_j)]^2}{\mu_{\rm eq}({\cal A}_j)} 
+ O(\delta\mu_t^3)
\ee
Substituting the wavenumber expansion (\ref{wvnbr.exp}) of the perturbations $\delta\mu_t({\cal A}_j)$,
using Eq. (\ref{Tkg.bis}) and the sum rules (\ref{sum.rule.1})-(\ref{sum.rule.2}), 
we finally obtain the fundamental result that 
the entropy production takes the value expected from nonequilibrium thermodynamics
\be
\boxed{\Delta_{\rm i}^{\tau}S \simeq \tau \, k_{\rm B} \, {\cal D} \, \frac{({\rm grad} \, n)^2}{n}}
  \label{entr.thermo}
\ee
where $n =  \mu_t({\cal M}_l)$ is the particle density \cite{G98,GDG00,DGG02}.
This result is a consequence of the singular character of the
nonequilibrium steady state.  It can indeed be shown that
the entropy production would vanish if the steady state had a density
given by a function instead of a non-trivial distribution.
In the present derivation from the hydrodynamic modes,
the singular character of the nonequilibrium steady state is directly
coming from the hydrodynamic modes.
If we consider the nonequilibrium steady state (\ref{NESS.prob.dens}) 
between two reservoirs of particles separated by a distance $L$,
the invariant measure is still absolutely continuous 
with respect to the Lebesgue measure, although very different
and much more complicated than the microcanonical equilibrium measure.
For finite $L$, there is a characteristic scale below which the
continuity of the measure manifests itself.
This scale decreases exponentially with the distance
with respect to the physical boundaries.  For a partition into cells
which are larger than this characteristic scale, the entropy production
which depends on the partition takes the thermodynamic value 
(\ref{entr.thermo}).  Instead, it vanishes for smaller cells.
In the limit $L\to\infty$, the invariant measure looses
its absolute continuity and becomes singular so that
the thermodynamic value is obtained for cells of arbitrarily
small size \cite{G98}.  In the present derivation starting from
the hydrodynamic modes, the system is infinite
so that the state is singular and the thermodynamic value
is also obtained for cells of arbitrarily small size \cite{GDG00,DGG02}.
This shows that the thermodynamic entropy production
does not finally depend on the chosen partition.

 \section{Fluctuation theorem}
 \label{FT}
 
 In the present section, we deal with systems in 
 nonequilibrium steady states.
Our purpose is to obtain their properties such as the mean values of the currents
in terms of the affinities (also known as the thermodynamic forces)
and their fluctuations.  These fluctuation properties
are expressed in terms of large-deviation functions
which obey symmetry relations known as the fluctuation theorem
\cite{ECM93,ES94,GC95,K98,C99,LS99,M99,G04a,AG04,AG05,AG06}.

\subsection{Description in terms of a master equation}

In order to maintain a nonequilibrium steady state,
the system should be in contact with at least two thermostats
or chemiostats at different temperatures
or chemical potentials.  Accordingly, the system is not isolated
and its dynamics depends on the degrees of freedom of the environment.
We expect, on the one hand, a dissipation of energy toward the
environment and, on the other hand, a reaction of the
environment back onto the subsystem in the form of fluctuations.
As a consequence, boundary conditions should be imposed
which introduce in general some stochasticity from the boundary
coming from the external degrees of freedom.
As aforementioned, we are led to a stochastic description 
in terms of some master equation
\be
\frac{d}{dt}P_t(\omega) = \sum_{\rho,\omega'(\neq\omega)} \left[P_t(\omega') 
W_{\rho}(\omega'\vert\omega) - P_t(\omega) W_{-\rho}(\omega\vert\omega')\right]
\label{master0}
\ee
for the probability $P_t(\omega)$ to find the system in the state $\omega$ by
the time $t$. $W_{\rho}(\omega\vert\omega')$ denotes the rate
of the transition $\omega\to\omega'$ for the elementary process $\rho$.
There is a transition rate associated with several 
possible elementary processes $\rho=\pm 1,\pm 2,...,\pm r$.  
The total probability is conserved $\sum_{\omega}P_t(\omega)=1$ for all times $t$.

A graph $G$ can be associated with the random process
in which each state $\omega$ constitutes a vertex while the edges represent the 
transitions $\omega\to\omega'$
allowed by the elementary processes $\rho=\pm1,...,\pm r$.
Schnakenberg has developed a graph analysis of the random processes
by considering the cycles $\vec{C}$ of the graph $G$ \cite{S76}.
The method is based on the definition of a maximal tree
$T(G)$ of the graph $G$, which should satisfy the following properties:

(1) $T(G)$ is a covering subgraph of $G$, i.e., $T(G)$ contains all the vertices of $G$
and all the edges of $T(G)$ are edges of $G$;

(2) $T(G)$ is connected;

(3) $T(G)$ contains no circuit (i.e., no cyclic sequence of edges).

\noindent In general a given graph $G$ has several maximal trees $T(G)$ (see Ref. \cite{S76}).

The edges $s_l$ of $G$ which do not belong to $T(G)$ are called the chords of $T(G)$.
If we add to $T(G)$ one of its chords $s_l$, the resulting subgraph $T(G)+s_l$
contains exactly one circuit, $C_l$, which is obtained from $T(G)+s_l$ by removing
all the edges which are not part of the circuit.  The set of circuits $\{C_1,C_2,...,C_l,...\}$
is called a fundamental set.  An arbitrary orientation can be assigned to each circuit $C_l$
to define a cycle $\vec{C}_l$.  A maximal tree $T(G)$ together with its associated
fundamental set of cycles $\{\vec{C}_1,\vec{C}_2,...,\vec{C}_l,...\}$ provides
a decomposition of the graph $G$.  We notice that the maximal tree $T(G)$
can be chosen arbitrarily because each cycle $\vec{C}_l$ can be redefined by
linear combinations of the cycles of the fundamental set.
Schnakenberg noticed that the product of the ratios of forward and backward transition
rate over a cycle is independent of the states of the cycle and can be expressed
in terms of the nonequilibrium constraints imposed on the system from
its environment \cite{S76}.  These nonequilibrium constraints are expressed in terms
of the affinities or generalized thermodynamic forces which we shall define
in specific cases here below.

For isothermal processes, the ratio of the
transition rates of a forward and backward elementary processes 
$\omega\underset{-\rho}{\overset{+\rho}{\rightleftharpoons}}\omega'$
typically obeys the relation:
\be
\frac{W_{+\rho}(\omega\vert\omega')}{W_{-\rho}(\omega'\vert\omega)}
= {\rm e}^{\beta(X_{\omega}-X_{\omega'})}
\label{ratio}
\ee
where $X_{\omega}$ denotes the value of the
thermodynamic potential corresponding to the situation (see Table II).
This relation corresponds to an assumption of local thermodynamic equilibrium.

\vskip 0.5 cm
\hrule \vskip1pt \hrule height1pt
\vskip 0.2 cm

\centerline{TABLE II. The statistical ensemble and thermodynamic potential to use}

\centerline{for the different possible conditions imposed on a system.}

\vskip 0.2 cm
\hrule 
\vskip 0.1 cm

$$
\vcenter{\openup1\jot 
\halign{#\hfil&\qquad#\hfil&\qquad#\hfil&\qquad#\hfil&\qquad#\hfil\cr
{\it fixed} &   &    &  &  \cr 
{\it quantities} & {\it system} &  {\it ensemble}  & {\it potential} & $X$ \cr & & & & \cr
$V,T$ & closed & isochoric-isothermal canonical  & Helmholtz free energy & $F=E-TS$ \cr 
$P,T$ & closed & isobaric-isothermal  &  Gibbs free enthalpy & $G=F+PV$ \cr 
$V,T,\mu$ & open & isochoric-isothermal-isopotential grand can.  &  grand potential & $J=F-\mu N$ \cr 
$P,T,\mu$ & open & isobaric-isothermal-isopotential & - & $K=G-\mu N$ \cr 
}}
$$

\vskip 0.1 cm
\hrule height1pt \vskip1pt \hrule 
\vskip 0.5 cm

If we take the product of these ratios over a cycle, we get
\be
\prod_{\omega\in\vec{C}}\frac{W_{+\rho}(\omega\vert\omega')}{W_{-\rho}(\omega'\vert\omega)}
=\exp\left[\beta\sum_{\omega\in\vec{C}}(X_{\omega}-X_{\omega'})\right]
= \exp\left(-\beta\oint \frac{\partial X}{\partial \omega}d\omega\right)
\equiv \exp A(\vec{C})
\label{ratio.product}
\ee
which defines the affinity of the cycle.

In order to fix the ideas, we here consider two simple examples
which are found in the biological nanosystem F$_1$-ATPase.
The first example concerns the rotational motion of the shaft of the motor
by the angle $\theta$ which is the state variable.
For a given chemical state, the rotational motion takes place in
a potential $V(\theta)$ which induces an internal torque
$\tau_{\rm int}=-\partial V/\partial \theta$ and under an external torque $\tau_{\rm ext}$. 
The internal potential $V(\theta)$ is periodic in the angle $\theta$:
$V(2\pi)=V(0)$.
Under laboratory experimental conditions,
the temperature and pressure are fixed so that the potential is
a free enthalpy.  The affinity of the cycle $\vec{C}$ corresponding
to the full rotation of the shaft by $\Delta\theta=2\pi$ is thus given by
\be
A(\vec{C}) = - \beta \oint \frac{\partial G}{\partial \theta}d\theta 
= \beta \oint \left(-\frac{\partial V}{\partial \theta}+\tau_{\rm ext}\right) d\theta = 2\pi\; \beta \; \tau_{\rm ext}
\ee
For this example, the affinity is proportional to the external torque which performs
some work on the nanosystem and maintains it out of equilibrium.
Under the effect of an external torque, the F$_1$-ATPase can synthesize ATP molecules
from ADP and P$_{\rm i}$.
The second example is the binding of ATP to the catalytic site of F$_1$-ATPase:
\be
{\rm ATP} + 0 \underset{-\rho}{\overset{+\rho}{\rightleftharpoons}} 1
\ee
The angle $\theta$ is supposed to remain fixed during this elementary process.
This is a chemical reaction in which the numbers of molecules
of the different species change and we have to use the potential $K$
if the process takes place at given temperature and pressure.
The chemical potential of ATP is fixed by the concentration [ATP]
in the environment of the F$_1$-ATPase according to
\be
\mu_{\rm ATP} = \mu_{\rm ATP}^0 + k_{\rm B}T \ln \frac{[{\rm ATP}]}{c^0} 
\ee
where $c^0$ is the reference concentration of one mole per liter
while $\mu_{\rm ATP}^0$ is the corresponding standard chemical potential.
The ratio (\ref{ratio}) of the transition rates of the forward and backward elementary processes
is given in terms of the potential $X=K=G-\mu N$ where $\mu=\mu_{\rm ATP}$
and $N=N_{\rm ATP}$ denote the chemical potential  and number of ATP molecules 
inside the F$_1$-ATPase. We find that
\be
\Delta K = K_0-K_1 = G_0-\mu N_0 - G_1 +\mu N_1 = G_0-G_1+\mu = V_0(\theta)-V_1(\theta)+\mu_{\rm ATP}
\ee
since $N_0=0$ and $N_1=1$.  Therefore, the ratio of the transitions rates of these elementary processes should satisfy
\be
\frac{W_{+\rho}(0\vert 1)}{W_{-\rho}(1\vert 0)}
= {\rm e}^{\beta\left[ V_0(\theta)-V_1(\theta) + \mu_{\rm ATP}^0 \right]} \; \frac{[{\rm ATP}]}{c^0} 
\label{ratio.ATP}
\ee
For a chemical reaction under isothermal and isobaric conditions, 
the product of these ratios over a cycle with $G(2\pi)=G(0)$ will in general be
\be
\prod_{\omega\in\vec{C}}\frac{W_{+\rho}(\omega\vert\omega')}{W_{-\rho}(\omega'\vert\omega)}
= \exp\left(-\beta\oint \frac{\partial K}{\partial\omega} d\omega\right)
= \exp\left( \beta \sum_{i=1}^c \mu_i \Delta N_i(\vec{C})\right) =\exp A(\vec{C})
\ee
so that the affinity of the cycle is here given by the chemical potential difference 
between the initial and final states of the cycle divided
by the thermal energy.  At equilibrium, this difference of chemical potential vanishes.
We conclude that nonequilibrium conditions can be maintained by some differences
of chemical potential between the reactants and the products.
This is the case in nanomotors fuelled by chemical nonequilibrium concentrations
such as the F$_1$-ATPase in a solution of ATP.

The master equation admits in general a time-independent stationary solution 
$dP_{\rm st}(\omega)/dt=0$.  Out of equilibrium [i.e., if $A(\vec{C})\neq 0$ for some cycle $\vec{C}$], 
this stationary solution describes the nonequilibrium steady state.
If all the nonequilibrium constraints vanish [i.e., if $A(\vec{C})= 0$ for all the cycles $\vec{C}$], 
the stationary solution represents the equilibrium state $P_{\rm eq}(\omega)$ 
which obeys the {\it detailed balance} conditions:
\be
P_{\rm eq}(\omega') W_{\rho}(\omega'\vert\omega) = P_{\rm eq}(\omega) W_{-\rho}(\omega\vert\omega')
\label{detailed.balance}
\ee
which should hold for all the elementary processes $\rho=\pm1, \pm 2,...,\pm r$.
These conditions are the consequences of Boltzmann equilibrium distribution and of
the identities (\ref{ratio}).  This can be checked in the canonical ensemble for instance.
Indeed, according to the Boltzmann distribution,
the equilibrium probability of a state $\omega$ is given by 
$P_{\rm eq}(\omega) = Z_{\omega}^{-1} {\rm e}^{-\beta E_{\rm \omega}} 
= {\rm e}^{-\beta F_{\rm \omega}}$
so that the detailed balance conditions (\ref{detailed.balance}) are satisfied because of
Eq. (\ref{ratio}) with $X_{\omega}=F_{\omega}$.  The master equation  (\ref{master0})
is said to obey detailed balance if it admits a stationary solution satisfying the
conditions (\ref{detailed.balance}).

\subsection{Entropy, entropy flow, and entropy production}

The entropy of the system at the current time $t$ can be defined as \cite{G04a}
\be
S_t = \sum_{\omega} S^0(\omega) P_t(\omega) - k_{\rm B} \sum_{\omega} P_t(\omega) \ln P_t(\omega) 
\ee
The first term should be included if the graining into the states $\omega$ is coarse enough
that an entropy $S^0(\omega)$ is naturally associated with them.
This is the case if $\omega$ represents the number of molecules of each species in the system for instance.

The time variation of the entropy can be decomposed into a term for the entropy flow to
the environment and the entropy production, in much the same way 
as done in Eqs. (\ref{entr.var})-(\ref{entr.prod.def}) for the deterministic description
\cite{S76,JVN84}:
\be
\frac{dS}{dt} = \frac{d_{\rm e}S}{dt} + \frac{d_{\rm i}S}{dt}
\ee
Introducing the {\it currents}
\be
J_{\rho}(\omega,\omega';t) \equiv 
P_t(\omega') W_{\rho}(\omega'\vert\omega) - P_t(\omega) W_{-\rho}(\omega\vert\omega')
\label{current.def}
\ee
and the {\it affinities}
\be
A_{\rho}(\omega,\omega';t) \equiv k_{\rm B} \ln \frac{P_t(\omega') W_{+\rho}(\omega'\vert\omega)}{P_t(\omega) W_{-\rho}(\omega\vert\omega')} 
\label{affinity.def}
\ee
the {\it entropy flow} is here given by
\be
\frac{d_{\rm e}S}{dt} = \sum_{\rho,\omega,\omega'} J_{\rho}(\omega,\omega';t) \left[ S^0(\omega) \; 
- \frac{k_{\rm B}}{2} \ln\frac{W_{+\rho}(\omega'\vert\omega)}{W_{-\rho}(\omega\vert\omega')} \right] 
\ee
while the {\it entropy production} is given by
\be
\frac{d_{\rm i}S}{dt} = \frac{k_{\rm B}}{2} \sum_{\rho,\omega,\omega'} 
\left[ P_t(\omega') W_{\rho}(\omega'\vert\omega) - P_t(\omega) W_{-\rho}(\omega\vert\omega')\right] \; 
\; \ln\frac{P_t(\omega') W_{+\rho}(\omega'\vert\omega)}{P_t(\omega) W_{-\rho}(\omega\vert\omega')} = \frac{1}{2} \sum_{\rho,\omega,\omega'} J_{\rho}(\omega,\omega';t) \;
A_{\rho}(\omega,\omega';t) \geq 0
\label{d_iS/dt}
\ee
The entropy production is always non-negative in agreement with the second law
of thermodynamics. In a system without nonequilibrium constraint, the probability
distribution $P_t(\omega)$ undergoes a relaxation toward the equilibrium state $P_{\rm eq}(\omega)$
for which the entropy production vanishes because of the detailed balance conditions (\ref{detailed.balance}).  During this relaxation, the entropy production is positive.
Therefore, the solution of a master equation satisfying detailed balance necessarily converges toward the equilibrium state.  In a system with nonequilibrium constraints, the relaxation proceeds
toward a nonequilibrium steady state in which the entropy production continues to be positive.

\subsection{Fluctuation theorem in nonequilibrium steady states}

The master equation rules the time evolution of a random process,
the trajectories of which undergo random jumps between the states $\omega$.
The jumps occur at successive times: 
$0 < t_1 < t_2 < \cdots < t_m < t$.  The waiting times $t_{j+1}-t_{j}$ on some state
$\omega(t)=\sigma_j$ are
exponentially distributed according to the transition rates.
Accordingly, a trajectory or path can be represented as
\be
\pmb{\omega}(t) = \sigma_0 {\overset{\rho_1}\longrightarrow} \sigma_1
{\overset{\rho_2}\longrightarrow} \sigma_2 {\overset{\rho_3}\longrightarrow} \cdots
{\overset{\rho_n}\longrightarrow} \sigma_n
\label{path}
\ee
where $\sigma_j=\omega(t)$ for $t_{j}<t<t_{j+1}$ and $\rho_j$ is the elementary process
causing the transition $\sigma_{j-1}\to\sigma_j$.  An alternative representation of the
trajectory is provided by sampling at regular time intervals $t_k=k\tau$:
\be
\pmb{\omega}(t) = \omega_0\omega_1\omega_2\cdots \omega_{n-2}\omega_{n-1}
\label{path.tau}
\ee
with $\omega_k=\omega(k\tau)$.  
The probability of a trajectory or path in the stationary state is given by
\be
\mu\left[\pmb{\omega}(t)\right] = P_{\rm st}(\omega_0) \;
P_{\tau}(\omega_0 {\overset{\tilde\rho_1}\longrightarrow} \omega_1) \;
P_{\tau}(\omega_1 {\overset{\tilde\rho_2}\longrightarrow} \omega_2) \;
\cdots \;
P_{\tau}(\omega_{n-2} {\overset{\tilde\rho_{n-1}}\longrightarrow} \omega_{n-1})
\label{proba.traj}
\ee
where $\tilde\rho_k=0,\pm 1,\pm 2,...,\pm r$ denotes either an elementary process
if $\tilde\rho_k \neq 0$ or the continuation of the current state if $\tilde\rho_k=0$.
The conditional probabilities of such transitions are given by
\begin{equation}
P_{\tau}(\omega{\overset{\tilde\rho}\longrightarrow} \omega') =
\left\{ 
\begin{array}{ll}
W_{\tilde\rho}(\omega\vert\omega') \; \tau + O(\tau^2) & \quad {\rm if} \ \tilde\rho \neq 0 \\ & \\
1-\sum_{\tilde\rho=\pm 1}^{\pm r} \sum_{\omega'} W_{\tilde\rho}(\omega\vert\omega') 
\; \tau + O(\tau^2) & \quad {\rm if} \ \tilde\rho = 0 \\
\end{array}
\right.
\label{transition.proba}
\ee
Accordingly, the probability of a trajectory (\ref{proba.traj}) can be approximated
as the product of the transition rates multiplied by $\tau$ of the actual transitions 
$\sigma_{j-1} {\overset{\rho_j}\longrightarrow} \sigma_j$:
\be
\mu\left[\pmb{\omega}(t)\right] \simeq P_{\rm st}(\sigma_0) \; 
W_{\rho_1}(\sigma_0\vert\sigma_1) \; \tau \; 
W_{\rho_2}(\sigma_1\vert\sigma_2) \; \tau\; 
\cdots
W_{\rho_{m}}(\sigma_{m-1}\vert\sigma_m) \; \tau
\label{proba.traj.approx}
\ee
with $\sigma_0=\omega_0$ and $\sigma_m=\omega_{n-1}$.
In the case $\tilde\rho_k=0$, no transition occurs in the corresponding time interval $\tau$,
which contributes by a factor of order one in $\tau$.  Similarly,
the probability of the time-reversed trajectory or path is given by
\be
\mu\left[\pmb{\omega}^{\rm R}(t)\right] \simeq P_{\rm st}(\sigma_m) \; 
W_{-\rho_{m}}(\sigma_m\vert\sigma_{m-1}) \; \tau
\cdots
W_{-\rho_2}(\sigma_2\vert\sigma_1) \; \tau\; 
W_{-\rho_1}(\sigma_1\vert\sigma_0) \; \tau \; 
\label{proba.rev.traj.approx}
\ee

In order to evaluate how detailed balance is satisfied along a trajectory,
we can introduce the fluctuating quantity
\be
{\cal Z}(t) \equiv \ln \frac{W_{\rho_1}(\sigma_0\vert\sigma_1) 
W_{\rho_2}(\sigma_1\vert\sigma_2) \cdots
W_{\rho_{m}}(\sigma_{m-1}\vert\sigma_m)}
{W_{-\rho_1}(\sigma_1\vert\sigma_0) 
W_{-\rho_2}(\sigma_2\vert\sigma_1) \cdots
W_{-\rho_{m}}(\sigma_{m}\vert\sigma_{m-1})}
\label{Z}
\ee
following Lebowitz and Spohn \cite{LS99,G04a}.
The denominator of Eq. (\ref{Z}) is the product of the rates of the transitions which are time reversed
with respect to those appearing in the numerator.
Therefore, the quantity (\ref{Z}) is defined 
in terms of the product of the ratios (\ref{ratio}) between the forward and backward processes
at each jump event along the trajectory ($0 < t_1 < t_2 < \cdots < t_m < t$).

If the process is isothermal and if Eq. (\ref{ratio}) holds, the quantity (\ref{Z})
is given by ${\cal Z}(t) =  \beta (X_{\sigma_0}-X_{\sigma_m})= \beta (X_0-X_t)$
where the potential $X$ is defined in Table II.
If the system is at equilibrium, we expect that $X_t$ is distributed as $X_0$ so that
$\langle{\cal Z}(t)\rangle=0$ in agreement with the fact that detailed balance
is satisfied at equilibrium.  In constrast, we expect a nontrivial behavior of
the fluctuating quantity ${\cal Z}(t)$ out of equilibrium.
Furthermore, we notice that this quantity can be written in the long-time limit as
\be
{\cal Z}(t) \simeq \ln \frac{\mu[\pmb{\omega}(t)]}{\mu[\pmb{\omega}^{\rm R}(t)]}
\label{Z.proba}
\ee
in terms of the multiple-time probabilities of the path $\pmb{\omega}(t)$ 
and time-reversed path $\pmb{\omega}^{\rm R}(t)$ entering the
definitions of the dynamical entropies (\ref{dyn.entr}) and (\ref{TR.dyn.entr}).
Indeed, the numerator in Eq. (\ref{Z.proba}) is the measure (\ref{proba.traj.approx})
of the path itself while the denominator is the measure (\ref{proba.rev.traj.approx})
of the time-reversed path and the boundary term 
$\ln\left[ P_{\rm st}(\sigma_0) /P_{\rm st}(\sigma_m) \right]$
becomes negligible in the long-time limit with respect to the sum of the other terms
which increases linearly with time.  Accordingly, we recover the definition (\ref{Z})
after neglecting the boundary term in the long-time limit.

In order to characterize the large-deviation properties of ${\cal Z}(t)$, its generating function can
be introduced as
\be
Q(\eta) \equiv \lim_{t\to\infty} -\frac{1}{t} \ln \langle {\rm e}^{-\eta {\cal Z}(t)}\rangle
\label{Q}
\ee
where the average is carried out in the nonequilibrium steady state.

According to the result (\ref{Z.proba}), we have the following succession of equalities
\be
\langle {\rm e}^{-\eta {\cal Z}(t)}\rangle \simeq
\sum_{\pmb{\omega}(t)} \mu[\pmb{\omega}(t)]
\left\{\frac{\mu[\pmb{\omega}^{\rm R}(t)]}{\mu[\pmb{\omega}(t)]}\right\}^{\eta} 
= \sum_{\pmb{\omega}(t)} \mu[\pmb{\omega}^{\rm R}(t)]
\left\{\frac{\mu[\pmb{\omega}(t)]}{\mu[\pmb{\omega}^{\rm R}(t)]}\right\}^{1-\eta}  
= \sum_{\pmb{\omega}^{\rm R}(t)} \mu[\pmb{\omega}(t)]
\left\{\frac{\mu[\pmb{\omega}^{\rm R}(t)]}{\mu[\pmb{\omega}(t)]}\right\}^{1-\eta}  
\simeq \langle {\rm e}^{-(1-\eta){\cal Z}(t)}\rangle 
\label{id}
\ee
where the third equality uses the fact that summing over the time-reversed paths
is the same as summing over the paths since both sums cover all the possible paths.
If we substitute the result (\ref{id}) in the generating function (\ref{Q}), we obtain the 
so-called {\it fluctuation theorem}
\begin{equation}
\boxed{Q(\eta) = Q(1-\eta)}
\label{FT.Q}
\end{equation}
A consequence of the fluctuation theorem (\ref{FT.Q}) and of the definition (\ref{Q}) is that
$Q(0) = Q(1) = 0$.

We now consider the large-deviation function defined as the rate of exponential decay of the probability 
that the quantity $\frac{{\cal Z}(t)}{t}$ takes its value in the interval $(\zeta,\zeta+d\zeta)$:
\begin{equation}
\mu\left[ \frac{{\cal Z}(t)}{t}\in(\zeta,\zeta+d\zeta)\right] \sim {\rm e}^{-R(\zeta) t} \; d\zeta
\label{R}
\end{equation}
This large-deviation function is related to the generating function (\ref{Q}) by 
the Legendre transform
\be
R(\zeta) = {\rm max}_{\eta} \left[ Q(\eta) - \zeta \; \eta \right]
\label{RmaxQ}
\ee

The fluctuation theorem (\ref{FT.Q}) has for corollary that the large-deviation function (\ref{R}) satisfies the identity
\be
\boxed{\zeta = R(-\zeta) - R(\zeta)}
\label{FT.R}
\ee
or equivalently that
\be
\boxed{\frac{\mu\left[ \frac{{\cal Z}(t)}{t}\in(\zeta,\zeta+d\zeta)\right]}
{\mu\left[ \frac{{\cal Z}(t)}{t}\in(-\zeta,-\zeta+d\zeta)\right]} 
\simeq {\rm e}^{\zeta t}  \qquad \mbox{for} \quad t\to +\infty}
\label{FT.prob.R}
\ee
In the form (\ref{FT.R}), the fluctuation theorem appears as
a large-deviation relationships very similar to the escape-rate formula (\ref{eq:escformula}).
Indeed, we find in the left-hand member an irreversible property (here $\zeta$)
and in the right-hand member the difference between two large-deviation quantities
which here are two decay rates of multiple-time probabilities.  
We shall further comment on this connection in the conclusions.

Now, the mean value of the fluctuating quantity ${\cal Z}(t)$ can be interpreted in terms of
the entropy production here defined by Eq. (\ref{d_iS/dt}).
Indeed, Eq. (\ref{Z}) can be transformed as follows:
\be
{\cal Z}(t) = \ln \frac{P(\sigma_0) W_{\rho_1}(\sigma_0\vert\sigma_1) 
P(\sigma_1)W_{\rho_2}(\sigma_1\vert\sigma_2) \cdots
P(\sigma_{m-1})W_{\rho_{m}}(\sigma_{m-1}\vert\sigma_m)}
{P(\sigma_1)W_{-\rho_1}(\sigma_1\vert\sigma_0) 
P(\sigma_2)W_{-\rho_2}(\sigma_2\vert\sigma_0) \cdots
P(\sigma_m)W_{-\rho_{m}}(\sigma_{m}\vert\sigma_{m-1})} - \ln \frac{P(\sigma_0)}{P(\sigma_m)}
\simeq \frac{1}{k_{\rm B}}\sum_{j=1}^m A_{\rho_j}(\sigma_{j-1},\sigma_{j},t_j)
\label{transf.Z}
\ee
in terms of the affinities (\ref{affinity.def}) associated with each random jumps.
Since these events occur on average at rates given by the currents (\ref{current.def}),
the statistical average of the quantity ${\cal Z}(t)$ is given by
\be
\langle{\cal Z}(t)\rangle \simeq \frac{1}{k_{\rm B}}\int_0^t dt' 
\frac{1}{2} \sum_{\rho,\omega,\omega'} J_{\rho}(\omega,\omega';t') \;
A_{\rho}(\omega,\omega';t') 
= \frac{1}{k_{\rm B}}\int_0^t dt' \; \frac{d_{\rm i}S}{dt'} = 
\frac{1}{k_{\rm B}} \Delta_{\rm i}^tS
\ee
according to Eq. (\ref{d_iS/dt}).
As a consequence, the mean rate of linear increase of the fluctuating quantity
$\cal Z$ gives us the entropy production rate in a nonequilibrium steady state:
\be
\frac{d_{\rm i}S}{dt}\Big\vert_{\rm st} = k_{\rm B} \; \lim_{t\to\infty} \frac{1}{t} \langle {\cal Z}(t)\rangle
= k_{\rm B} \; \frac{dQ}{d\eta}(0)
\label{d_iS.Z}
\ee

\subsection{Fluctuation theorem for the currents and consequences}

A nonequilibrium steady state is maintained by currents
between the system and reservoirs (thermostats or chemiostats).
The number of independent currents is equal to the number of
independent differences of chemical potentials between the chemiostats.
Schnakenberg graph analysis and the equality (\ref{ratio.product})
show that the nonequilibrium constraints are hidden in the
product of the ratios of forward and backward transition rates
along the cycles of the graph associated with the process \cite{S76}.
A graph has many possible
cycles $\vec{C}$ but the corresponding affinities take a number
of different values limited by the number
of independent nonequilibrium constraints,
$A(\vec{C})=A_{\gamma}$ for $\vec{C} \in \gamma$.
For an isothermal reactive process between $a$ chemiostats,
we have only $a-1$ different values $A_{\gamma}=\Delta\mu_{\gamma}/(k_{\rm B}T)$ \cite{AG04}.
We denote by $j_{\gamma}(t)$ the instantaneous current corresponding
to the nonequilibrium constraint $\gamma$.  The mean value 
of this instantaneous current is related to the currents (\ref{current.def}) by
\be
J_{\gamma} \equiv \langle j_{\gamma}\rangle =\lim_{t\to\infty} \frac{1}{t} \int_0^t dt' j_{\gamma}(t') = \sum_{\vec{C}\in\gamma} J(\vec{C}) = 
\sum_{\vec{C}\in\gamma} \sum_{\rho,\omega,\omega' \in \vec{C}} J_{\rho}(\omega,\omega';t)\Big\vert_{\rm st}
\ee

The instantaneous currents $j_{\gamma}(t)$ fluctuate in time and their
fluctuation properties can be studied thanks to their generating function
defined as
\be
Q(\{\eta_{\gamma}\};\{A_{\gamma}\}) \equiv \lim_{t\to\infty} - \frac{1}{t} \ln 
\langle {\rm e}^{-\sum_{\gamma} \eta_{\gamma} A_{\gamma} \int_0^t j_{\gamma}(t') dt'}\rangle
\label{gen.macro}
\ee
If we define $\lambda_{\gamma}\equiv A_{\gamma} \eta_{\gamma}$,
the generating function (\ref{gen.macro}) becomes
\begin{equation}
q(\{\lambda_{\gamma}\};\{{A}_{\gamma}\}) \equiv Q(\{\lambda_{\gamma}/{A}_{\gamma}\};\{{A}_{\gamma}\})=
 \lim_{t\to\infty} - \frac{1}{t} \ln \langle {\rm e}^{-\sum_{\gamma} \lambda_{\gamma} \int_0^t j_{\gamma}(t') dt'}\rangle
\label{new.gen.macro}
\end{equation}
which shows that the statistical moments of the instantaneous currents 
$j_{\gamma}(t)$ can be generated
by successive differentiations.
In particular, the mean flux or rate of the overall process $\gamma$ is given by
\begin{equation}
J_{\gamma} = \frac{\partial q}{\partial \lambda_{\gamma}}\Big\vert_{\lambda_{\gamma}=0}
\label{q.macro.flux}
\end{equation}

Now, the generating function obeys the {\it fluctuation theorem for the currents} \cite{AG04,AG05,AG06}:
\be
\boxed{q(\{\lambda_{\gamma}\};\{{A}_{\gamma}\}) = q(\{{A}_{\gamma}-\lambda_{\gamma}\};\{{A}_{\gamma}\})}
\label{FT.gen.macro}
\ee
This result can be obtained by giving the exponent $\eta_j=\eta_{\gamma}$ each time a random jump 
$\sigma_{j-1}{\overset{\rho_j}\longrightarrow}\sigma_j$ concerns a transition in a cycle $\vec{C}\in\gamma$ and a vanishing exponent $\eta_j=0$ otherwise.
In this way, a calculation similar to Eq. (\ref{id}) can be performed as follows:
\bea
&& \langle {\rm e}^{-\sum_{\gamma} \eta_{\gamma} A_{\gamma} \int_0^t j_{\gamma}(t') dt'}\rangle
\simeq \sum_{\pmb{\omega}(t)} \mu[\pmb{\omega}(t)] \prod_j \left[ 
\frac{ W_{-\rho_j}(\sigma_j\vert\sigma_{j-1})}{ W_{\rho_j}(\sigma_{j-1}\vert\sigma_j)}\right]^{\eta_j}
= \sum_{\pmb{\omega}(t)} \mu[\pmb{\omega}(t)] \prod_j \left[ 
\frac{W_{\rho_j}(\sigma_{j-1}\vert\sigma_j)}{W_{-\rho_j}(\sigma_j\vert\sigma_{j-1})}\right]^{1-\eta_j}
\prod_j 
\frac{ W_{-\rho_j}(\sigma_j\vert\sigma_{j-1})}{ W_{\rho_j}(\sigma_{j-1}\vert\sigma_j)}
\nonumber\\
&& \simeq\sum_{\pmb{\omega}(t)} \mu[\pmb{\omega}^{\rm R}(t)] \prod_j \left[ 
\frac{W_{\rho_j}(\sigma_{j-1}\vert\sigma_j)}{W_{-\rho_j}(\sigma_j\vert\sigma_{j-1})}\right]^{1-\eta_j}
= \sum_{\pmb{\omega}^{\rm R}(t)} \mu[\pmb{\omega}(t)] \prod_j \left[ 
\frac{W_{-\rho_j}(\sigma_j\vert\sigma_{j-1})}{W_{\rho_j}(\sigma_{j-1}\vert\sigma_j)}\right]^{1-\eta_j}
\simeq \langle {\rm e}^{-\sum_{\gamma} (1-\eta_{\gamma}) A_{\gamma} \int_0^t j_{\gamma}(t') dt'}\rangle
\eea
which establishes the results (\ref{FT.gen.macro}).
A full proof is given in Ref. \cite{AG05}.

The fluctuation theorem has important consequences on the coefficients
of the expansion of the average currents as Taylor series
of the associated affinities
\be
J_{\alpha} = \sum_{\beta} L_{\alpha\beta} {A}_{\beta} 
+ \frac{1}{2} \sum_{\beta,\gamma} M_{\alpha\beta\gamma} {A}_{\beta}  {A}_{\gamma} 
+ \frac{1}{6} \sum_{\beta,\gamma,\delta} N_{\alpha\beta\gamma\delta} 
{A}_{\beta}  {A}_{\gamma} {A}_{\delta} + \cdots
\label{macro.fluxes-affinities}
\ee
The linear response of the fluxes ${J}_{\alpha}$ 
to a small perturbation in the affinities ${A}_{\beta}$ 
is characterized by the Onsager coefficients $L_{\alpha\beta}$, 
and the nonlinear response by the higher-order coefficients
$M_{\alpha\beta\gamma}$, $N_{\alpha\beta\gamma\delta}$,...

The Onsager reciprocity relations can be derived from the fluctuation theorem
as follows. The Onsager coefficients are given by differentiating the generating function:
\be
L_{\alpha\beta} = \frac{\partial {J}_{\alpha}}{\partial {A}_{\beta}}\Big\vert_{{A}_{\gamma}=0}
= \frac{\partial^2 q}{\partial \lambda_{\alpha}\partial {A}_{\beta}}(0;0)
\label{L}
\ee
If we differentiate the expression (\ref{FT.gen.macro}) of the fluctuation theorem with respect to $\lambda_{\alpha}$ and ${A}_{\beta}$ and set $\{\lambda_{\gamma}=0\}$ and $\{{A}_{\gamma}=0\}$,
we find the Green-Kubo relations and the Onsager reciprocity relations:
\be
L_{\alpha\beta}= - \frac{1}{2}\frac{\partial^2 q}{\partial \lambda_{\alpha}\partial \lambda_{\beta}}(0;0)
=\frac{1}{2}\int_{-\infty}^{+\infty} \langle \left[ j_{\alpha}(t)-\langle j_{\alpha}\rangle\right]
\left[ j_{\beta}(0)-\langle j_{\beta}\rangle\right]\rangle_{\rm eq} \; dt = L_{\beta\alpha}
\label{ORR}
\ee

If we carry out a similar reasoning for the higher-order coefficients, we obtain
the third-order response tensor \cite{AG04,AG05,AG06}
\begin{equation}
M_{\alpha\beta\gamma} \equiv  
\frac{\partial^3 q}{\partial \lambda_{\alpha}\partial {A}_{\beta}\partial{A}_{\gamma}}(0;0)
= \frac{1}{2}\left(R_{\alpha\beta,\gamma} + R_{\alpha\gamma,\beta}\right)
\label{M.diff3}
\end{equation}
with the quantities
\be
R_{\alpha\beta, \gamma}= - \frac{\partial^3 q}{\partial \lambda_{\alpha}\partial \lambda_{\beta}\partial{A}_{\gamma}}(0;0) = 
\frac{\partial}{\partial{A}_{\gamma}}\int_{-\infty}^{+\infty} \langle \left[ j_{\alpha}(t)-\langle j_{\alpha}\rangle\right]
\left[ j_{\beta}(0)-\langle j_{\beta}\rangle\right]\rangle_{\rm st} \; dt\Big\vert_{\{{A}_{\delta}=0\}}
\label{tensor.R}
\end{equation}
which are the derivatives of the nonequilibrium generalizations of the
Onsager coefficients $L_{\alpha\beta}$.  Similar results hold at fourth and higher orders \cite{AG04,AG05,AG06}.

\subsection{Connections to the nonequilibrium work relations}

Beside the fluctuation theorem for nonequilibrium steady states,
other relations have been obtained such as Jarzynski's nonequilibrium work theorem \cite{Jarzynski}.
These relations concern the work performed on a system by
varying in time some external control parameter $\lambda(t)$ from an initial value
$\lambda(0)=\lambda_A$ to a final one $\lambda(t_{\rm final})=\lambda_B$.
To be specific, the system can be supposed to be initially in the
canonical ensemble  
\be
 p_0(\bG) = \frac{1}{Z_A} \; 
\exp\left[ -\beta H(\bG,\lambda_A)\right]
\ee
where $H(\bG,\lambda)$ is the Hamiltonian of the total system including the small subsystem
of interest (e.g. a protein or a RNA molecule attached to beads or to an AFM) 
and its environment (e.g. the surrounding solution).
The work performed on the system is defined by
\be
{\cal W} \equiv  \int_A^B dt \; \dot\lambda \; \partial_{\lambda}H\left[ \bG(t),\lambda(t)\right] =
H(\bG_B,\lambda_B) - H(\bG_A,\lambda_A)
\label{work}
 \ee
 where $\bG_A$ and $\bG_B$ denote the phases of the system at the initial and final times 
 along the Hamiltonian trajectory.
Under such assumptions and following Jarzynski \cite{Jarzynski}, 
it is possible to derive, from Liouville's equation
for Hamiltonian systems,
the {\it nonequilibrium work theorem}:
  \be
\langle {\rm e}^{-\beta {\cal W}}\rangle = {\rm e}^{-\beta \Delta F}
\label{Jarzynski}
\ee
in terms of the difference $\Delta F=F_B-F_A= - k_{\rm B}T \ln(Z_B/Z_A)$ of the free energies calculated
at equilibrium for $\lambda=\lambda_B$ and $\lambda=\lambda_A$ respectively.
A remarkable result is that Jarzynski's equality (\ref{Jarzynski}) implies
Clausius inequality 
\be
\langle {\cal W} \rangle \geq \Delta F
\ee
because of the general inequality $\langle {\rm e}^x\rangle \geq {\rm e}^{\langle x\rangle}$.
If the work dissipated by the irreversible processes 
is defined as ${\cal W}_{\rm diss} \equiv {\cal W}-\Delta F$,
Eq. (\ref{Jarzynski}) takes the form
\be
\langle {\rm e}^{-\beta {\cal W}_{\rm diss}}\rangle = 1
\label{Jarzynski.Wdiss}
\ee

Crooks has obtained the other relation \cite{C99}:
\be
\frac{p_{\rm F}({\cal W})}{p_{\rm R}(-{\cal W})} = {\rm e}^{\beta({\cal W}-\Delta F)} 
= {\rm e}^{\beta{\cal W}_{\rm diss}} 
\label{Crooks}
\ee
where $p_{\rm F}$ is the probability density that the work (\ref{work}) takes the value $\cal W$
during the forward process $A\to B$, while $p_{\rm R}$ is the corresponding probability density 
for the reversed process $B\to A$. Equation (\ref{Crooks})
can be derived in the same Hamiltonian framework as Jarzynski's equality.
Indeed, by using Liouville's theorem $d\bG_A=d\bG_B$ and
by noticing that the work performed during the reversed process
is given by Eq. (\ref{work}) with $A$ and $B$ exchanged, we find that
\bea
p_{\rm F}({\cal W}) &=& \int d\bG_A \; 
\frac{1}{Z_A} \; {\rm e}^{-\beta H_A} \; \delta\left[{\cal W}-(H_B-H_A)\right] = \int d\bG_A \; 
\frac{1}{Z_A} \; {\rm e}^{-\beta (H_B-{\cal W})} \; \delta\left[-{\cal W}-(H_A-H_B)\right] \nonumber\\
&=& \frac{Z_B}{Z_A} \; {\rm e}^{\beta {\cal W}} \; \int d\bG_B \; \frac{1}{Z_B} \; 
{\rm e}^{-\beta H_B} \; \delta\left[-{\cal W}-(H_A-H_B)\right] = {\rm e}^{-\beta \Delta F} \; {\rm e}^{\beta {\cal W}} \; p_{\rm R}(-{\cal W})
\eea
Jarzynski's nonequilibrium work theorem is a consequence of Crooks' relation because
the average in Eq. (\ref{Jarzynski}) can be expressed in terms of the
probability density of the work during the forward process according to
\be
\langle {\rm e}^{-\beta{\cal W}}\rangle = \int d{\cal W} \; p_{\rm F}({\cal W}) \; {\rm e}^{-\beta{\cal W}}
= {\rm e}^{-\beta\Delta F} \underbrace{\int d{\cal W} \; p_{\rm R}(-{\cal W})}_{=1} =  {\rm e}^{-\beta\Delta F}
\ee
hence Eq. (\ref{Jarzynski}). 
Similar relations can be obtained for the other statistical ensembles
of Table II.  The microscopic Hamiltonian $H$ should be replaced by $H+PV$
in the isobaric-isothermal ensemble, by $H-\mu N$ in the grand canonical ensemble,
and by $H+PV-\mu N$ in the isobaric-isothermal-isopotential ensemble.
These relations can be used to obtain the differences 
of the corresponding thermodynamic potential from the statistics of the work
performed on the system during general nonequilibrium transformations.

Crooks' fluctuation theorem is closely related to
the so-called transient fluctuation theorem by Evans and Searles \cite{ES94,ES02}.

Both Eqs. (\ref{Jarzynski}) and (\ref{Crooks}) are also related to
the fluctuation theorem for nonequilibrium steady states of isothermal processes.
Indeed, we have shown that the quantity (\ref{Z}) has its generating function which satisfies $Q(1)=0$
so that we have the property that $\langle {\rm e}^{-{\cal Z}(t)}\rangle$ has a subexponential behavior
for $t\to\infty$.  Comparing with Jarzynski's identity (\ref{Jarzynski.Wdiss}),
we can thus identify the quantity ${\cal Z}(t)$
with the dissipated work in the long-time limit:
\be
{\cal Z} \simeq \beta \; {\cal W}_{\rm diss}
\ee
so that Eq. (\ref{d_iS.Z}) implies
\be
\frac{d_{\rm i}S}{dt}\Big\vert_{\rm st} = \frac{1}{T} \; \lim_{t\to\infty} \frac{\langle {\cal W}_{\rm diss}(t)\rangle}{t}
\ee
as expected, where $T$ denotes the temperature.
For the same reason, Crooks formula corresponds to
the stationary fluctuation theorem in the form (\ref{FT.prob.R})
with $\zeta t \simeq \beta \; {\cal W}_{\rm diss}(t)$
in systems where the external constraints
are stationary and the forward process is the same
as the reversed one.
In conclusion, Jarzynski's and Crooks' relations are
general results on the statistics of the work
dissipated by a system during a
nonequilibrium process and they are closely
related to the fluctuation theorem for
nonequilibrium steady state.  They
constitute constraints on the nonequilibrium
properties but they do not determine by themselves
the characteristic times of the nonequilibrium
processes as rates can do.  These nonequilibrium work relations
have important applications in the study of nanosystems.
They have recently been used to measure experimentally 
the folding free energies of proteins
and RNA molecules \cite{Phys.Today,CRJSTB05}.

 \section{Entropy production and dynamical randomness}
 \label{Info}
 
 A most remarkable recent result is that entropy production can be
 related to the forward and backward dynamical randomness of the trajectories or paths
 of the system as characterized by the entropies per unit time
 (\ref{dyn.entr}) and (\ref{TR.dyn.entr}) \cite{G04b}.  It is possible to show
 that the entropy production in a nonequilibrium steady state is given by
 the difference between these two dynamical entropies:
 \be
\boxed{\frac{d_{\rm i}S}{dt} = k_{\rm B} \; \left( h^{\rm R}-h\right) \geq 0}
 \label{fund.entr.prod}
 \ee
 for a given partition $\cal P$.  The right-hand side is always
 non-negative because the difference of both dynamical entropies is
 a relative entropy which is known to be non-negative \cite{W78}.
 
 Here below, this further large-deviation relationship is demonstrated for
 processes ruled by the master equation (\ref{master0}).
 If the trajectories of this process are sampled with the time interval $\tau$,
 the multiple-time probability of a path (\ref{path.tau}) is given by Eq. (\ref{proba.traj}).
 Using the definition (\ref{dyn.entr}), the entropy per unit time becomes
 \be
 h(\tau) = \left(\ln\frac{\rm e}{\tau}\right) \sum_{\rho,\omega,\omega'} P_{\rm st}(\omega) W_{\rho}(\omega\vert\omega') - \sum_{\rho,\omega,\omega'} P_{\rm st}(\omega) W_{\rho}(\omega\vert\omega') \ln W_{\rho}(\omega\vert\omega') + O(\tau)
 \label{h.tau}
 \ee
 This is a function of the sampling time $\tau$ because the process is continuous
 in time and the waiting times between the jumps are exponentially distributed \cite{G98,GW93}.
 This function characterizes the dynamical randomness of the random paths
 followed in the forward direction.
 
The time-reversed entropy per unit (\ref{TR.dyn.entr}) is given by
 an expression similar to Eq. (\ref{h.tau}) but with the exchange $\omega \leftrightarrow\omega'$
 in the transition rate appearing in the logarithm \cite{G04b}:
 \be
 h^{\rm R}(\tau) = \left(\ln\frac{\rm e}{\tau}\right) \sum_{\rho,\omega,\omega'} P_{\rm st}(\omega) W_{\rho}(\omega\vert\omega') - \sum_{\rho,\omega,\omega'} P_{\rm st}(\omega) W_{\rho}(\omega\vert\omega') \ln W_{\rho}(\omega'\vert\omega) + O(\tau)
 \label{hR.tau}
 \ee
 This other dynamical entropy characterizes the dynamical randomness of the time-reversed
 paths.  
 
 The difference between both dynamical entropies is 
\be
h^{\rm R}(\tau)  - h(\tau)  = \frac{1}{2} \sum_{\rho,\omega,\omega'} 
\left[ P_{\rm st}(\omega') W_{\rho}(\omega'\vert\omega) 
- P_{\rm st}(\omega) W_{-\rho}(\omega\vert\omega')\right] \; 
\; \ln\frac{P_{\rm st}(\omega') W_{+\rho}(\omega'\vert\omega)}{P_{\rm st}(\omega) W_{-\rho}(\omega\vert\omega')} +O(\tau)
\label{hR-h}
\ee
where we observe the same expression as in the definition (\ref{d_iS/dt}) of the
entropy production.  Hence we find Eq. (\ref{fund.entr.prod})
in the limit $\tau\to 0$ \cite{G04b}.

This fundamental result establishes a connection between the entropy production
and the dynamical randomness of the stochastic process.
We notice that both dynamical entropies should be equal, $h^{\rm R}=h$,
if the system was at equilibrium.  In contrast, a difference exists between them
in nonequilibrium steady states.  This shows that the dynamical randomness
of the time-reversed trajectories is higher than the one of the forward trajectories.
This breaking of time-reversal symmetry at the level of the dynamical randomness
has its origin in the boundary conditions imposed to maintain the nonequilibrium steady
state. As we have pointed out with Eq. (\ref{mu.noneq}) and explicitly showed
in Subsec. \ref{NESS}, these boundary conditions typically break the
time-reversal symmetry of the measure describing a nonequilibrium steady state.
This breaking manifests itself in the dynamical randomness of the process
as shown by Eq. (\ref{fund.entr.prod}).

This result can also be derived for deterministic systems where the entropy per unit
time is equal to the sum of positive Lyapunov exponents according to Pesin's formula
while the time-reversed entropy per unit time appears to be equal to
minus the sum of negative Lyapunov exponents.  In dynamical systems
which model thermostated systems, we thus recover the result that minus
the sum of all the Lyapunov exponents, i.e., the phase-space contraction rate,
should be interpreted as the entropy production.  However, the introduction
of the time-reversed dynamical entropy and Eq. (\ref{fund.entr.prod}) 
also apply to systems without Lyapunov exponents (for instance
in the so-called non-chaotic Hamiltonian systems made of polygonal billiards 
provided that they are open
and crossed by fluxes of particles between reservoirs).

Moreover, the formula (\ref{fund.entr.prod}) establishes a
fundamental relationship between entropy production
and the way information is processed by a nonequilibrium system.
According to the Shannon-McMillan-Breiman theorem \cite{CFS82}, 
the multiple-time probability of a path decays as
\be
\mu(\pmb{\omega})=\mu(\omega_0\omega_1...\omega_{n-1}) \sim {\rm e}^{-n \tau h({\cal P}) }
\label{SMB.thm}
\ee
for almost all the trajectories if the process is ergodic.  
In contrast, for a typical path of the process, the probability to find
the time-reversed path decays faster as
\be
\mu(\pmb{\omega}^{\rm R})=\mu(\omega_{n-1}...\omega_1\omega_0) 
\sim {\rm e}^{-n \tau h^{\rm R}({\cal P}) }
\label{TR.SMB.thm}
\ee
According to Eq. (\ref{fund.entr.prod}), the ratio of both multiple-time probabilities 
typically decays as
\be
\frac{\mu(\pmb{\omega}^{\rm R})}{\mu(\pmb{\omega})}
=\frac{\mu(\omega_{n-1}...\omega_1\omega_0)}{\mu(\omega_0\omega_1...\omega_{n-1})}
\sim {\rm e}^{-n \tau \left[h^{\rm R}({\cal P})-h({\cal P})\right] }
= {\rm e}^{-n \tau \frac{1}{k_{\rm B}}\frac{d_{\rm i}S}{dt} }
= {\rm e}^{-\frac{1}{k_{\rm B}}\Delta_{\rm i}^{n\tau}S }
\label{decay.ratio}
\ee
at a rate given by the entropy production.  This result shows that
nonequilibrium constraints on a system perform a {\it selection of trajectories}.
The probability of the typical trajectories decays more slowly than 
for the time-reversed trajectories.
Therefore, the process observed forward in time turns out to be more ordered than the
process observed backward because of the inequality $h < h^{\rm R}$
under nonequilibrium conditions.  It is remarkable that this effect of 
nonequilibrium temporal ordering is active already close to equilibrium and
does not need a threshold in the nonequilibrium constraints as
it is the case for the nonequilibrium ordering into 
the Glansdorff-Prigogine dissipative structures.

Moreover, the entropy per unit time $h$ can be interpreted as a rate
of production of information by the process.  Indeed, it is the rate
at which information should be recorded during the observation
of the trajectories of the system in order to reproduce its time evolution.
We can conceive that this connection has implications for processes
involving biological information as in genetics.
In this regard, Eq. (\ref{fund.entr.prod}) already shows that
the nonequilibrium conditions naturally lead to a selection 
in the path or history of the system and to the generation of temporal order.
This temporal order manifests itself at the level of the information
about the time evolution of the paths or histories of the system
as referred to in Eq. (\ref{fund.entr.prod}).
This opens the way to possible feedback if a mechanism of polymerization
internal to the system can transfer temporal information
 into the spatial information contained 
in an aperiodic polymeric chain and vice versa.
For this to happen, nonequilibrium conditions are required
beside the existence of the aforementioned mechanism allowing the
transfer of information between a spatial encoding and 
the temporal development of the process.
Further consequences of Eq. (\ref{fund.entr.prod})
about the processing of information in physical processes
have been discussed in Ref. \cite{G04b}.

In this perspective, Eq. (\ref{fund.entr.prod}) provides
for the first time an interpretation of the second law
as a principle of ordering and this even for
systems close to equilibrium.
The usual interpretation of the second law is that
the spatial order tends to decrease during the time evolution
since the entropy increases and is interpreted as a measure
of the spatial disorder in the system.
Here, in contrast, we see that entropy production leads to
temporal order in the sense that the entropy per unit time
is smaller than the time-reversed one.
This temporal order finds its origin in the decrease of spatial order
in the total system composed of the subsystem and its environment
in agreement with the usual interpretation
of the second law.  The remarkable result is that the concepts
of dynamical entropies coming from dynamical systems theory
and chaos theory provide a framework in which
this new understanding of the second law 
as a principle of temporal ordering can be established.

 \section{Applications to nanosystems}
 \label{Nano}
 
 In this section, we apply the previous methods of nonequilibrium statistical mechanics
 to nanosystems.
 
 \subsection{Friction in sliding carbon nanotubes}
 
 Two carbon nanotubes embedded one inside the other can slide in a telescoping
 motion.  Indeed, they interact by van der Waals attractive forces which causes
 the system of double-walled nanotubes to increase the number of van der Waals
 bonds by decreasing their total length.  This leads to an oscillatory motion
 in a V-shaped potential for the distance between the centers of mass
 of both nanotubes.  However, the double-walled nanotube system
 has many degrees of freedom beside the distance between the centers of mass.
 Therefore, the dissipation of energy is possible from the translational motion
 in the aforementioned one-dimensional potential to the many vibrational degrees of freedom
 of each nanotubes.  This dissipation is caused by friction between both nanotubes
 \cite{Servantie}.
 
 \begin{figure}[ht]
\begin{center}
\includegraphics[scale=0.7]{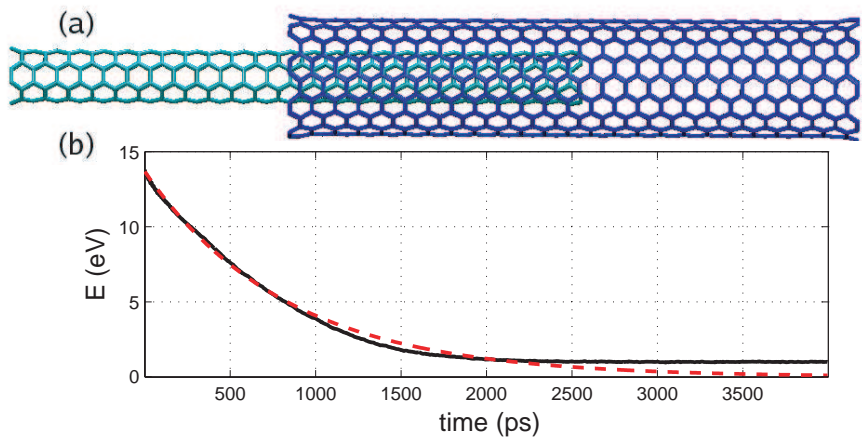}
\caption{(a) The double-walled carbon nanotubes (4,4)@(9,9) with $N_1=400$ and $N_2=900$ carbon atoms; (b) Dissipation of the energy $E$ of the one-degree-of-freedom translational motion of
the centers of mass with respect to the resting configuration at $T=300$ K.  
The nonvanishing final energy is due to the heating. The dashed line is the exponential decay
of the energy at the rate $2\zeta/(3\mu)$ predicted by the model (\ref{Langevin.eq})
for a V-shaped potential $V(r)\simeq F\vert r\vert$ \cite{Servantie}.}
\label{fig8}
\end{center}
\end{figure}

 The dynamics of the whole system and the irreversible property of friction can be
 studied by molecular-dynamics simulation from the total Hamiltonian
 \be
H=T^{(1)}+T^{(2)}+U_{\rm TB}^{(1)}+U_{\rm TB}^{(2)}+ \sum_{i=1}^{N_1}\sum_{j=1}^{N_2} 
U_{\rm LJ}\left(\Vert\mathbf{r}_i^{(1)}-\mathbf{r}_j^{(2)}\Vert\right)
\ee
 where $T^{(1)}$ and $T^{(2)}$ are respectively the kinetic energies of the inner and outer nanotubes 
while $U_{\rm TB}^{(1)}$ and $U_{\rm TB}^{(2)}$ 
are the Tersoff-Brenner potentials of both nanotubes \cite{Brenner}.
The positions and momenta of the carbon atoms of both nanotubes are denoted by 
$\{ \mathbf{r}_i^{(a)}\}_{i=1}^{N_a}$ and $\{ \mathbf{p}_i^{(a)}\}_{i=1}^{N_a}$ with $a=1$ (resp. $a=2$) 
for the inner (resp. outer) tube.
The kinetic energies are given by $T^{(a)}=(1/2m)\sum_{i=1}^{N_a} (\mathbf{p}_i^{(a)})^2$,
where $m=12$ amu is the mass of a carbon atom.
The intertube potential is modeled by a 6-12 Lennard-Jones potential.
Systems containing about 1300 carbon atoms can be simulated (see Fig. \ref{fig8}).
The study \cite{Servantie} shows that the translational motion is well described by a Langevin-type 
Newtonian equation
\be
\mu \; \ddot{r} = -\frac{dV(r)}{dr} -\zeta \; \dot{r} + F_{\rm fluct}(t)
\label{Langevin.eq}
\ee
where $r$ is the distance between the centers of mass and $\mu=N_1N_2m/(N_1+N_2)$ is their relative mass. $V(r)$ is the van der Waals potential of interaction between both nanotubes which has
a V-shape. The friction coefficient is given by Kirkwood formula
\be
\zeta \simeq \beta \int_0^{\tau} dt \; 
\left[\langle F_{\rm vdW}(t) F_{\rm vdW}(0) \rangle - \langle F_{\rm vdW} \rangle^2\right]
\label{Kirkwood}
\ee
in terms of the time integral of the time autocorrelation function of the van der Waals
force between both nanotubes.  This autocorrelation function should be integrated
in time until a cutoff time $\tau$ where the time integral reaches a plateau value
\cite{Kapral,Espanol,Levesque}.

Moreover, a fluctuating force is required to describe the last stages of energy dissipation
when the vibrational degrees of freedom cause small amplitude fluctuations in the 
motion of the centers of mass.  On time scales longer than the characteristic time of the
vibration (about 50 fs) the fluctuating force can be assumed to be a Gaussian white noise
so that the process can be described by the Fokker-Planck equation (\ref{Fokker-Planck.eq})
restricted to one degree of freedom and with the friction coefficient (\ref{Kirkwood}).
The Langevin equation (\ref{Langevin.eq})
and the corresponding Fokker-Planck equation (\ref{Fokker-Planck.eq}) 
describe the damping of the oscillations.
The period of oscillations is about 10 ps while the relaxation time of the order of 1000 ps.
Accordingly, the oscillations are underdamped in this system.
If the friction was larger, the inertial term in the left-hand side of Eq.
(\ref{Langevin.eq}) would be negligible and the motion would be overdamped
and ruled by the equation, $\zeta \; \dot{r}= -(dV/dr) + F_{\rm fluct}(t)$,
which would be the case if the system was embedded in a liquid for instance.

It is quite striking that such a nanosystem can already present an irreversible
property such as the relaxation toward a microcanonical state of equilibrium
under the effect of some kind of friction.  This friction is here present in a system of
about a thousand atoms.  

The preceding nanosystem undergoes a simple nonequilibrium process of
 relaxation toward a state of equilibrium because of friction.
 The reason is that energy is not injected to drive its motion
 as it is the case in the nanomotor of Ref. \cite{Zettl} which uses
 multiple-walled carbon nanotubes as shaft and electrostatic forces
 for driving.  In this case, rotational friction dissipates the energy injected by
 the external driving.

 \subsection{Biological nanomotors}
 
 Energy can also be injected by chemical fuel as in biological
 nanomotors such as F$_1$-ATPase which is able to drive the rotational motion
 of an actin filament or beads glued to its shaft \cite{Noji,Yasuda}.
 Such a nanosystem may be found in different chemical states depending
 on whether ATP is bonded to one of the catalytic sites of the protein complex
 or has been converted into its products ADP and inorganic phosphate P$_{\rm i}$.
 In each chemical state, the shaft of the motor is submitted to
 an internal torque deriving from some free-enthalpy potential $V_{\sigma}(\theta)$
 which depends on the angle $\theta$.  
 The motion of this angle is described
 by a Langevin-type equation
 \be
 \zeta \frac{d\theta}{dt} = -\frac{\partial V_{\sigma}}{\partial \theta} + \tau_{\rm ext} 
 + \tau_{\rm fluct}(t)
 \ee
 where $\tau_{\rm ext} $ is some external torque while $\tau_{\rm fluct}(t)$
 is the fluctuating torque due to the environment and which can be taken
 as a Gaussian white noise related to the friction $\zeta$ by the fluctuation-dissipation
 theorem.  Since the motor can be in different chemical states which change
 at each reactive event, the potential $V_{\sigma}$
 randomly jumps with the chemical state $\sigma$.
 The mechanical motion of the shaft
 is thus coupled to the chemical reaction in such mechano-chemical processes.
The master equation is here a set of coupled Fokker-Planck equations
 including terms for the description the random jumps between the discrete chemical states $\sigma$:
 \be
 \partial_t \; p_{\sigma}(\theta,t) + \partial_{\theta}J_{\sigma} = \sum_{\rho,\sigma'} 
 \left[ p_{\sigma'}(\theta,t) \; W_{\rho,\sigma',\sigma}(\theta) - 
 p_{\sigma}(\theta,t) \; W_{-\rho,\sigma,\sigma'}(\theta) \right]
 \ee
 where $p_{\sigma}(\theta,t)$ is the probability density to find the motor
 in the chemical state $\sigma$ and the angle $\theta$ at time $t$ and
 \be
 J_{\sigma} = -{\cal D} \; \partial_{\theta}p_{\sigma} + \frac{1}{\zeta}\left(-\partial_{\theta}V_{\sigma}+\tau_{\rm ext}\right) p_{\sigma}
 \ee
 with the diffusion coefficient ${\cal D}=k_{\rm B}T/\zeta$ \cite{Juelicher}.
 
 \begin{figure}[ht]
\begin{center}
\includegraphics[scale=0.55]{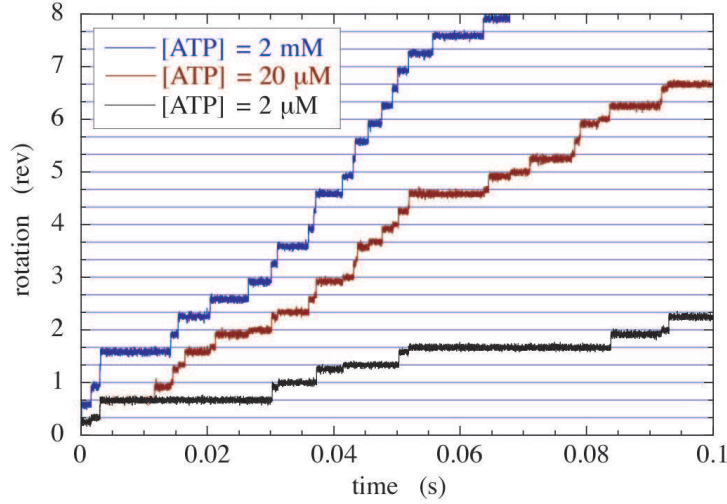}
\caption{Examples of trajectories showing the rotation of the
 shaft of the F$_1$-ATPase motor with a glued bead of diameter $d=40$ nm
 for different concentrations of ATP \cite{Gerritsma}.}
\label{fig9}
\end{center}
\end{figure}

 The transition rates can be determined by Arrhenius law
 so that Eq. (\ref{ratio.ATP}) can be satisfied.
 Such master equations fall into the framework described here above,
 which allows us to obtain the thermodynamics of the nanomotor,
 in particular, the entropy production and the average currents 
 which are here the mean rotation rate of the shaft
 and the mean rate of ATP consumption or synthesis.
 These average currents depend on the affinities or generalized
 thermodynamic forces which are here the difference of chemical potentials
 between the reactants and products as well as the external torque.
 The equilibrium state is defined as the situation where the ATP hydrolysis
 \be
 {\rm ATP} \rightleftharpoons {\rm ADP} \; + \; {\rm P}_{\rm i}
 \ee
 has reached equilibrium.  This corresponds to the equilibrium concentrations
 obeying
 \be
 \frac{[{\rm ATP}]_{\rm eq}}{[{\rm ADP}]_{\rm eq}[{\rm P}_{\rm i}]_{\rm eq}} = \exp\frac{\Delta\mu^0}{k_{\rm B}T} \simeq 4.5 \; 10^{-6} \; {\rm M}^{-1}
 \ee
with the standard chemical potential difference
$\Delta\mu^0 = -30.5 \; {\rm kJ/mole}=-7.3 \; {\rm kcal/mole}=-50 \; {\rm pN \; nm}$.
 For instance, the absence of the product immediately drives the system out of equilibrium
 and causes the rotation of the shaft.  
Figure \ref{fig9} depicts the nonequilibrium rotation of the shaft 
 of the F$_1$-ATPase motor with a glued bead of diameter $d=40$ nm
 simulated by a model with six chemical states \cite{Gerritsma}.  We observe that
 the rotation proceeds by apparent jumps.
 The fluctuations of the rotation of the motor are ruled by the fluctuation theorem,
 as shown elsewhere \cite{AG06b}.
   
\subsection{Fluctuation theorem in nonequilibrium chemical reactions}

\begin{figure}[ht]
\begin{center}
\includegraphics[width=18cm]{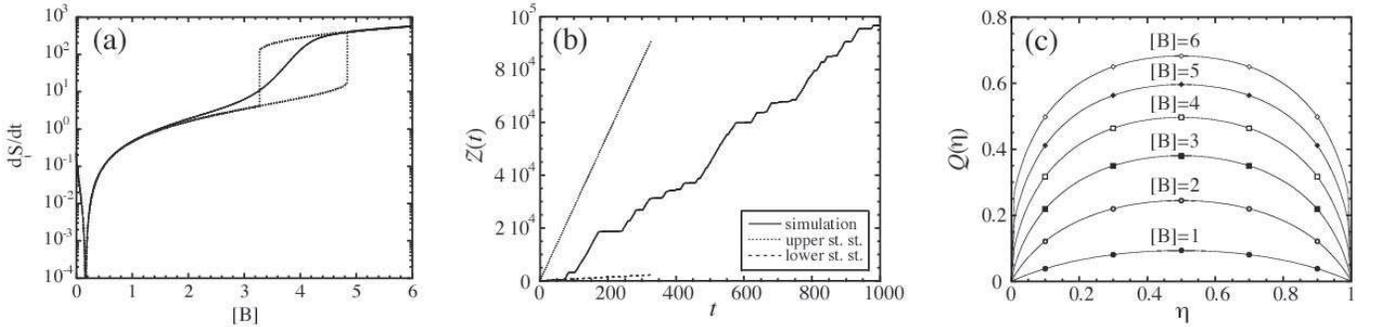}
\caption{(a) Entropy production versus the control concentration $[{\rm B}]$  
in the Schl\"ogl model (\ref{Schloegl.model}) 
 for $k_{+1}[{\rm A}]=0.5$, $k_{-1}=3$, and $k_{+2}=k_{-2}=1$, 
 obtained from the macroscopic kinetic equation (dashed lines) 
 and in the stochastic description with $\Omega=10$ (solid line).
(b) Stochastic time evolution of the quantity ${\cal Z}(t)$ 
in the Schl\"ogl model (\ref{Schloegl.model}) with 
$k_{+1}[{\rm A}]=0.5$, $k_{-1}=3$, $k_{+2}=k_{-2}=1$, $[{\rm B}]=4$, and $\Omega=10$.  
The increase of ${\cal Z}(t)$ fluctuates between the entropy production rate 
of the lower (long-dashed line) 
and upper (dashed line) stationary macroscopic concentrations.
(c) Generating function (\ref{Q}) of the fluctuating quantity ${\cal Z}(t)$ versus $\eta$ 
in the Schl\"ogl model (\ref{Schloegl.model}) for 
$k_{+1}[{\rm A}]=0.5$, $k_{-1}=3$, $k_{+2}=k_{-2}=1$, $[{\rm B}]=1,2,...,6$, and $\Omega=10$.
The thermodynamic equilibrium is located at 
$[{\rm B}]_{\rm eq}=\frac{1}{6}$ where $Q(\eta)=0$ \cite{G04a}.}
\label{fig10}
\end{center}
\end{figure}

Here, we illustrate the application of the fluctuation theorem to nonequilibrium chemical reactions
\cite{G04a}.
For this purpose, we consider Schl\"ogl's trimolecular reaction \cite{Schl71,Schl72}
\be
{\rm A} \, \underset{k_{-1}}{\overset{k_1}{\rightleftharpoons}}\, {\rm X} \ , \qquad
3 \, {\rm X} \, \underset{k_{-2}}{\overset{k_2}{\rightleftharpoons}}\, 2\, {\rm X} + {\rm B}
\label{Schloegl.model}
\ee
The reaction (\ref{Schloegl.model}) presents a phenomenon of bistability at the macroscopic level.
In the macroscopic description, the mean concentration of X shows
a hysteresis between two macroscopic steady states at low and high concentrations.  
In contrast, the stochastic description at
 the nanoscale takes into account the molecular fluctuations so that
 the number $X$ is randomly distributed over a range of values.
 The probability distribution $P_{\rm st}(X)$
in the nonequilibrium steady state corresponding to bistability presents two maxima
reminiscent of the two macroscopic steady states.
Accordingly, the mean value $\langle X\rangle$ interpolates between the two
macroscopic values.  A similar behavior happens for the entropy production 
depicted in Fig. \ref{fig10}a.
The quantity (\ref{Z}) can be computed along a random trajectory.
In the bistable regime, ${\cal Z}(t)$ alternatively increases
at the entropy production rates given by the lower and upper states
as seen in Fig. \ref{fig10}b.  Its generating function $Q(\eta)$
has the symmetry of the fluctuation theorem as shown in
Fig. \ref{fig10}c.  Close to equilibrium, the quantity
${\cal Z}(t)$ wildly fluctuates in time with only a slow increase
in contrast to far-from-equilibrium situations where the fluctuations
are negligible with respect to the increase, as seen 
in Fig. \ref{fig10}b.  The faster the increase
the larger the entropy production rate.  

Fully irreversible
reactions happen when the nonequilibrium constraints
are so high that some reversed reactions never occur.
This is the case if one of the reaction constant 
vanishes, $k_{-\rho}=0$, although the constant of
the reversed reaction does not, $k_{\rho}\neq0$.
In such fully irreversible reactions, the entropy production
is infinite, which can be understood as the limiting case
where the quantity ${\cal Z}(t)$ steeply increases very far from
the equilibrium.

 \subsection{Nonequilibrium chemical clocks at the nanoscale}
 
\begin{figure}[ht]
\centerline{\includegraphics[width=18cm]{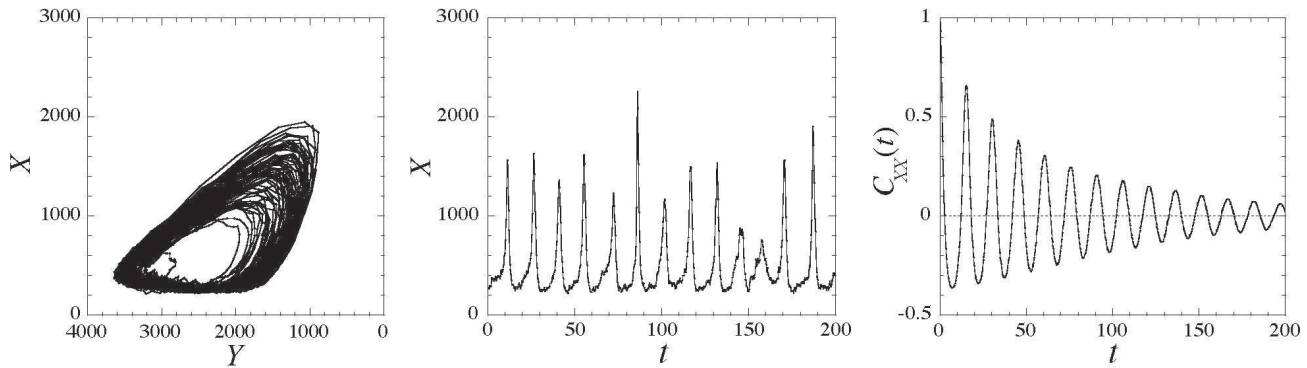}}
\caption{Simulations of the Brusselator in the oscillatory regime by Gillespie's algorithm for
the parameter values $k_1=0.5$, $k_2=1.5$, $k_3=k_4=1$ and the extensivity
parameter $\Omega=1000$.  The left-hand plot shows the phase portrait
in the plane of the number of molecules of species X and Y.  
The middle plot shows the number of molecules X as a
function of time.  The right-hand plot shows the autocorrelation function of the number of
molecules X and its decay due to the molecular fluctuations \cite{G02}.}
\label{fig11}
\end{figure}

 In this subsection, we consider a far-from-equilibrium chemical reactions which is fully irreversible
 so that the entropy production is here infinite.  This reaction is the abstract model of chemical clock
 known under the name of the {\it Brusselator} \cite{PrigLef68}:
\be
{\overset{k_1}{\longrightarrow}}  \ {\rm X}
\ , \qquad {\rm X} \ {\overset{k_2}{\longrightarrow}}
 \ {\rm Y} \ , \qquad 2\, {\rm X}+{\rm Y} \ {\overset{k_3}{\longrightarrow}}  \ 3\, {\rm X} \ , \qquad
{\rm X} \ {\overset{k_4}{\longrightarrow}}
\label{Brusselator}
\ee
 This reaction is supposed to evolve in a homogeneous system.
 Because of the molecular fluctuations, it is described by Nicolis' master equation
which can be simulated by Gillespie's numerical algorithm \cite{Gillespie76,Gillespie77}.
 At the macroscopic level, the Brusselator presents an oscillatory regime
 beyond a Hopf bifurcation giving birth to a limit cycle,
 which is a periodic solution of the macroscopic kinetic equations.
 At the nanoscale, such a limit cycle is noisy so that the successive oscillations
 are no longer perfectly correlated in time.  In order to measure the effect
 of noise, we can compute the time autocorrelation of the number of molecules
 of one species as depicted in Fig. \ref{fig11} \cite{G02}.
 We observe that the autocorrelation function presents oscillations
 which are damped exponentially at a rate which is inversely proportional to
 the extensivity parameter $\Omega$.  The larger the system, the longer the correlation
 time of the oscillations.  Accordingly, the oscillations should not be expected
 to remain correlated if the system is too small.  
 It turns out that the oscillations remain
 correlated down to systems containing about one hundred molecules,
 which corresponds to the nanoscale \cite{G02}.  Therefore, we may expect that
 chemical clocks exist down to the nanoscale in agreement
 with the experimental observation of such chemical clocks
 in heterogeneous catalysis at the nanoscale \cite{Kruse}.
 
 \section{Conclusions}
 \label{Concl}
 
 As the previous examples show, nonequilibrium statistical mechanics fruitfully applies
 to nanosystems in order to determine their thermodynamic properties.
 These new developments are challenging for statistical mechanics
 and are bringing fundamentally new concepts for the understanding of
 the irreversible processes and the second law.
 
 Nanosystems can be isolated, closed, or open.  In isolated nanosystems, the equilibrium state
 is microcanonical as it is the case in dynamical systems theory of Hamiltonian systems.
 However, many nanosystems are in contact with heat or particle reservoirs.
 In these cases, they are submitted to the motion of the degrees of freedom
 external to the system and, moreover, particles can enter and exit open systems.
 In this latter case, Liouville's equation leads to a hierarchy of equations for
 the different numbers of possible particles in the system as we have shown in
 Subsec. \ref{hierarchy}.
 
 The new advances concern the large-deviation dynamical relationships which have
 been obtained during the last fifteen years and the possibility of {\it ab initio}
 derivation of the entropy production from the underlying Hamiltonian dynamics
 \cite{G98,GDG00,DGG02}.
 
 First of all, the relaxation toward the state of thermodynamic equilibrium can be
 understood with dynamical systems theory and the concept of Pollicott-Ruelle
 resonance which naturally shows how the time-reversal symmetry can be broken
 at the statistical level of the description in systems with sensitivity to initial conditions.
 The Pollicott-Ruelle resonances give the relaxation rates which are intrinsic
 to the dynamics.  The associated eigenstates can describe the hydrodynamic modes
 which have only been considered at the phenomenological level of description and
 within kinetic theory.  The use of the new concepts allows us to avoid approximations
 and to convince oneself of the dynamical origin of these modes.
 The surprise has been that these modes are singular which justifies the use
 of coarse graining.  In particular, the diffusive modes can only be represented
 by their cumulative functions which form fractal curves.  The fractal dimension
 is determined in terms of the diffusion coefficient.  The singular character of
 the hydrodynamic modes turns out to be fundamental to obtain a non-vanishing
 entropy production.  Remarkably, the direct calculation starting from Gibbs'
 coarse-grained entropy leads to the entropy production expected
 from nonequilibrium thermodynamics and this even in systems
 with many particles \cite{DGG02}.  This cannot be a simple coincidence
 and suggests its generalization toward the other transport properties such as
 viscosity and heat conductivity.
 
Of great importance is the discovery of a series of fundamental large-deviation 
dynamical relationships in different approaches:
\begin{itemize}
\item the escape-rate theory \cite{GR89,GN90,GB95,DG95,VG03b};
\item the theory of the hydrodynamic modes of diffusion \cite{GDG01,GCGD01,CG02};
\item the fluctuation theorem \cite{ECM93,ES94,GC95,K98,C99,LS99,M99,G04a,AG04,AG05,AG06};
\item the relationship between entropy production and dynamical randomness 
in nonequilibrium steady states
\cite{G04b}.
\end{itemize}
These relationships all share the same structure that
an irreversible property is given as the difference between two large-deviation quantities
of the microscopic or mesoscopic dynamics.

In the escape-rate formalism which is concerned 
by nonequilibrium systems with absorbing boundary conditions,
the leading Pollicott-Ruelle resonance is the escape rate
which is proportional to the transport coefficient and given by Eq. (\ref{eq:escformula}) as the difference
between the sum of positive Lyapunov exponents and the Kolmogorov-Sinai entropy.
For diffusion, we get \cite{GN90,GB95}
\be
{\mathcal D}\left(\frac{\pi}{L}\right)^2 \simeq \gamma = \
\left( \sum_{\lambda_i>0} \lambda_i - h_{\rm KS}\right)_{L} 
\label{diff.esc.formula.bis}
\ee
and for viscosity or other transport coefficients \cite{DG95,VG03b}
\be
\eta\left(\frac{\pi}{\chi}\right)^2 \simeq \gamma = \
\left( \sum_{\lambda_i>0} \lambda_i - h_{\rm KS}\right)_{\chi} 
\ee
Very similar relationships are obtained for the hydrodynamic modes
of diffusion in the two-dimensional Lorentz gases 
where the dispersion relation $s_{\bf k}$ is also the leading
Pollicott-Ruelle resonance \cite{GCGD01}
\be
 {\mathcal D}k^2 \simeq -{\rm Re}\; s_{\bf k} = 
\lambda(D_{\rm H}) - \frac{ h_{\rm KS}(D_{\rm H})}{D_{\rm H}}  
\label{mode.formula}
\ee
by Eqs. (\ref{formula}) and (\ref{h(beta)}).
In these relations, the transport coefficient is given in terms
of the difference between the positive Lyapunov exponents and
the Kolmogorov-Sinai entropy per unit time which characterizes
the dynamical randomness.

For nonequilibrium steady states, the fluctuation theorem (\ref{FT.R})
\be
\zeta = R(-\zeta) - R(\zeta)
\label{FT.R.bis}
\ee
has again the same structure as Eqs. (\ref{diff.esc.formula.bis})-(\ref{mode.formula}) 
with an irreversible quantity such as $\zeta$ in the left-hand side
and the difference between two decay rates of probabilities in the right-hand side \cite{ECM93,GC95,K98,LS99,M99,G04a,AG04,AG05,AG06}.

Finally, the entropy production in nonequilibrium steady states 
can be given in terms of the difference between
two quantities characterizing the dynamical randomness of the underlying dynamics,
namely, the entropy per unit time of Kolmogorov and Sinai
and the newly introduced time-reversed entropy per unit time \cite{G04b}
\be
\frac{d_{\rm i}S}{dt} = k_{\rm B} \left( h^{\rm R} - h\right) 
\label{entr.prod.hR.h.bis}
\ee
Here again, the structure is analogue to the one of the
relations (\ref{diff.esc.formula.bis})-(\ref{mode.formula}) 
because the irreversible property which is the entropy production
is given as the difference between two dynamical entropies per unit time.
Furthermore, in Eq. (\ref{entr.prod.hR.h.bis}),
the entropy per unit time is at the same place as the Kolmogorov-Sinai entropy
in Eq. (\ref{diff.esc.formula.bis})-(\ref{mode.formula}).
The structure is also the same as in the fluctuation theorem (\ref{FT.R.bis}).

The relationships (\ref{diff.esc.formula.bis})-(\ref{entr.prod.hR.h.bis}) are indeed
large-deviation dynamical formulas for the statistical time evolution in nonequilibrium conditions.
They have the common feature of giving an irreversible property as the difference
of two large-deviation quantities such as the decay rates of multiple-time probabilities
or the growth rates of phase-space volumes.  Moreover, they are compatible
with Liouville's theorem.  This is also the case for Jarzynski nonequilibrium work theorem \cite{Jarzynski}
and the related Crooks fluctuation theorem \cite{C99}.  
The discovery of all these new large-deviation properties is 
a major advance in nonequilibrium statistical mechanics during the last fifteen years.

\vspace{0.3cm}

{\bf Acknowledgments.} 
The author thanks Bob Dorfman and Gr\'egoire Nicolis for support and encouragement in
this research, as well as David Andrieux and 
Jean-Sabin McEwen for their helpful comments. 
This research is financially supported
by the ``Communaut\'e fran\c caise de Belgique" 
(``Actions de Recherche Concert\'ees", contract No.~04/09-312), and by
the F.~N.~R.~S. Belgium (F.~R.~F.~C., contract No.~2.4577.04).

%%%%%%%%%%%%%%%%%%%%%%%%%%%%%%%%%%%%%%%%%%%%%%%%%%%%%%%%%%%%%%%%%%%%%

\end{document}